\documentclass[useAMS,usenatbib]{mn2e}

\usepackage{verbatim} 
\usepackage{natbib} 
\usepackage{amsmath} 
\usepackage{amsbsy}
\usepackage{amssymb}
\usepackage{mathrsfs} 
\usepackage{lscape} 
\usepackage{graphicx}
\usepackage{epstopdf}
\usepackage{deluxetable}
\usepackage{ctable} 
\usepackage{fixltx2e}

\title[Neutral hydrogen in $z=2-4$ galaxy halos]{Neutral hydrogen in galaxy halos at the peak of the cosmic star formation history}
\author[Faucher-Gigu\`ere et al.]{Claude-Andr\'e Faucher-Gigu\`ere$^{1}$, Philip F. Hopkins$^{2}$, Du\v{s}an Kere\v{s}$^{3}$, Alexander L.\newauthor Muratov$^{3}$, 
Eliot Quataert$^{4}$, and Norman Murray$^{5,6}$\vspace*{6pt}\\
$^{1}$Department of Physics and Astronomy and Center for Interdisciplinary Exploration and Research in Astrophysics (CIERA),\\ Northwestern University, 2145 Sheridan Road, Evanston, IL 60208, USA: cgiguere@northwestern.edu.\\
$^{2}$TAPIR, Mailcode 350-17, California Institute of Technology, Pasadena, CA 91125, USA.\\
$^{3}$Department of Physics, Center for Astrophysics and Space Science, University of California, San Diego, 9500 Gilman Drive,\\ La Jolla, CA 92093.\\
$^{4}$Department of Astronomy and Theoretical Astrophysics Center, University of California, Berkeley, CA 94720-3411, USA.\\
$^{5}$Canadian Institute for Theoretical Astrophysics, 60 St. George Street, University of Toronto, ON M5S 3H8, Canada.\\
$^{6}$Canada Research Chair in Astrophysics.
}

\begin{document}
\maketitle


\begin{abstract}
We use high-resolution cosmological zoom-in simulations from the FIRE project  to make predictions for the covering fractions of neutral hydrogen around galaxies at $z=2-4$. 
These simulations resolve the interstellar medium of galaxies and explicitly implement a comprehensive set of stellar feedback mechanisms. 
Our simulation sample consists of 16 main halos covering the mass range $M_{\rm h}\approx10^{9}-6\times10^{12}$ M$_{\odot}$ at $z=2$, including 12 halos in the mass range $M_{\rm h}\sim10^{11}-10^{12}$ M$_{\odot}$ corresponding to Lyman break galaxies (LBGs). 
We process our simulations with a ray tracing method to compute the ionization state of the gas. 
Galactic winds increase the HI covering fractions in galaxy halos by direct ejection of cool gas from galaxies and through interactions with gas inflowing from the intergalactic medium. 
Our simulations predict HI covering fractions for Lyman limit systems (LLSs) consistent with measurements around $z\sim2-2.5$ LBGs; these covering fractions are a factor $\sim2$ higher than our previous calculations without galactic winds. 
The fractions of HI absorbers arising in inflows and in outflows are on average $\sim50\%$ but exhibit significant time variability, ranging from $\sim10\%$ to $\sim90\%$. 
For our most massive halos, we find a factor $\sim3$ deficit in the LLS covering fraction relative to what is measured around quasars at $z\sim2$, suggesting that the presence of a quasar may affect the properties of halo gas on $\sim100$ kpc scales. 
The predicted covering fractions, which decrease with time, peak at $M_{\rm h}\sim10^{11}-10^{12}$ M$_{\odot}$, near the peak of the star formation efficiency in dark matter halos. 
In our simulations, star formation and galactic outflows are highly time dependent; HI covering fractions are also time variable but less so because they represent averages over large areas.  
\end{abstract}

\begin{keywords}
galaxies: formation --- galaxies: evolution --- galaxies: haloes --- quasars: absorption lines --- intergalactic medium --- cosmology: theory\vspace{-0.5cm}
\end{keywords}

\section{Introduction}
Inflows of gas from the intergalactic medium (IGM) are required to sustain the star formation rates that are measured across cosmic time \citep[e.g.,][]{2009ApJ...696.1543P, 2010ApJ...717..323B}. 
At the same time, powerful outflows are ubiquitous around star-forming galaxies at all redshifts \citep[e.g.,][]{2003ApJ...588...65S, 2005ApJ...621..227M, 2009ApJ...692..187W, 2010ApJ...717..289S, 2012ApJ...761...43N, 2012ApJ...751...51J, 2014ApJ...794..156R} and are believed to play a major role in slowing down the growth of galaxies, at least below the knee of the galaxy stellar mass function \citep[e.g.,][]{2003MNRAS.339..289S, 2005ApJ...618..569M, 2006MNRAS.373.1265O,2008MNRAS.387.1431D, 2011MNRAS.415...11D, 2013MNRAS.428.2966P, 2013MNRAS.436.3031V, 2014MNRAS.437.3529K}. 
In the last few years, probing these inflows and outflows has been one of the central goals of spectroscopic measurements of absorption by hydrogen and metals in the circumgalactic medium (CGM), both at low redshift and around the peak of the star formation history at $z\gtrsim2$ \citep[e.g.,][]{2006ApJ...651...61H, 2010ApJ...717..289S, 2011Sci...334..948T, 2012ApJ...750...67R, 2012ApJ...747L..26R, 2012ApJ...760..127M, 2013ApJ...776..136P, 2013ApJ...776L..18C, 2014arXiv1404.6540L}. 

In this work, we use a new suite of high-resolution cosmological zoom-in simulations 
to make predictions for neutral hydrogen around $z=2-4$ galaxies. This redshift interval includes the peak of the cosmic star formation history \citep[e.g.,][]{2007ApJ...670..928B}, where cosmological gas inflows and galactic winds are expected to be particularly important. 
Our simulations are part of the FIRE (Feedback in Realistic Environments) project\footnote{See project website: http://fire.northwestern.edu.} and are distinguished by the predictive power enabled by explicitly resolving the main structures in the interstellar medium (ISM) of individual galaxies. 
They explicitly model several of the main stellar feedback processes, including photoionization by massive stars, stellar winds from young and evolved stars, supernovae of Type I and II, and radiation pressure on dust grains. 
In a previous paper \citep[][]{2014MNRAS.445..581H}, we showed that our implementation of stellar feedback successfully reproduces the observationally-inferred relationship between stellar mass and dark matter halo mass (the $M_{\star}-M_{\rm h}$ relation) and the time-averaged star formation histories of galaxies below $\sim L^{\star}$ at all redshifts for which observations are currently available. 
Importantly, this success was achieved without tuning parameters of the calculations. This establishes the viability of our feedback model and is in itself a significant achievement relative to previous simulations that were tuned to reproduce the basic properties of galaxies at $z=0$ but failed at higher redshifts or outside the halo mass range for which the models were tuned \citep[e.g.,][but see Stinson et al. 2013\nocite{2013MNRAS.428..129S} and Ceverino et al. 2014\nocite{2014MNRAS.442.1545C} for other models that compare favorably to observations]{2007ApJ...667..170S, 2010Natur.463..203G, 2011ApJ...742...76G, 2011ApJ...728...51B, 2014MNRAS.tmp...38T}. 
\cite{2014arXiv1404.2613A} recently obtained broadly consistent results after implementing an approximate treatment of the same physics using a numerically distinct method. 

\begin{footnotesize}
\ctable[
  caption={{\normalsize Parameters for Simulations from Hopkins et al. (2014)}\label{tbl:sims_Hopkins}},center,star
  ]{lcccccccl}{
\tnote[ ]{Parameters describing the initial conditions for our simulations (units are physical): \\
{\bf (1)} Name: Simulation designation. Simulations ${\bf mx}$ have a main halo mass $\sim10^{x}$ M$_{\odot}$ at $z=0$.\\
{\bf (2)} $M_{\rm h}(z=2)$: Mass of the main halo at $z=2$. \\
{\bf (3)} $m_{\rm b}$: Initial baryonic (gas and star) particle mass in the high-resolution region. \\
{\bf (4)} $\epsilon_{\rm b}$: Minimum baryonic force softening (fixed in physical units past $z\sim10$; minimum SPH smoothing lengths are comparable or smaller). Force softening lengths are adaptive (mass resolution is fixed).\\
{\bf (5)} $m_{\rm dm}$: Dark matter particle mass in the high-resolution region.\\ 
{\bf (6)} $\epsilon_{\rm dm}$: Minimum dark matter force softening. As for baryons, the dark matter force softening lengths are adaptive and the minimum is fixed in physical units past $z\sim10$.
}
}{
\hline\hline
\multicolumn{1}{c}{Name} &
\multicolumn{1}{c}{$M_{\rm h}(z=2)$} &
\multicolumn{1}{c}{$m_{b}$} & 
\multicolumn{1}{c}{$\epsilon_{\rm b}$} & 
\multicolumn{1}{c}{$m_{\rm dm}$} & 
\multicolumn{1}{c}{$\epsilon_{\rm dm}$} & 
\multicolumn{1}{c}{Merger} & 
\multicolumn{1}{c}{Notes} \\ 
\multicolumn{1}{c}{\ } &
\multicolumn{1}{c}{M$_{\sun}$} &
\multicolumn{1}{c}{M$_{\sun}$} &
\multicolumn{1}{c}{pc} &
\multicolumn{1}{c}{M$_{\sun}$} &
\multicolumn{1}{c}{pc} &
\multicolumn{1}{c}{History} & 
\multicolumn{1}{c}{\ } \\ 
\hline
{\bf m09} & 1.3e9 & 2.6e2 & 1.4 & 1.3e3 & 30 & normal & isolated dwarf \\ 
{\bf m10} & 3.8e9 & 2.6e2 & 3 & 1.3e3 & 30 & normal & isolated dwarf \\
{\bf m11} & 3.8e10 & 7.1e3 & 7 & 3.5e4 & 70 & quiescent & -- \\
{\bf m12v} & 2.0e11 & 3.9e4 & 10 & 2.0e5 & 140 & violent & several $z<2$ mergers \\ 
{\bf m12q} & 5.1e11 & 7.1e3 & 10 & 2.8e5 & 140 & late merger & -- \\ 
{\bf m12i}  & 2.7e11 & 5.0e4 & 14 & 2.8e5 & 140 & normal & large ($\sim10\,R_{\rm vir}$) box \\ 
{\bf m13}  & 8.7e11 & 3.7e5 & 20 & 2.3e6 & 210 & normal & small group mass at $z=0$ \\
{\bf m14}  & 5.9e12 & 4.4e6 & 70 & 2.3e7 & 700 & normal & galaxy cluster at $z=0$ \\
\hline\hline\\
}
\end{footnotesize}

The integrated star formation properties of galaxies constitute only one test of galaxy formation theories. It is critical to compare model predictions with other observational constraints. 
Systematic Bayesian analyses of semi-analytic models (SAMs) constrained by the galaxy mass function reveal large degeneracies in best-fitting model parameters \citep[e.g.,][]{2013MNRAS.431.3373H, 2014MNRAS.443.1252L}, particularly for parameters that describe star formation and feedback processes \citep[e.g.,][]{2012MNRAS.421.1779L}. 
Cosmological simulations that rely on parametrized sub-resolution prescriptions for these processes (and that are therefore in some respects similar to SAMs) also hint at significant problems related to the degeneracies identified in SAMs. 
For example, the large-volume parameterized cosmological volume simulations of \cite{2013MNRAS.436.3031V} are in good agreement with the observationally-inferred cosmic star formation history, stellar mass function, stellar mass-halo mass relation, galaxy luminosity functions, and Tully-Fisher relation at $z=0$ and (for several of these quantities) at higher redshifts \citep[][]{2014MNRAS.tmp...38T}. 
However, the same simulations fail to reproduce the slope of the observed mass-metallicity relations of galaxies even after adding a `metal loading' factor to the galactic wind model \citep[][]{2013MNRAS.436.3031V, 2014MNRAS.tmp...38T}. 
Encouragingly, Ma et al. (in prep.) show that the explicit stellar feedback model in the FIRE simulations results in much better agreement with the observed stellar mass-stellar metallicity and stellar mass-gas phase metallicity relations.

Another limitation of parameterized simulations of galaxy formation simulations is that their feedback implementations have often relied on somewhat \emph{ad hoc} approximations. 
These include temporarily suppressing hydrodynamical interactions \citep[e.g.,][]{2003MNRAS.341.1253H, 2006MNRAS.373.1265O, 2013MNRAS.436.3031V, 2013MNRAS.432...89F, 2014MNRAS.442.3745M} or gas cooling \citep[e.g.,][]{2006MNRAS.373.1074S, 2007MNRAS.374.1479G, 2010MNRAS.407.1581S, 2011ApJ...742...76G} as kinetic or thermal energy is injected to model stellar feedback. 
Historically, approximations of this sort were necessary because the ISM of simulated galaxies and stellar feedback were not resolved in cosmological simulations. 
While of limited predictive power, these parameterized simulations were instrumental in demonstrating the range of problems in galaxy formation that stellar feedback can potentially solve, including the suppression of the galaxy stellar mass function \citep[e.g.,][]{2003MNRAS.339..312S, 2011MNRAS.415...11D} and the dispersal of heavy elements in the intergalactic medium \citep[e.g.,][]{2006MNRAS.373.1265O, 2010MNRAS.407.1581S}. 
Recent large-volume simulations \citep[][]{2014MNRAS.444.1518V, 2015MNRAS.446..521S} with parameterized feedback models also contain for the first time a mixture of spiral and elliptical galaxies in qualitative agreement with observations (but with significant quantitative discrepancies; Genel et al. 2014\nocite{2014MNRAS.445..175G}). 
Furthermore, it will remain impractical for years to come to resolve the ISM and explicitly implement stellar feedback in very large volume cosmological simulations, which are ultimately necessary to realize the promise of percent-level cosmology \citep[e.g.,][]{2011MNRAS.415.3649V} and make full use of large galaxy surveys as will provided by the Large Synoptic Survey Telescope\footnote{http://www.lsst.org/lsst\_home.shtml}, EUCLID\footnote{http://sci.esa.int/euclid} and WFIRST\footnote{http://wfirst.gsfc.nasa.gov}.

The issues with parameterized galaxy formation models nevertheless call for an approach that avoids parameter tuning and instead attempts to directly predict star formation and the effects of feedback processes. 
The FIRE simulations build on earlier work simulating star formation and stellar feedback in isolated galaxies. These calculations have been used to study the origin of the Kennicutt-Schmidt relation \citep[][]{2011MNRAS.417..950H}, the structure of the ISM and the properties of giant molecular clouds \citep[GMCs;][]{2012MNRAS.421.3488H}, galactic winds driven by stellar feedback \citep[][]{2012MNRAS.421.3522H}, gas inflow in gas rich disks \citep[][]{2012MNRAS.427..968H} and during galaxy mergers \citep[][]{2013MNRAS.430.1901H}, and a range of other problems. As part of the FIRE project, we have extended the methods introduced in this earlier work to zoom-in cosmological simulations. 
The explicit implementation of different stellar feedback processes 
is particularly important for studies of the CGM because spectroscopic measurements are sensitive to the phase structure of the CGM, and therefore to \emph{how} galactic winds are accelerated \citep[e.g.,][]{2012MNRAS.421.3522H, 2013ApJ...770...25A, 2013ApJ...777L..16B}.  
Thus, CGM observations offer the potential to break key degeneracies in galaxy formation models. 

Our focus in this paper is HI around high-redshift galaxies, motivated by the availability of observations to compare with.
Building on previous work by \cite{2003ApJ...584...45A} and \cite{2010ApJ...717..289S} on quasar-galaxy pairs, \cite{2012ApJ...750...67R} reported systematic measurements of neutral hydrogen in the CGM of $z\sim2-3$ LBGs in the Keck Baryonic Structure Survey (KBSS), including covering fractions for different ranges of HI columns. 
They found a covering fraction $f_{\rm cov}(>10^{17.2} ~{\rm cm^{-2}},~<R_{\rm vir})=30\pm14\%$ for Lyman limit systems (LLSs; $N_{\rm HI}>10^{17.2}$ cm$^{-2}$) within a projected virial radius $R_{\rm vir}$. 
\cite{2013ApJ...762L..19P} carried out an analogous experiment but in which the foreground galaxy is also a quasar (with dark matter halos more massive on average by a factor $\sim3$ than the LBGs in Rudie et al.'s sample) and found a surprisingly large covering fraction for LLSs of $\sim65\%$ within a projected virial radius of luminous quasars at $z\sim2$. 
Also using quasar-quasar pairs, \cite{2014arXiv1411.6016R} find that the covering fraction of LLSs within 100 proper kpc of foreground damped Ly$\alpha$ absorbers (DLAs) is consistent with that found around LBGs. 
While we do not consider the cosmological distribution of DLAs in this work, we will show that the broad peak in predicted LLS covering fraction within 100 proper kpc of foreground halos around $M_{\rm h}\sim10^{12}$ M$_{\odot}$ is consistent with DLAs arising in relatively massive halos, consistent with recent clustering measurements \citep[][]{2012JCAP...11..059F, 2014MNRAS.440.2313B}. 
In future papers, we will use our simulations to study other CGM statistics, including metal lines \citep[][]{2012ApJ...751...94R, 2014arXiv1403.0942T}, absorber kinematics \citep[][]{2010ApJ...717..289S, 2012ApJ...750...67R}, and the azimuthal dependence of absorber properties relative to galactic disks \citep[][]{2012ApJ...760L...7K, 2014ApJ...794..130B}.

Cosmological simulations of galaxy formation have previously been used to predict covering fractions of neutral hydrogen around high-redshift galaxies. 
\cite{2011MNRAS.412L.118F} used smooth particle hydrodynamics (SPH) simulations of two LBG-mass halos without galactic winds and focused on LLSs and DLAs. 
\cite{2011MNRAS.418.1796F} carried out a similar analysis of seven adaptive mesh refinement (AMR) cosmological simulations. 
The simulations of \cite{2011MNRAS.418.1796F} included weak stellar feedback, but that feedback was insufficient to drive strong galactic outflows. 
\cite{2011MNRAS.413L..51K} and \cite{2012MNRAS.424.2292G} obtained broadly consistent results also using AMR simulations. 
The CGM measurements of \cite{2012ApJ...750...67R} showed that the simulations of \cite{2011MNRAS.412L.118F} and \cite{2011MNRAS.418.1796F} under-predicted the covering fraction of LLSs within $R_{\rm vir}$ of $z\sim2$ LBGs by a factor $\sim2$ (though with marginal statistical significance, given the observational uncertainties on this measurement). 
\cite{2013ApJ...765...89S} presented a comprehensive analysis of the Eris2 simulation at $z\sim3$. 
In contrast to previous simulation studies of the CGM, Eris2 included stellar feedback strong enough to reproduce many properties of Milky Way progenitor galaxies \citep[][]{2011ApJ...742...76G}. 
Shen et al. reported a covering fraction of LLSs within $R_{\rm vir}$ of $27\%$. 
This value is in good agreement with the \cite{2012ApJ...750...67R} measurement around LBGs, though Shen et al. analyzed a redshift at the upper end of the redshift range $z\sim2-3$ probed by Rudie et al.'s observations. 
\cite{2014ApJ...780...74F} recently reported a study of  21 AMR simulations (with halo masses $M_{\rm h}\sim2\times10^{11}-4\times10^{12}$ M$_{\odot}$) focusing on LLSs at $z\sim2-3$. 
Although the simulations analyzed by \cite{2014ApJ...780...74F} included thermal feedback from stellar winds and supernovae, the authors note that their simulations over-predict the stellar masses of galaxies by a factor of $\sim2$ and result in winds with low wind mass loading factors. 
We compare our simulations in more details with those of \cite{2014ApJ...780...74F} in \S \ref{coverings_quant}; while our conclusions regarding HI covering fractions are qualitatively consistent with theirs, we find systematically higher covering fractions in our simulations with stronger stellar feedback. 
The present analysis improves on previous studies by simultaneously including a statistical sample of simulated halos as well as strong stellar feedback which reproduces a broad range of observed galaxy properties without the need for parameter tuning.

We begin by describing our simulations and analysis methods in \S \ref{sec:simulations}. 
In \S \ref{sec:covering_fractions}, we compile our main results on the covering fractions of neutral hydrogen for different intervals of column density and for different definitions of the maximum impact parameters. 
We also discuss the time and orientation variance of our predictions. 
We focus on our most massive halos and the evidence that luminous active galactic nuclei (AGN) may affect halo gas on $\sim 100$ kpc scales in \S \ref{sec:quasars}. 
We conclude and summarize our main results in \S \ref{sec:discussion}.
Appendices summarize convergence tests and the effects of feedback on covering fractions.

\begin{footnotesize}
\ctable[
  caption={{\normalsize Simulation Parameters for New High-Redshift Halos}\label{tbl:sims_new}},center,nostar
  ]{lccccccl}{
\tnote[ ]{
Parameters are defined as in Table \ref{tbl:sims_Hopkins}.
}
}{
\hline\hline
\multicolumn{1}{c}{Name} &
\multicolumn{1}{c}{$M_{\rm h}(z=2)$} &
\multicolumn{1}{c}{$m_{\rm b}$} & 
\multicolumn{1}{c}{$\epsilon_{\rm b}$} & 
\multicolumn{1}{c}{$m_{\rm dm}$} & 
\multicolumn{1}{c}{$\epsilon_{\rm dm}$}\\ 
\multicolumn{1}{c}{\ } &
\multicolumn{1}{c}{M$_{\sun}$} & 
\multicolumn{1}{c}{M$_{\sun}$} &
\multicolumn{1}{c}{pc} &
\multicolumn{1}{c}{M$_{\sun}$} &
\multicolumn{1}{c}{pc} \\ 
\hline
{\bf z2h830} & 5.4e11 & 5.9e4 & 9 & 2.9e5 & 143 \\
{\bf z2h650} & 4.0e11 & 5.9e4 & 9 & 2.9e5 & 143 \\
{\bf z2h600} & 6.7e11 & 5.9e4 & 9 & 2.9e5 & 143 \\
{\bf z2h550} & 1.9e11 & 5.9e4 & 9 & 2.9e5 & 143 \\
{\bf z2h506} & 1.2e12 & 5.9e4 & 9 & 2.9e5 & 143 \\
{\bf z2h450} & 8.7e11 & 5.9e4 & 9 & 2.9e5 & 143 \\
{\bf z2h400} & 7.9e11 & 5.9e4 & 9 & 2.9e5 & 143 \\
{\bf z2h350} & 7.9e11 & 5.9e4 & 9 & 2.9e5 & 143 \\
\hline\hline\\
}
\end{footnotesize}

\begin{figure*}
\begin{center}
\includegraphics[width=1.0\textwidth]{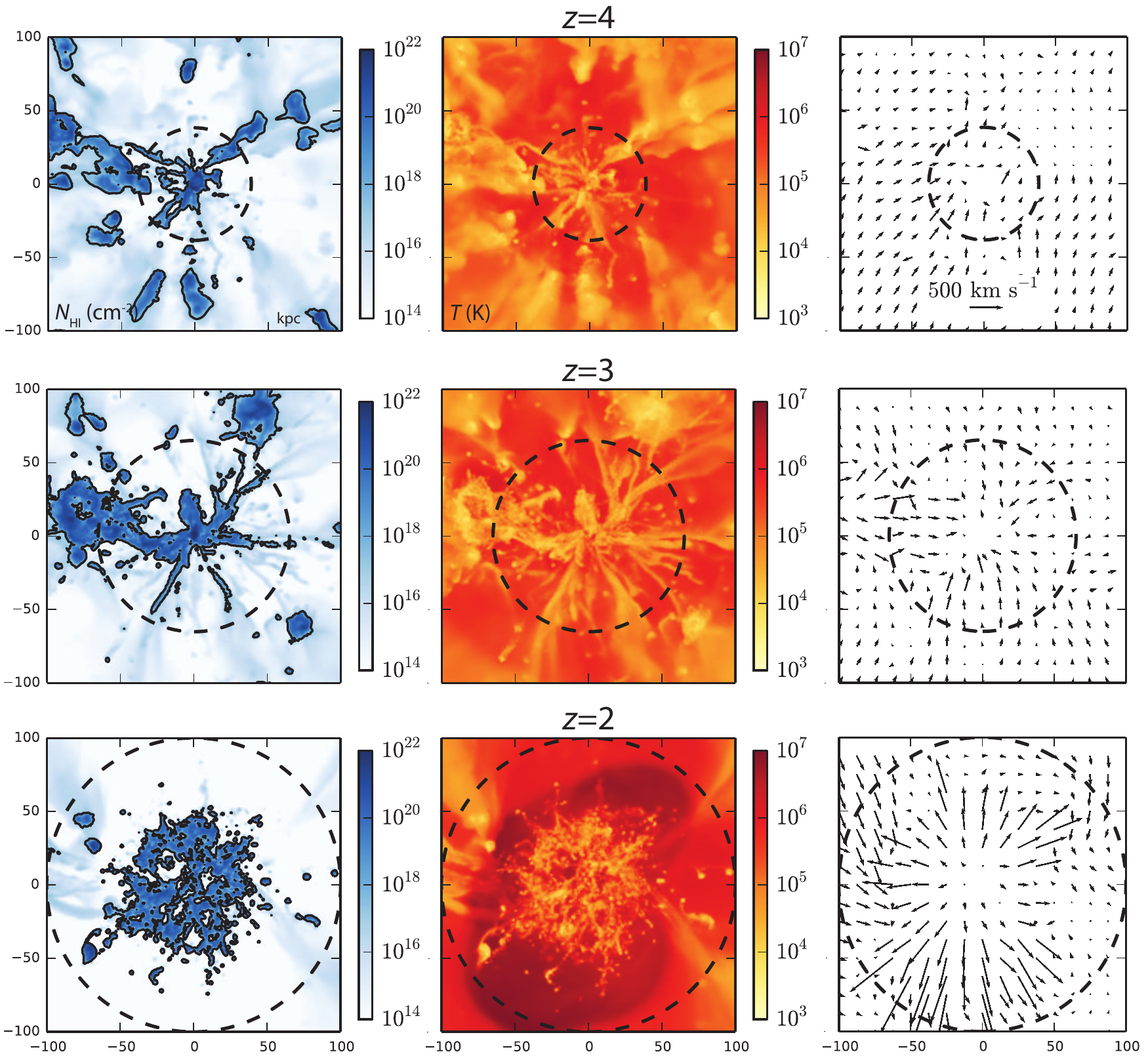}
\end{center}
\caption[]{Simulation {\bf z2h350} centered on the main halo from $z=4$ (top) to $z=2$ (bottom). 
\emph{Left:} Neutral hydrogen column density map. \emph{Center}: Temperature map (weighted by gas density squared to emphasize dense gas), \emph{Right:} Gas kinematics. 
The virial radius of the halo is indicated in each panel by the dashed circles and 
Lyman limit systems ($N_{\rm HI}>10^{17.2}$ cm$^{-2}$) are indicated by solid contours. 
At $z=4$ and $z=3$, the gas around the central galaxy is systematically inflowing from the IGM. 
At $z=2$, the galaxy is driving a powerful outflow with gas velocities $\sim500-1,000$ km s$^{-1}$, illustrating the time variability of CGM gas properties resulting from the time variability of the star formation history of our simulated halos. 
Length scales are consistent across rows and columns.
}
\label{fig:halo350_summary} 
\end{figure*}

\begin{figure*}
\begin{center}
\includegraphics[width=1.0\textwidth]{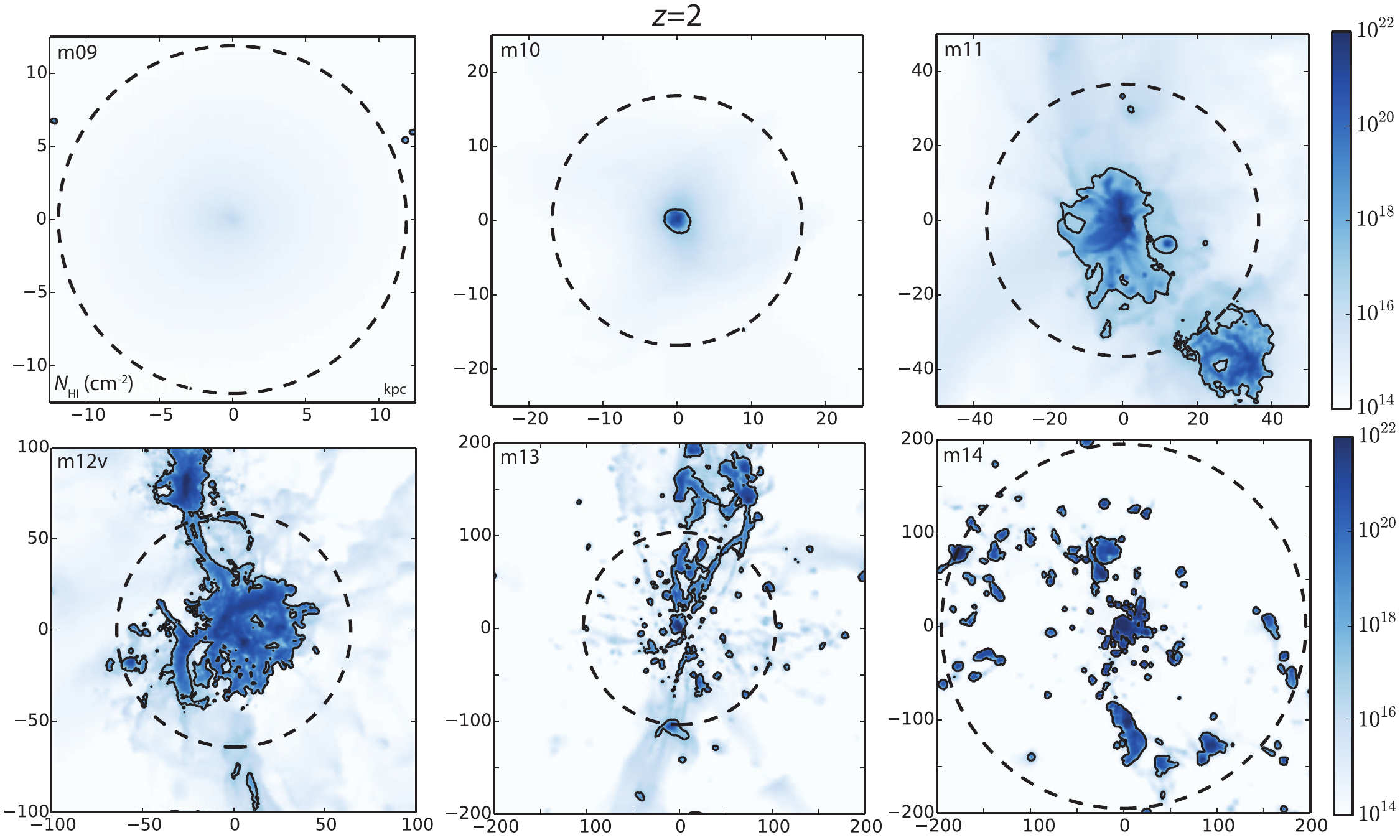}
\end{center}
\caption[]{HI maps for six of the simulations of Table \ref{tbl:sims_Hopkins} at $z=2$, spanning the halo mass range $M_{\rm h}=1.3\times10^{9}-5.9\times10^{12}$ M$_{\odot}$. 
The virial radius of each halo is indicated by the dashed circle and Lyman limit systems ($N_{\rm HI}>10^{17.2}$ cm$^{-2}$) are indicated by solid contours. 
The lowest mass halos, ${\bf m09}$ and ${\bf m10}$, have very small areas covered by dense HI.}
\label{fig:z2_HI_summary} 
\end{figure*}

\begin{figure*}
\begin{center}
\mbox{
\includegraphics[width=0.5\textwidth]{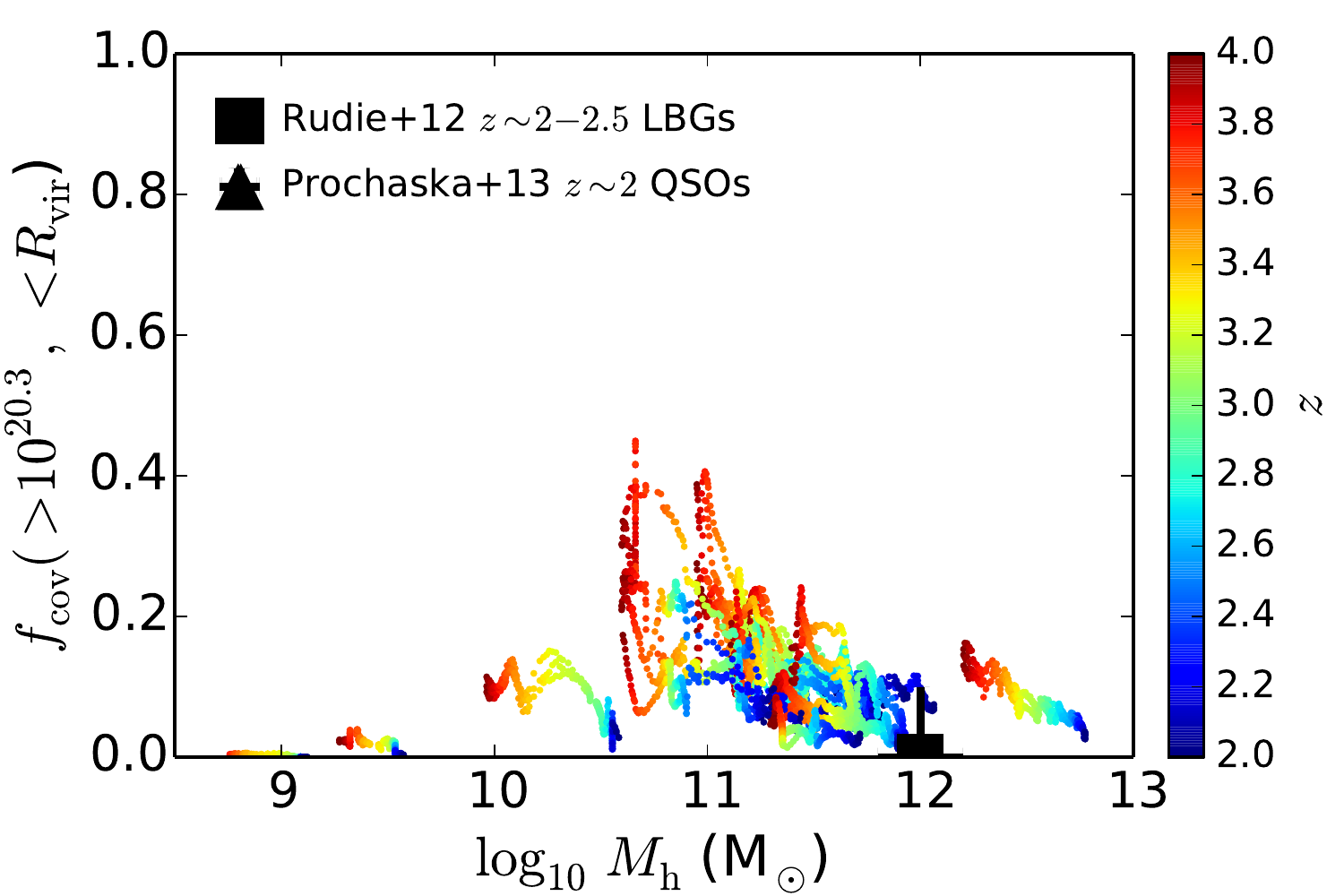}
\includegraphics[width=0.5\textwidth]{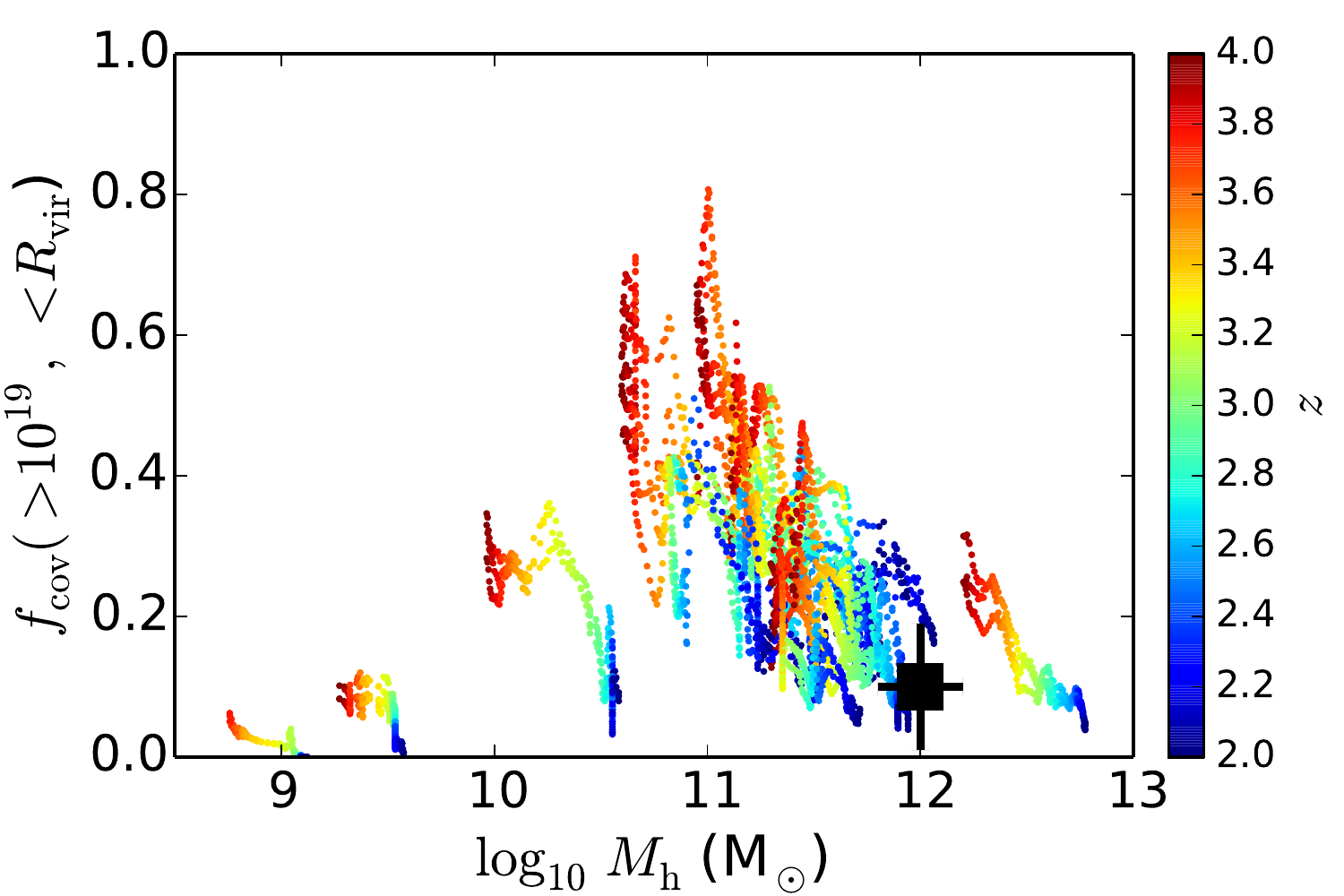}
}
\mbox{
\includegraphics[width=0.5\textwidth]{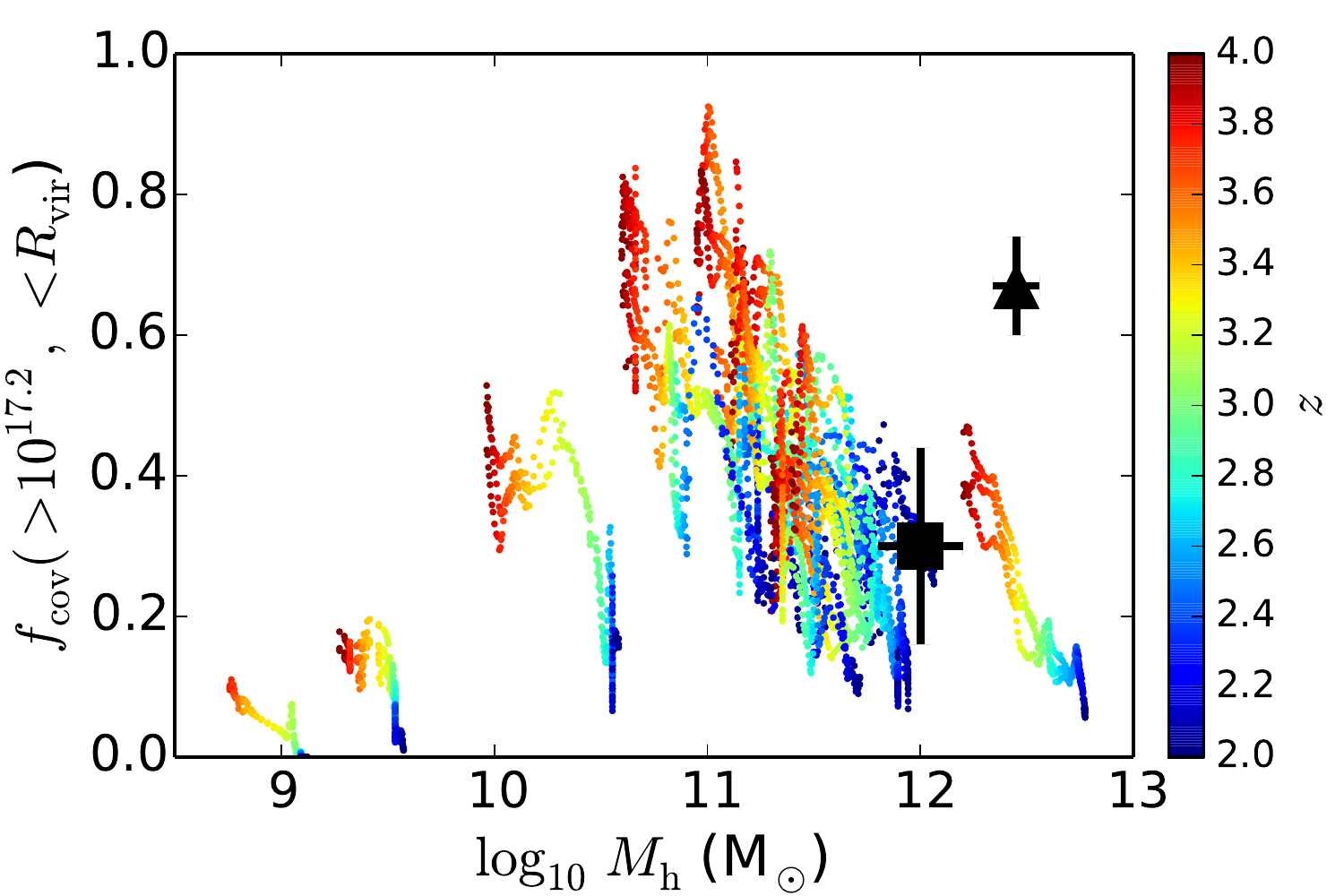}
\includegraphics[width=0.5\textwidth]{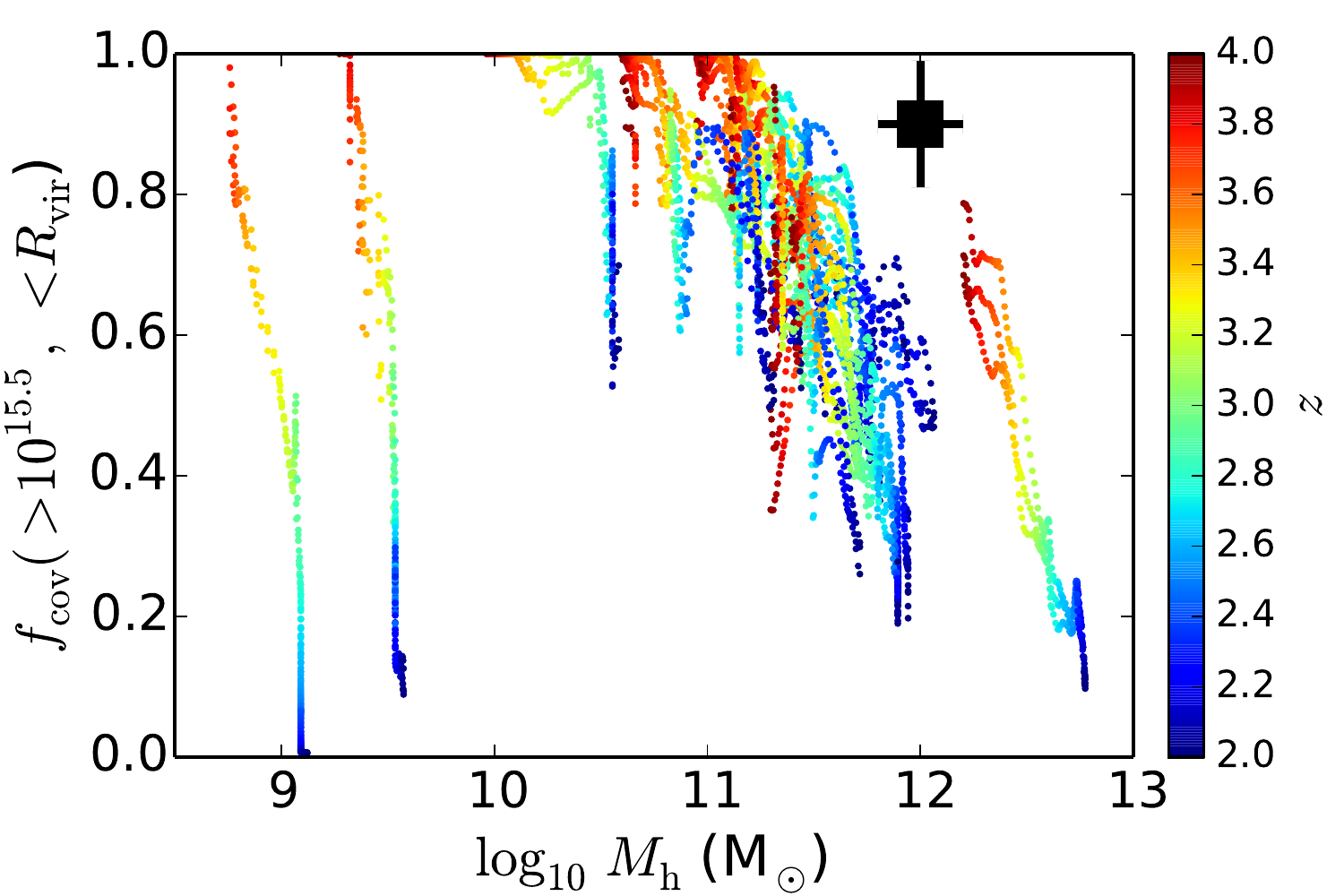}
}
\end{center}
\caption[]{HI covering fractions within a projected virial radius from our simulations for four different minimum $N_{\rm HI}$ values: $N_{\rm HI}>10^{20.3}$ cm$^{-2}$ (DLAs; top left), $N_{\rm HI}>10^{19}$ cm$^{-2}$ (SLLSs; top right), $N_{\rm HI}>10^{17.2}$ cm$^{-2}$ (LLSs; bottom left), and $N_{\rm HI}>10^{15.5}$ cm$^{-2}$ (bottom right). 
For each simulation, covering fractions are plotted as a function of dark matter halo mass for 100 time slices from $z=4$ to $z=2$, and for each time slice we plot the covering fractions for three orthogonal sky projections. 
Point color indicates the redshift of the snapshot. 
The oscillations in point tracks reflect the time variability of halo gas covering fractions (see \S \ref{sec:time_variability}). 
In each panel, the large black square symbol indicates the measurement around LBGs at $z\sim2-2.5$ from \cite{2012ApJ...750...67R}; the black triangle shows the LLS covering fraction measured around $\sim2$ quasars by \cite{2013ApJ...762L..19P}. 
For DLAs, SLLSs, and LLSs, the agreement between the model predictions and observations is excellent. 
For LLSs in particular, our predictions are a factor $\gtrsim2$ higher than in our previous calculations without strong stellar feedback \citep[][]{2011MNRAS.412L.118F}. 
For lower columns $N_{\rm HI}>10^{15.5}$ cm$^{-2}$, the observations appear to exceed the simulations by a factor of $\sim2$ (see text). 
The covering fraction $f_{\rm cov}(>10^{17.2}~{\rm cm}^{-2};~<R_{\rm vir})\approx64\%$ measured around luminous quasars at $z\sim2$ by \cite{2013ApJ...762L..19P} is much larger than predicted by any of our simulations with stellar feedback only. 
We discuss the implications for quasar-hosting halos in \S \ref{sec:quasars}.
}
\label{fig:covering_fractions_summary} 
\end{figure*}

\begin{figure*}
\begin{center}
\includegraphics[width=0.99\textwidth]{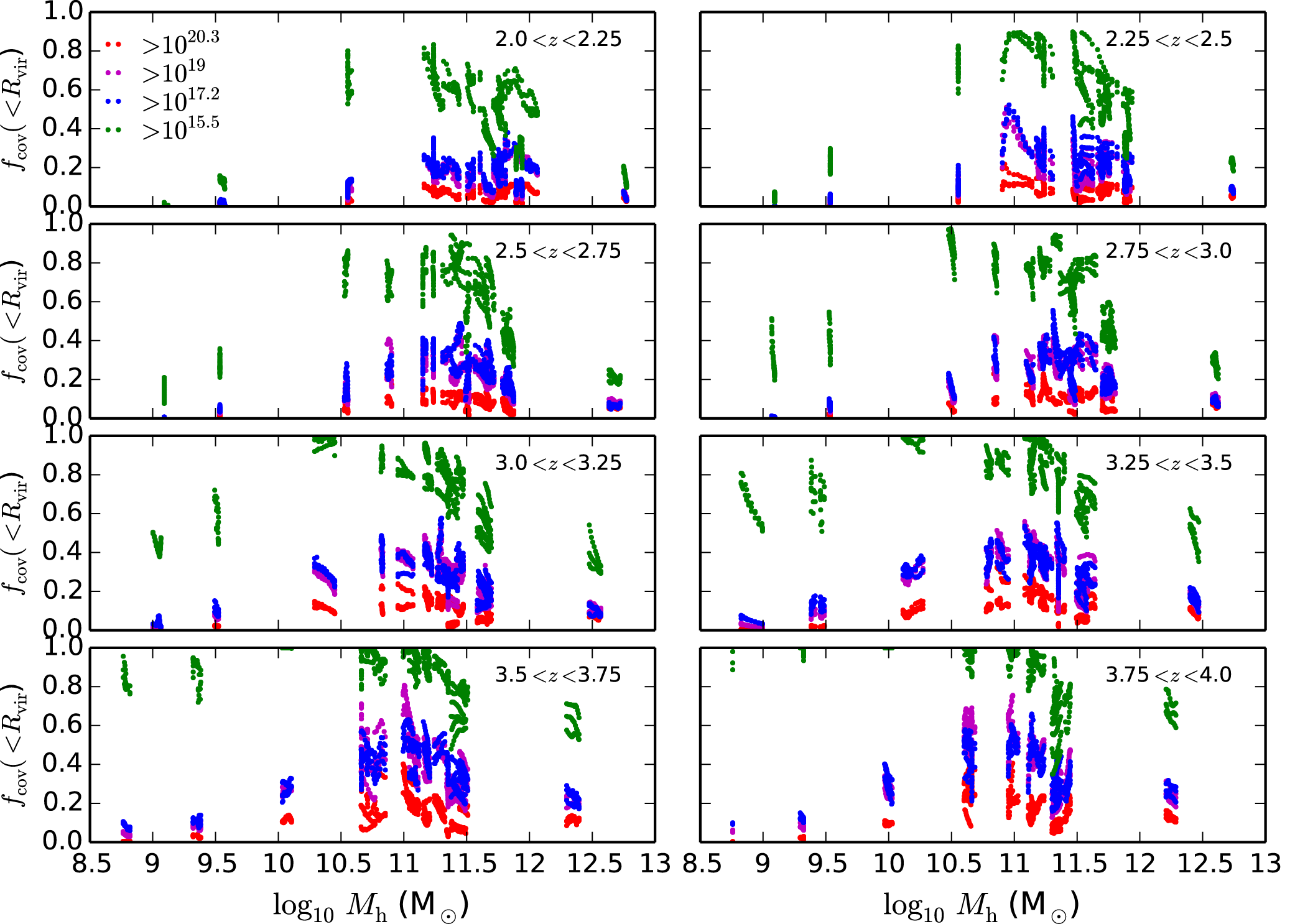}
\end{center}
\caption[]{Simulated covering fractions as in Figure \ref{fig:covering_fractions_summary} (within $R_{\rm vir}$) but in different panels for different redshift intervals to isolate the dependence on halo mass. 
For LLSs, the halo mass dependence of the covering fractions is relatively weak for $M_{\rm h}\sim10^{11}-10^{12}$ M$_{\odot}$, but the covering fractions of dense HI absorbers are distinctly lower for lower-mass halos ($M_{\rm h}\lesssim10^{10.5}$ M$_{\odot}$) at all redshifts considered.
}
\label{fig:fcov_vs_Mh_1Rvir} 
\end{figure*}

\begin{figure*}
\begin{center}
\includegraphics[width=0.99\textwidth]{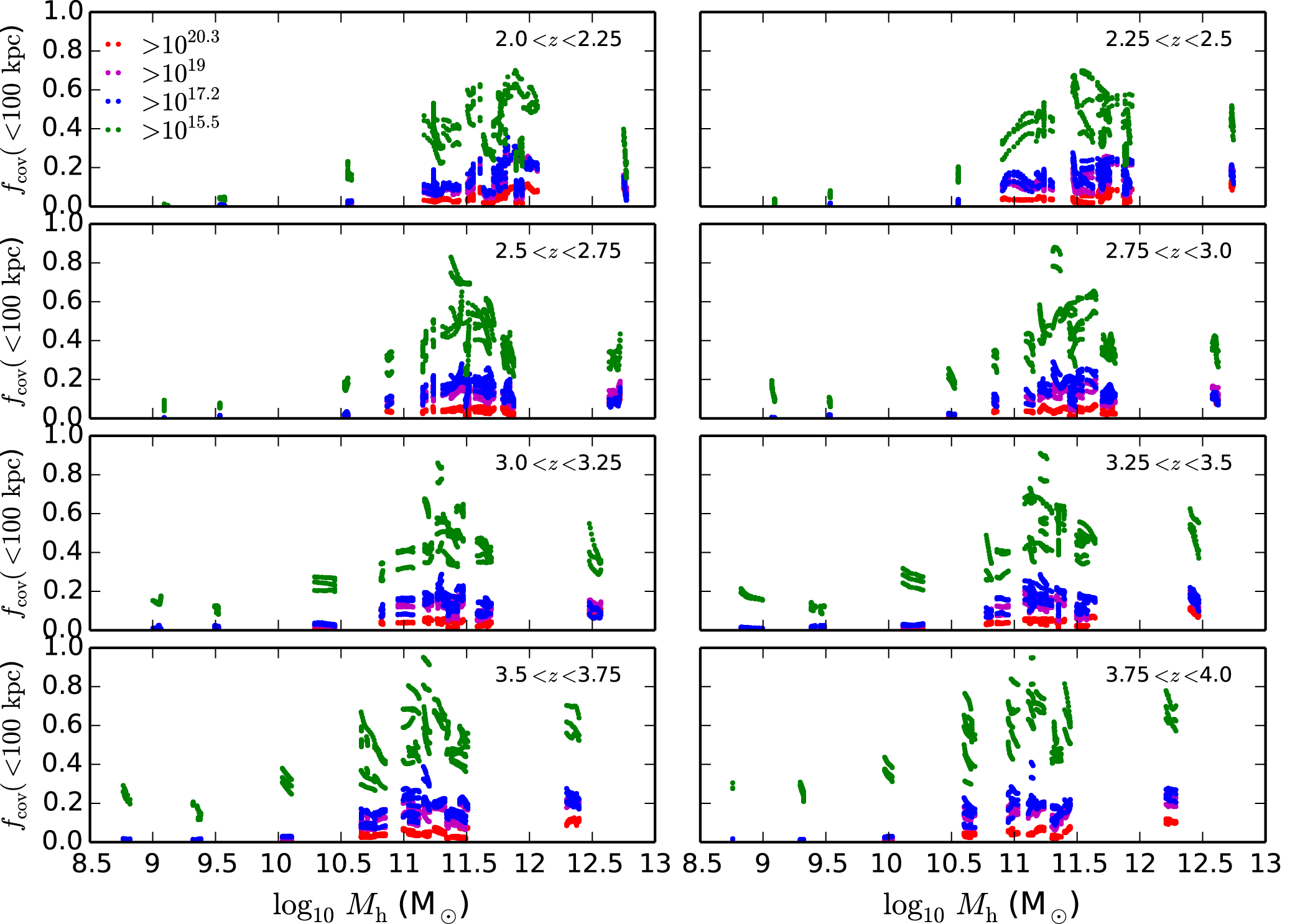}
\end{center}
\caption[]{Same as Figure \ref{fig:fcov_vs_Mh_1Rvir} but for covering fractions within a fixed impact parameter of 100 proper kpc.}
\label{fig:fcov_vs_Mh_Rmax} 
\end{figure*}

\begin{figure*}
\begin{center}
\mbox{
\includegraphics[width=0.5\textwidth]{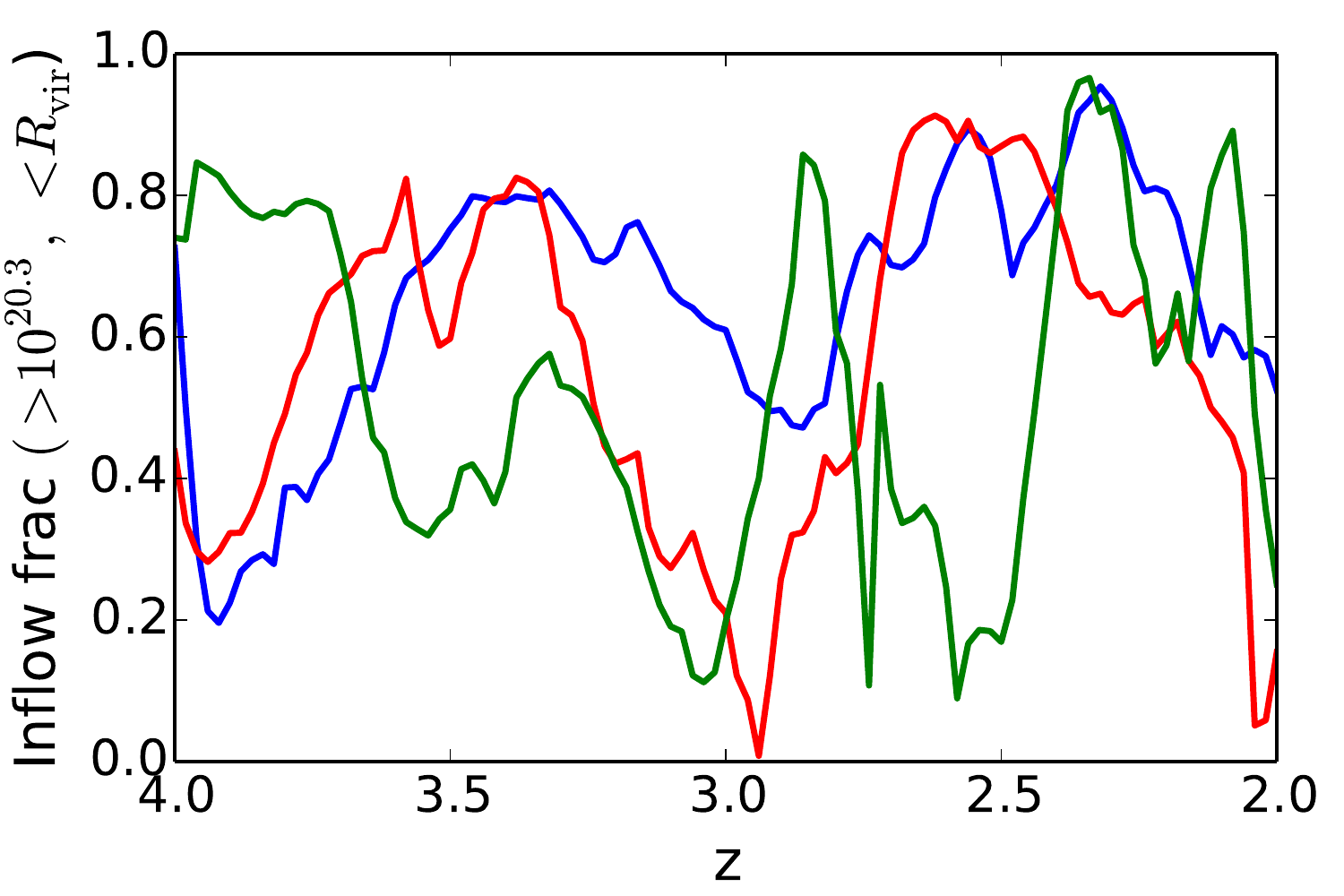}
\includegraphics[width=0.5\textwidth]{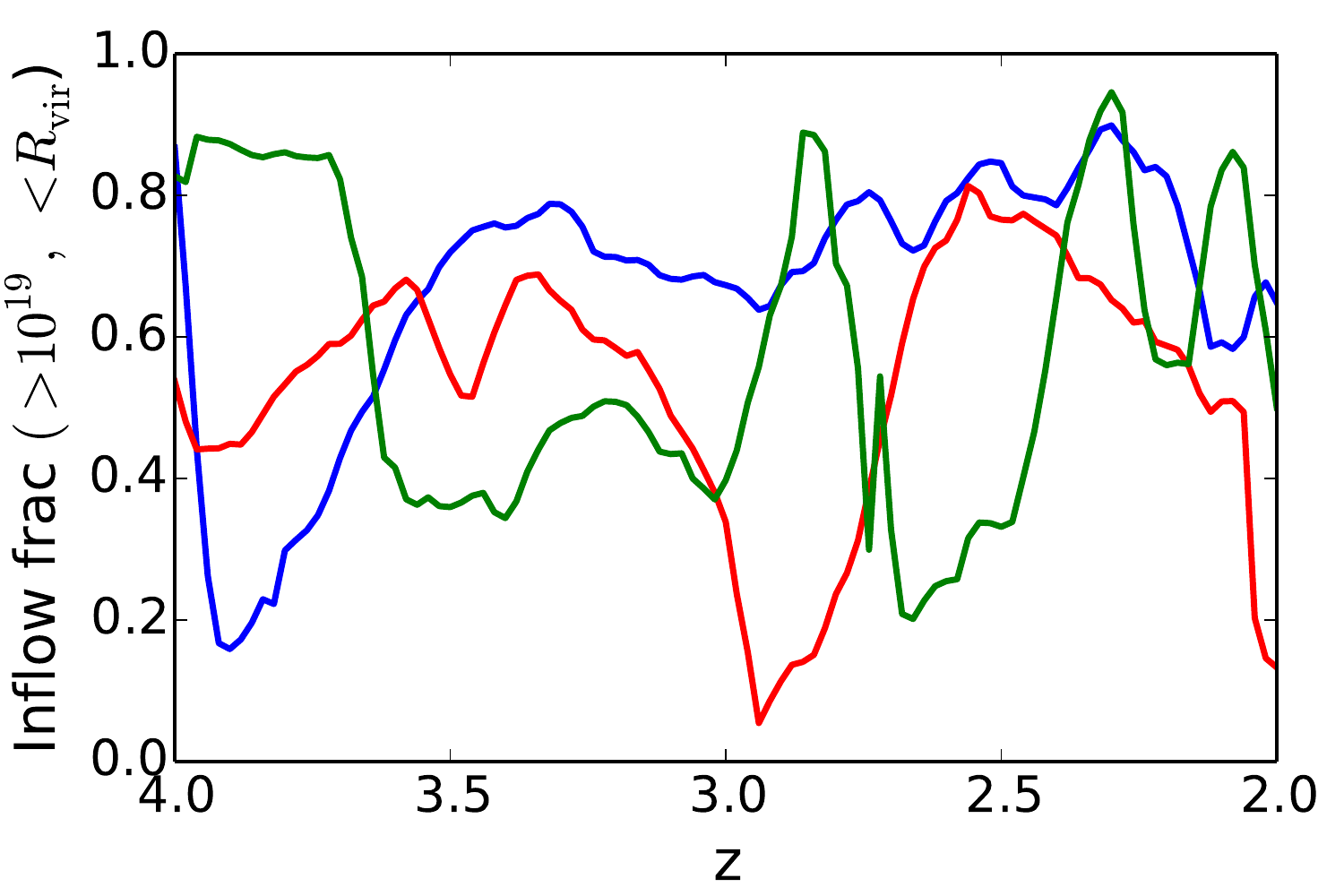}
}
\mbox{
\includegraphics[width=0.5\textwidth]{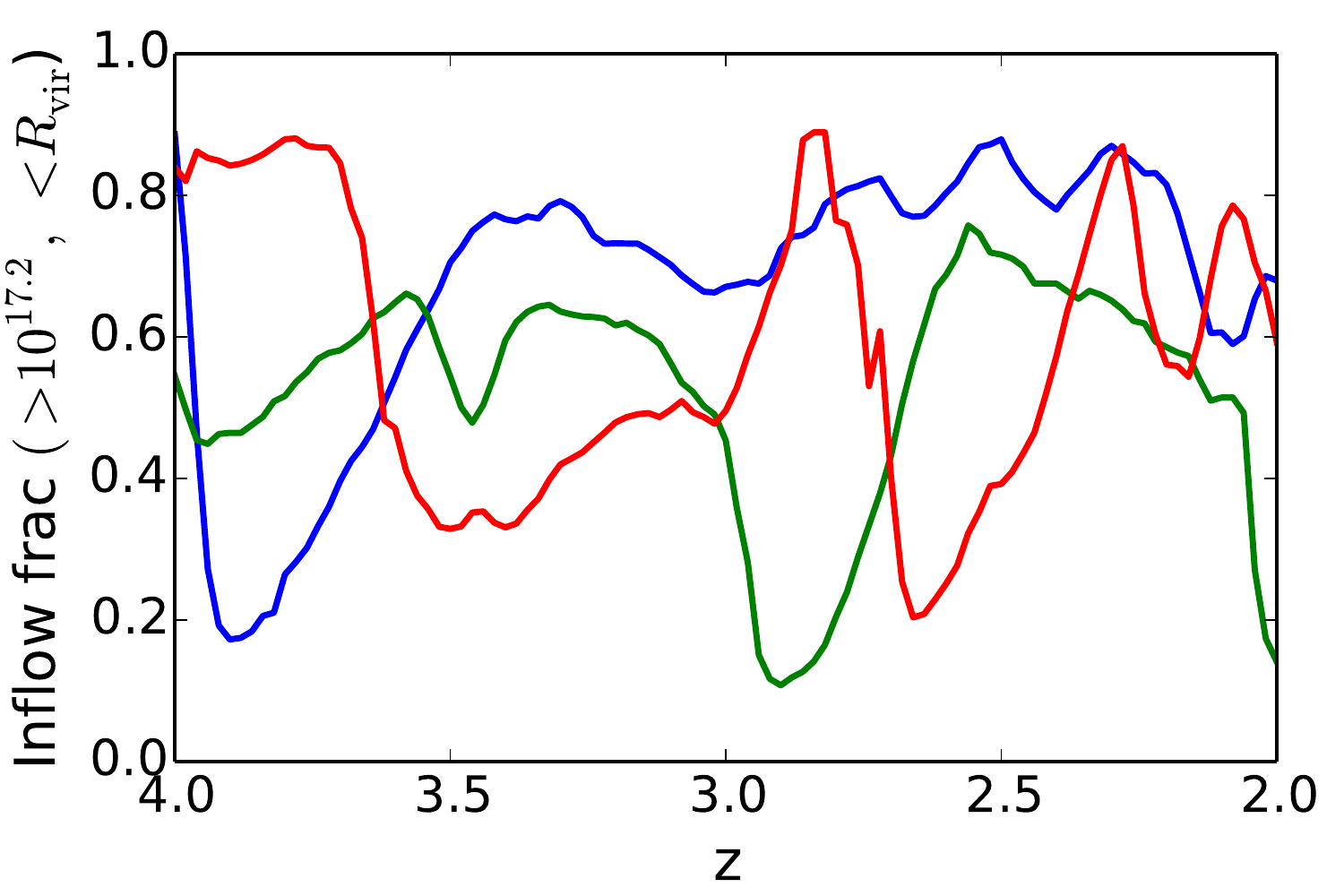}
\includegraphics[width=0.5\textwidth]{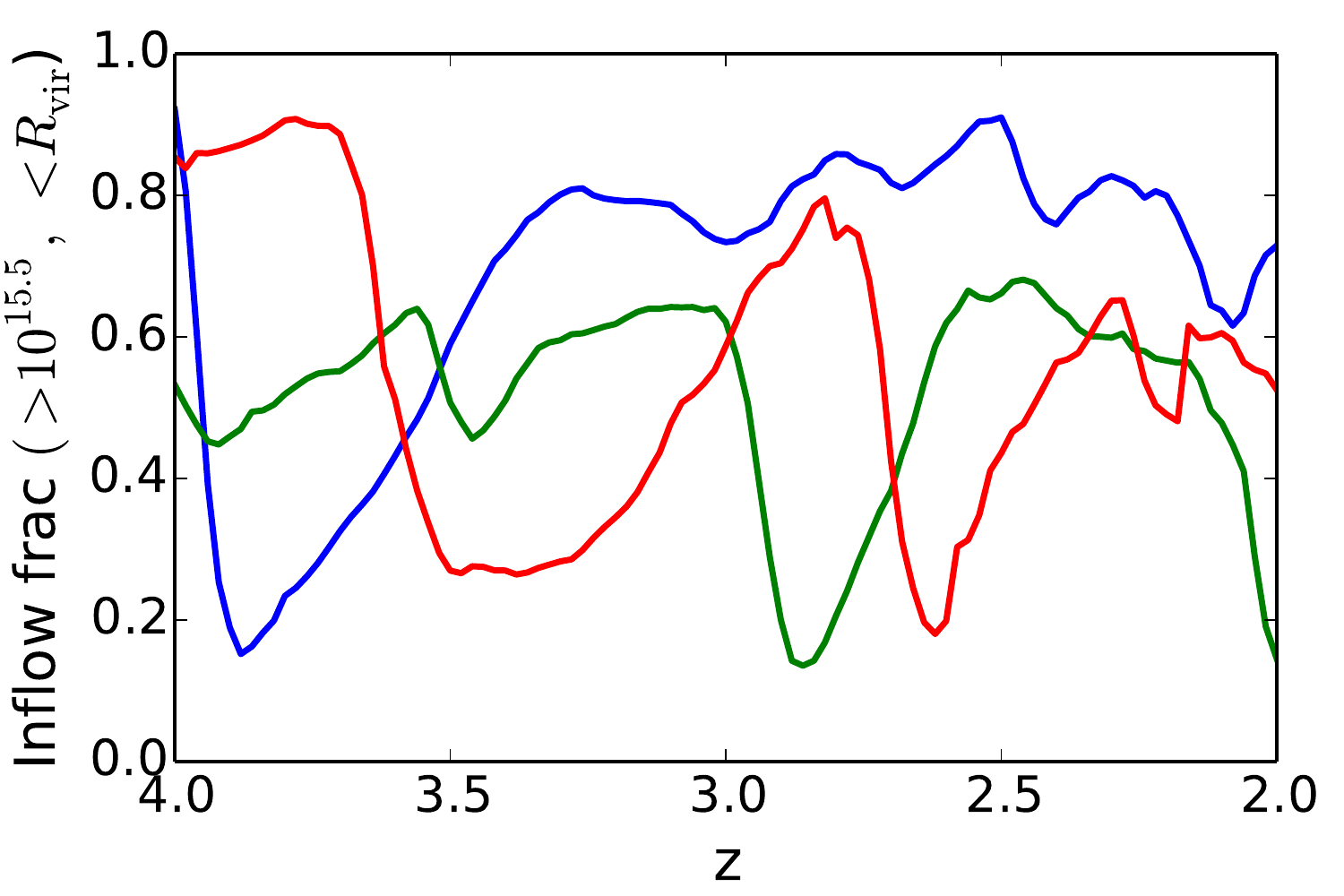}
}
\end{center}
\caption[]{
Fraction of covering fractions corresponding to inflowing gas (as determined by the radial component of the mass-weighted velocity for each pixel) within a projected virial radius as a function of redshift for the same four column density intervals as in Figure \ref{fig:covering_fractions_summary}.  
For clarity, we show only one sky projection for each simulation and show only representative simulations {\bf m12i}, {\bf z2h506}, and {\bf z2h350} (green, blue, and red, respectively).
}
\label{fig:inflow_fraction} 
\end{figure*}

\section{Cosmological simulations with stellar feedback}
\label{sec:simulations}
\subsection{Simulation details}
A full description of the numerical methods and physics included in our simulations is given in \cite{2014MNRAS.445..581H}; we summarize their main features here.

All simulations use the newly developed GIZMO\footnote{http://www.tapir.caltech.edu/$\sim$phopkins/Site/GIZMO.html} simulation code \citep[][]{2014arXiv1409.7395H} in ``P-SPH'' mode \citep{2013MNRAS.428.2840H}. 
P-SPH uses a pressure-entropy formulation of the smooth particle hydrodynamics (SPH) equations that eliminates artificial surface tension at contact discontinuities found in traditional density-based SPH formulations \citep[e.g.,][]{2007MNRAS.380..963A} and resolves the major historical differences between SPH and grid-based codes (included adaptive and moving mesh methods), particularly with respect to fluid mixing instabilities \citep[see also][]{2013ApJ...768...44S}. 
P-SPH also manifestly conserves momentum, energy, angular momentum, and entropy. 
We use the ``inviscid'' artificial viscosity prescription of \cite{2010MNRAS.408..669C}, which allows for excellent shock-capturing while reducing the viscosity to nearly zero away from shocks. 
We also implement entropy mixing following \cite{2008JCoPh.22710040P}, using the higher-order dissipation switches described in \cite{2010MNRAS.408..669C}. 
To avoid errors due to particles with long time steps interacting with particles with much shorter time steps, we use the time step limiter proposed by \cite{2012MNRAS.419..465D}. 
Our simulations employ an adaptive fifth-order spline kernel with a neighbor number $N_{\rm ngb}\approx60$ designed to optimally resolve sound waves down to a wavelength of order the gas smoothing length \citep[][]{2012MNRAS.425.1068D}. 
The gravity solver in GIZMO is a modified version of the {\small GADGET-3} gravity algorithm \citep{2005MNRAS.364.1105S} which implements the adaptive softening method of \cite{2007MNRAS.374.1347P} and a modified softening kernel that represents the exact solution for the potential of the SPH smoothing kernel following \cite{2012MNRAS.425.1104B}. 

Our P-SPH implementation has been tested extensively and found to give good agreement with analytic solutions and well-tested grid codes on a broad suite of test problems. Many of these tests are presented in \citet{2013MNRAS.428.2840H}. Kere\v{s} et al. (in prep.) also test P-SPH in cosmological simulations (with a sub-resolution ISM model and without galactic winds) for Milky Way-mass halos and show that the main discrepancies between the halo gas properties between traditional density-based SPH and grid-based simulations \citep[e.g.,][]{1999ApJ...525..554F, 2012MNRAS.425.2027K} have been eliminated. 
In particular, the artificial dense cool clumps found in SPH simulations of Milky Way-mass halos \citep[][]{2006ApJ...644L...1S,2006MNRAS.370.1612K, 2009ApJ...700L...1K} and absent in grid-based codes are also absent in P-SPH. 
Furthermore, the density and temperature profiles of hot halo gas in P-SPH are close to those found in grid-based codes and substantially different than in density-based SPH simulations, resulting in enhanced hot gas accretion rates onto galaxies \citep[as found, e.g., in the mesh-based calculations of][]{2009MNRAS.397L..64A, 2013MNRAS.429.3353N}. 
These findings are in agreement with other direct comparisons of improved SPH formulations with the results of grid codes \citep[e.g.,][]{2014MNRAS.440.3243P}. 
These tests support the reliability of P-SPH for simulations of the CGM. 

We use multi-scale ``zoom-in'' initial conditions with maximum spatial and mass resolution centered around halos of interest \citep[e.g.,][]{1985PhDT.........7P, 1993ApJ...412..455K}. 
Initial conditions were generated with the MUSIC code \citep[][]{2011MNRAS.415.2101H} using second-order Lagrangian perturbation theory. 
Simulations {\bf m11}, {\bf m12q}, {\bf m12i}, and {\bf m13} were chosen to match initial conditions from the AGORA project \citep[][]{2014ApJS..210...14K}, which will enable future comparisons with a wide range of different codes and physics implementations. 
The {\bf m12v} initial conditions are the same as the `{\bf B1}' run used by \cite{2011MNRAS.412L.118F} to study the covering fraction of neutral hydrogen in simulations without galactic winds. 
In \cite{2014MNRAS.445..581H}, we showed that the {\bf mxx} series of simulations reproduces several key integrated star formation properties of real galaxies, including the $M_{\star}-M_{\rm h}$ relationship at all redshifts where it is observationally constrained. 
For the present paper, we add one simulation to the {\bf mxx} series, {\bf m14}, which corresponds to a small galaxy cluster at $z=0$. 
We did not report this simulation in \cite{2014MNRAS.445..581H} because its lower resolution implies that GMCs are not well resolved at $z=0$. 
However, GMCs (identified as `giant clumps'; e.g. Genzel et al. 2011\nocite{2011ApJ...733..101G}) are both larger and much more massive at $z=2-4$, so that this simulation is adequate for our high-redshift analysis. 
We also expanded our simulation suite with 8 new simulations run to $z=2$ with main halos chosen to enable comparisons with CGM measurements around LBGs at $z\sim2-3$ ({\bf z2hxxx}). 
At $z=2$, the main halos in these runs have mass $M_{\rm h}=1.9\times10^{11}-1.2\times10^{12}$ M$_{\odot}$. 
These halos were selected based on mass only and are therefore not biased in terms of large-scale environment or merger history. 
All halos included in our analysis are main halos, i.e. we do not include satellites.   
For comparison, LBGs from the CGM analysis of \cite{2012ApJ...750...67R}  are at $z=2-2.5$ and reside in dark matter halos of average mass $M_{\rm h}\sim10^{12}$ M$_{\odot}$ \citep[][]{2005ApJ...619..697A, 2012ApJ...752...39T}. 
Tables \ref{tbl:sims_Hopkins} and \ref{tbl:sims_new} summarize the parameters of the simulations analyzed in this work. 

In our simulations, gas is allowed to cool to molecular cloud temperatures via atomic and molecular line emission, in addition to the standard processes described by \cite{1996ApJS..105...19K}. 
Star formation in our simulation proceeds only in dense regions ($n_{\rm H}\gtrsim10-100\,{\rm cm^{-3}}$) that are also locally self-gravitating (i.e., where the local virial parameter $\alpha \equiv \delta v^{2} \delta r / G m_{\rm gas}(<\delta r) < 1$ on the smallest available scale, $\delta r$ being the force softening or smoothing length) and we ensure that the most massive gravitational-bound structures in the ISM of our simulated galaxies (the Toomre-mass GMCs) are well resolved.\footnote{Except at very early times, when galaxy progenitors cannot be resolved.} 
\cite{2013MNRAS.432.2647H} showed that the local self-gravity criterion is necessary to obtain the correct spatial distribution of star formation in galaxies. 
For ${\bf m14}$, we adopt a lower density threshold $n_{\rm H}=1\,{\rm cm^{-3}}$ because of the lower resolution of the simulation. 
Numerical simulations \citep[][]{2011MNRAS.417..950H, 2013ApJ...770...25A} and analytic models \citep[e.g.,][]{2013MNRAS.433.1970F} that are successful at reproducing the observed Kennicutt-Schmidt law \citep[][]{1998ApJ...498..541K, 2010MNRAS.407.2091G} indicate that the formation of GMCs is the rate-limiting step for star formation in galactic disks and that the star formation-gas density relation on small-scales has a weak effect on the global star formation properties within galaxies as long as stellar feedback (which balances gravity in the disk) is included. 
Thus, our simulations are able to make meaningful predictions for star formation in galaxies without resolving the formation of individual stars. 
Stellar feedback is modeled by implementing energy, momentum, mass, and metal return from radiation, supernovae, stellar winds, and photoionization following the predictions of the STARBURST99 stellar population synthesis model \citep[][]{1999ApJS..123....3L}.

We explicitly follow chemical abundances of nine metal species (C, N, O, Ne, Mg, Si, S, Ca, and Fe), with enrichment following each source of mass return individually. 
During the course of the hydrodynamical calculation, ionization balance of all tracked elements is computed using the ultraviolet background model of \cite{2009ApJ...703.1416F}\footnote{Publicly available at http://galaxies.northwestern.edu/uvb.} and we apply an on-the-fly correction for dense, self-shielded gas. 
Self-shielding is accounted for with a local Jeans-length approximation (integrating the local density at a given particle out to a Jeans length to determine a surface density $\Sigma$), then attenuating the background seen at that point by $\exp{(-\kappa \Sigma)}$ (where $\kappa$ is the opacity). 
Confirmation of the accuracy of this approximation in radiative transfer experiments can be found in \cite{2010ApJ...725..633F} and \cite{2013MNRAS.430.2427R}. 

All our simulations assume a ``standard'' flat $\Lambda$CDM cosmology with $h\approx0.7$, $\Omega_{\rm m} = 1-\Omega_{\Lambda} \approx 0.27$ and $\Omega_{\rm b}\approx0.046$. Minor variations about these fiducial values are adopted for some simulations to match the parameters of simulations from the AGORA project and from our previous work. Uncertainties in our calculations are dominated by baryonic physics and the small variations in cosmological parameters do not introduce significant effects in our analysis.

\subsection{HI calculations}
\label{sec:HI_calculations}
For the neutral hydrogen predictions of this paper, we post-process all our simulation snapshots with the ionizing radiative transfer code developed in \cite{2010ApJ...725..633F}. 
Our methodology is the same as for our previous calculations of covering fractions on simulations without galactic winds \citep{2011MNRAS.412L.118F}. 
Briefly, we select halos in our zoom-in simulations and interpolate the gas density distribution around the halo center onto a Cartesian grid of side length $L$ and with $N$ grid points along each dimension. 
For most of our calculations, $L=200$ proper kpc and $N=256$, but $L$ is varied roughly in proportion to the virial radii of halos at $z=2$ (e.g., for $L=25,~50$ proper kpc for {\bf m09}, {\bf m10}, and $L=400$ proper kpc for {\bf m13} and ${\bf m14}$). 
The radiative transfer calculations model the cosmic ionizing background in a plane-parallel approximation by tracing rays inward from each of face of the grid and iteratively solving for ionizing balance, taking into account both photoionization and collisional ionization. 
To speed up the calculations, we assume that all photons have energy 13.6 eV. 
We have compared results of this approximation with the results of more exact calculations assuming a truncated power-law photon energy distribution and found good agreement. 
We find that the neutral fractions calculated with our radiative transfer code in post-processing are generally in good agreement with those calculated during the course of the hydrodynamical simulation. 
This is reassuring since the self-shielding approximation implemented in our simulations was calibrated to similar radiative transfer calculations. 
In this work, we neglect ionization of CGM gas by local sources. 
This tends to overestimate covering fractions of neutral hydrogen, but only slightly \citep[][]{2011MNRAS.412L.118F, 2011MNRAS.418.1796F}. 
To model the conversion of atomic hydrogen to H$_{\rm 2}$, we set the maximum $n_{\rm HI}$ to 1 cm$^{-3}$ \citep[e.g.,][]{2001ApJ...562L..95S}. 

For each simulation, we analyze 100 time slices from $z=4$ to $z=2$ and for each time slice, we consider three orthogonal sky projections, labeled by $xy$, $xz$, and $yz$. 
The covering fraction for an $N_{\rm HI}$ interval is evaluated by counting the fraction of cells in a projected radiative transfer grid that fall within the specified range. 
Halos are identified using Amiga's Halo Finder \citep[AHF;][]{2009ApJS..182..608K} and we adopt the virial overdensity definition of \cite{1998ApJ...495...80B}. 

\section{HI covering fractions}
\label{sec:covering_fractions}
\subsection{Visual examples}
Figure \ref{fig:halo350_summary} provides a graphical overview of the CGM gas properties (neutral hydrogen column, temperature, and kinematics) in one of our zoom-ins, {\bf z2h350}, from $z=4$ to $z=2$. 
At $z=4$ and $z=3$, filamentary cool gas infalling from the IGM is seen to penetrate significantly into the halo. 
By $z=2$, the situation is dramatically different: a powerful outflow with velocities $\sim500-1,000$ km s
$^{-1}$ is driven by the central galaxy. 
These kinematic differences, which correlate with a change in the geometry of the cool neutral gas from filamentary to quasi-spherical and clumpy, are a manifestation of the time variability of the star formation and galactic outflow rates in our simulations. 
The dense cool gas corresponding to infalling filaments and outflowing clumps are well traced by Lyman limit system contours. 

Figure \ref{fig:z2_HI_summary} focuses on HI maps at $z=2$ and shows how the HI distribution varies as function of halo mass. 
For {\bf m12v} and {\bf m13}, which have $z=2$ halo masses $M_{\rm h}=2-5.9\times10^{11}$ M$_{\odot}$, the dense neutral hydrogen roughly traces intergalactic filaments that join the halo (and which interact with outflows).
For the lower-mass halos, the HI distribution is more quasi-spherical, consistent with such halos being embedded in large-scale structure filaments rather than at the intersection of them. 
The least massive halos, ${\bf m09}$ and ${\bf m10}$, have much smaller covering fractions of dense HI, which we will quantify in the next section. 
Gas in these low-mass halos is particularly sensitive to heating by the ionizing background \citep[e.g.,][]{1992MNRAS.256P..43E, 1996ApJ...465..608T, 2011MNRAS.417.2982F} and to the effects of stellar feedback \citep[e.g.,][]{1986ApJ...303...39D}. 
Our most massive halo, ${\bf m14}$, exhibits clumps of dense HI embedded in a hot halo but also has reduced covering fractions for LLSs and DLAs. 
We will return to this halo when we compare our predictions with quasar-hosting halos in \S \ref{sec:quasars}.

\subsection{Covering fractions}
\label{coverings_quant}
For our quantitative predictions, we consider four ranges of HI column density, $N_{\rm HI} >10^{15.5}$ cm$^{-2}$, $N_{\rm HI} >10^{17.2}$ cm$^{-2}$ 
(LLSs), $N_{\rm HI} >10^{19}$ cm$^{-2}$ (super Lyman limit systems [SLLSs]), and $N_{\rm HI} >10^{20.3}$ cm$^{-2}$ (DLAs). 
\cite{2012ApJ...750...67R} report the covering fractions for these column densities around 
 LBGs in their survey, allowing us to compare our simulations with observations. 

Figure \ref{fig:covering_fractions_summary} shows the covering fractions within a projected virial radius for these four column density ranges from $z=4$ to $z=2$ for all the halos in Tables \ref{tbl:sims_Hopkins} and \ref{tbl:sims_new}. 
The simulations are compared to the measurements of \cite{2012ApJ...750...67R} for LBGs at $z\sim2-2.5$. 
For DLAs, SLLSs, and LLSs, the agreement between the model predictions and the observations is excellent. 
For lower columns $N_{\rm HI}>10^{15.5}$ cm$^{-2}$, the observed covering fractions appear to exceed the simulations by a factor of $\sim2$. 
One possibility is that this could be partially an artifact of our analysis procedure: while the \cite{2012ApJ...750...67R} measurements shown in Figure \ref{fig:covering_fractions_summary} include all systems within $|\Delta v|<700$ km s$^{-1}$ of the foreground galaxy, the simulation data points include only systems within $\pm L/2$ of the central galaxy owing to the finite size of the radiative transfer grid from which the covering fractions are evaluated (for most of our halos, $L=200$ proper kpc but $L$ is scaled approximately with the virial radius; \S \ref{sec:HI_calculations}). 
We have however repeated our analysis with radiative transfer grids double the size of our fiducial calculations and found that most of the difference for $N_{\rm HI}>10^{15.5}$ cm$^{-2}$ covering fractions persists.\footnote{For the higher column density systems (such as LLSs and DLAs), our results are very well converged with depth of the radiative transfer box since these dense systems originate primarily within the viral radius of galaxies.} 
Furthermore, \cite{2012ApJ...750...67R} also report that the covering fractions measured within $R_{\rm vir}$ are not significantly changed if only systems within $|v|<300$ km s$^{-1}$ of the foreground galaxy are included. 
Both of these results suggest that our simulations underpredict the amount of low-density neutral gas in LBG halos. 
This could point to missing physics in our current calculations, such as magnetic fields and cosmic rays, which could play a significant role in injecting cool gas into halos \citep[e.g.,][]{2013ApJ...777L..16B, 2014MNRAS.437.3312S} and in maintaining the integrity of cool gas in galactic winds \citep[][]{2014arXiv1409.6719M}. 
Alternatively, it could be that the resolution of our simulations is not sufficient to accurately capture the ejection of low-density gas from galaxies. 
At $z\sim3$, \cite{2013ApJ...765...89S} report a covering fraction $\approx70\%$ for $N_{\rm HI}>10^{15.5}$ cm$^{-2}$ absorbers within $R_{\rm vir}$ for the Eris2 simulation, consistent with our predictions and also somewhat below Rudie et al.'s measurement. 
Figure \ref{fig:covering_fractions_summary} also shows that the covering fraction $f_{\rm cov}(>10^{17.2}~{\rm cm}^{-2};~<R_{\rm vir})\approx64\%$ measured around luminous quasars at $z\sim2$ by \cite{2013ApJ...762L..19P} is much larger than predicted by any of our simulations with stellar feedback only. 
We discuss the implications for quasar-hosting halos in \S \ref{sec:quasars}.

The oscillations visibble in point tracks in Figure \ref{fig:covering_fractions_summary} reflect the time variability of halo gas covering fractions stemming from time variability in galactic inflows and outflows (see \S \ref{sec:time_variability}). 
These oscillations in covering fractions are as much as a factor of $\sim2$ in amplitude. 
The scatter between different sky projections (typically $\lesssim20\%$) is much smaller and is reflected in the dispersion at fixed halo mass in each point track. 
In many of the other figures in this paper, we will therefore focus on a single sky projection for clarity.

In Figure \ref{fig:fcov_vs_Mh_1Rvir}, we compile the same simulated covering fractions but in different panels for different redshift intervals to isolate the dependence on halo mass. 
The halo mass dependence of the covering fractions is relatively weak for $M_{\rm h}\sim10^{11}-10^{12}$ M$_{\odot}$, but the covering fractions are distinctly lower for lower-mass halos ($M_{\rm h}\lesssim10^{10.5}$ M$_{\odot}$) at all redshifts considered, including for LLSs. 
Figure \ref{fig:fcov_vs_Mh_Rmax} compiles the covering fractions versus halo mass within a fixed impact parameter of 100 proper kpc. 
We find similar qualitative trends as for within a projected virial radius, but overall the covering fractions are lower because most of our halos have a virial radius $R_{\rm vir}<100$ proper kpc. 
The approximate constancy of LLS covering fractions within $100$ proper kpc for $M_{\rm h}\sim10^{11}-10^{12}$ M$_{\odot}$ \citep[see also][]{2014ApJ...780...74F} is consistent with observations that the covering fraction of LLSs around DLAs \citep[likely spanning a broad range of halo masses $\sim10^{10.5}-10^{13}$ M$_{\odot}$, e.g.][]{2012JCAP...11..059F, 2014MNRAS.440.2313B} is similar to that around LBGs within an impact parameter $\sim100$ proper kpc at $z\sim2$ \citep[][]{2014arXiv1411.6016R}. 
The much smaller covering fractions that we find for $M_{\rm h}\lesssim10^{10.5}$ M$_{\odot}$ however indicate that the overall DLA population cannot be dominated by such low-mass halos. 
Observational studies of MgII absorbers at $z\lesssim1$, which are believed to trace dense neutral hydrogen, have also revealed that such absorbers trace halos with characteristic mass $M_{\rm h}\sim10^{12}$ M$_{\odot}$ in spite of the much greater abundance of lower-mass halos \citep[][]{2008ApJ...679.1218T, 2011MNRAS.417..304L, 2014MNRAS.439..342G}. 
Interestingly, the covering fractions of cool gas within a virial radius peak for halos of mass comparable to that of maximum star formation efficiency in dark matter halos \citep[e.g.,][]{2010ApJ...717..379B, 2010ApJ...710..903M}, providing further evidence of the connection between halo gas properties and galaxy formation. 
Figures \ref{fig:covering_fractions_summary}-\ref{fig:fcov_vs_Mh_Rmax} show that covering fractions generally decrease with time, either when evaluated within $R_{\rm vir}$ or within a fixed transverse distance of $100$ proper kpc.

It is interesting to compare in more detail our predicted covering fractions with the results of \cite{2014ApJ...780...74F}, who used AMR simulations from \cite{2010MNRAS.404.2151C}, \cite{2012MNRAS.420.3490C}, and \cite{2013MNRAS.435..999D}. 
\cite{2014ApJ...780...74F} report LLS covering fractions within a virial radius for $M_{\rm h} \sim 10^{11}-10^{12}$ M$_{\odot}$ at $z=2$ that scatter between 0.1 and 0.2, with an average around 0.15. 
Figure \ref{fig:fcov_vs_Mh_1Rvir} shows that in our simulations the same quantity scatters between 0.1 and 0.4, with an average around 0.2. 
Thus our predicted LLS covering fractions within a virial radius are systematically higher than \cite{2014ApJ...780...74F}, a fact that is also true at $z=3$. 
These differences likely originate in the different feedback models. 
The simulations analyzed by \cite{2014ApJ...780...74F} implement only feedback from stellar winds and supernovae, and do so with thermal energy injection. 
This feedback implementation results in winds with low mass loading factors, $\beta \equiv \dot{M}_{\rm out} / {\rm SFR }\sim 0.3$ at $0.5 R_{\rm vir}$. 
\cite{2014ApJ...780...74F} also note that their simulations overpredict the stellar masses of galaxies by a factor of $\sim2$ and underpredict their gas mass fractions by a factor $\sim 2$. 
In our simulations, stellar feedback is stronger and produces galactic winds with mass loading factors $\sim 10$ for similar galaxies (Muratov et al. 2015\nocite{2015arXiv150103155M}). 
Our simulations are also in good agreement with the measured stellar mass function for this range of halo mass \citep[][]{2014MNRAS.445..581H}. 
These results indicate that the feedback in the FIRE simulations is more effective at expelling gas from galaxies than the feedback in \cite[][]{2014ApJ...780...74F}, a fact which is reflected in larger predicted covering fractions of cool halo gas. 
This further underscores how CGM measurements can distinguish between different feedback models. 
\cite{2013ApJ...765...89S} found $f_{\rm cov}(>10^{17.2}~{\rm cm^{-2}};~<R_{\rm vir})=0.27$ for a $M_{\rm h}=2.6\times10^{11}$ M$_{\odot}$  at $z\approx3$, consistent with our simulations.

\begin{figure*}
\begin{center}
\mbox{
\includegraphics[width=0.5\textwidth]{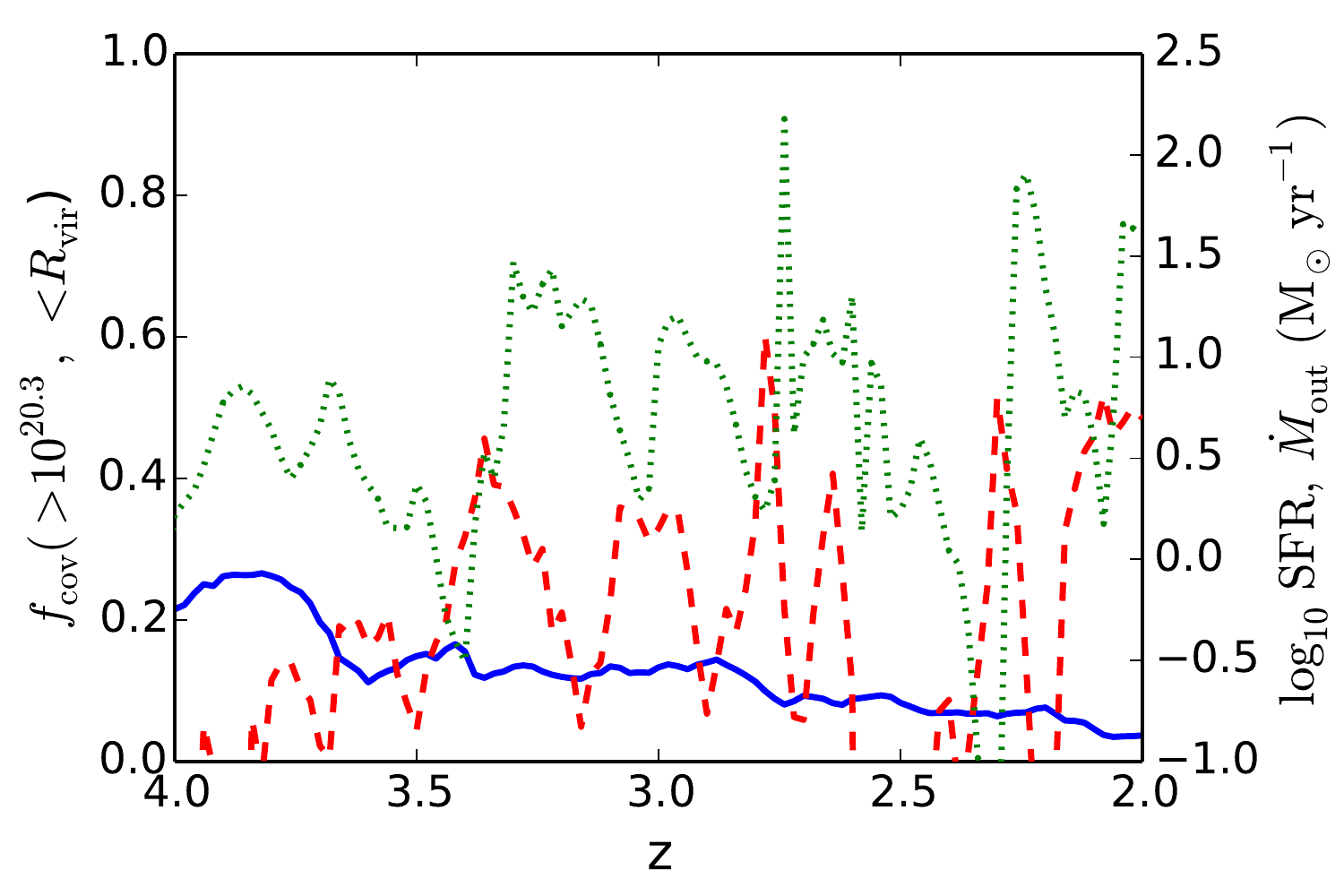}
\includegraphics[width=0.5\textwidth]{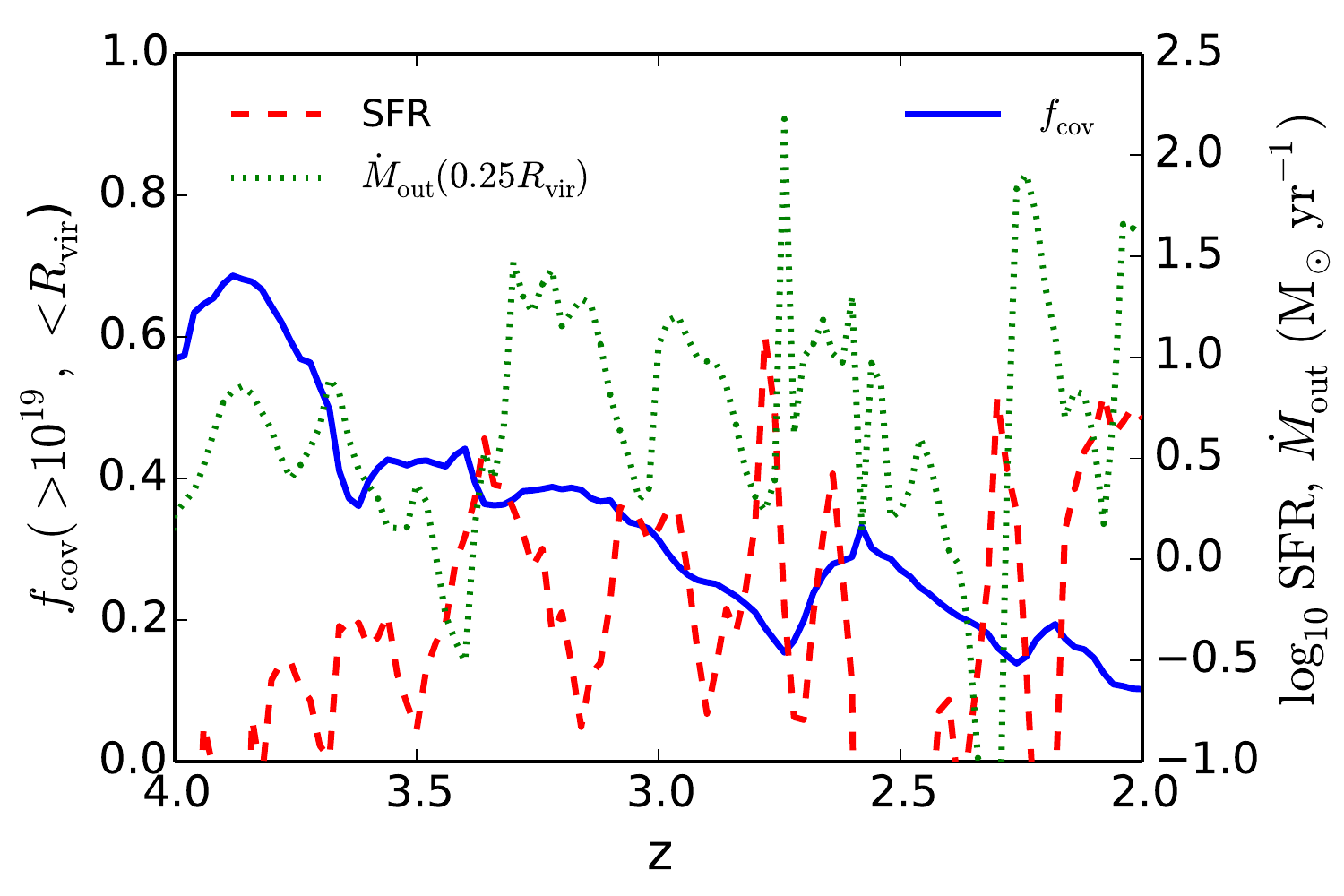}
}
\mbox{
\includegraphics[width=0.5\textwidth]{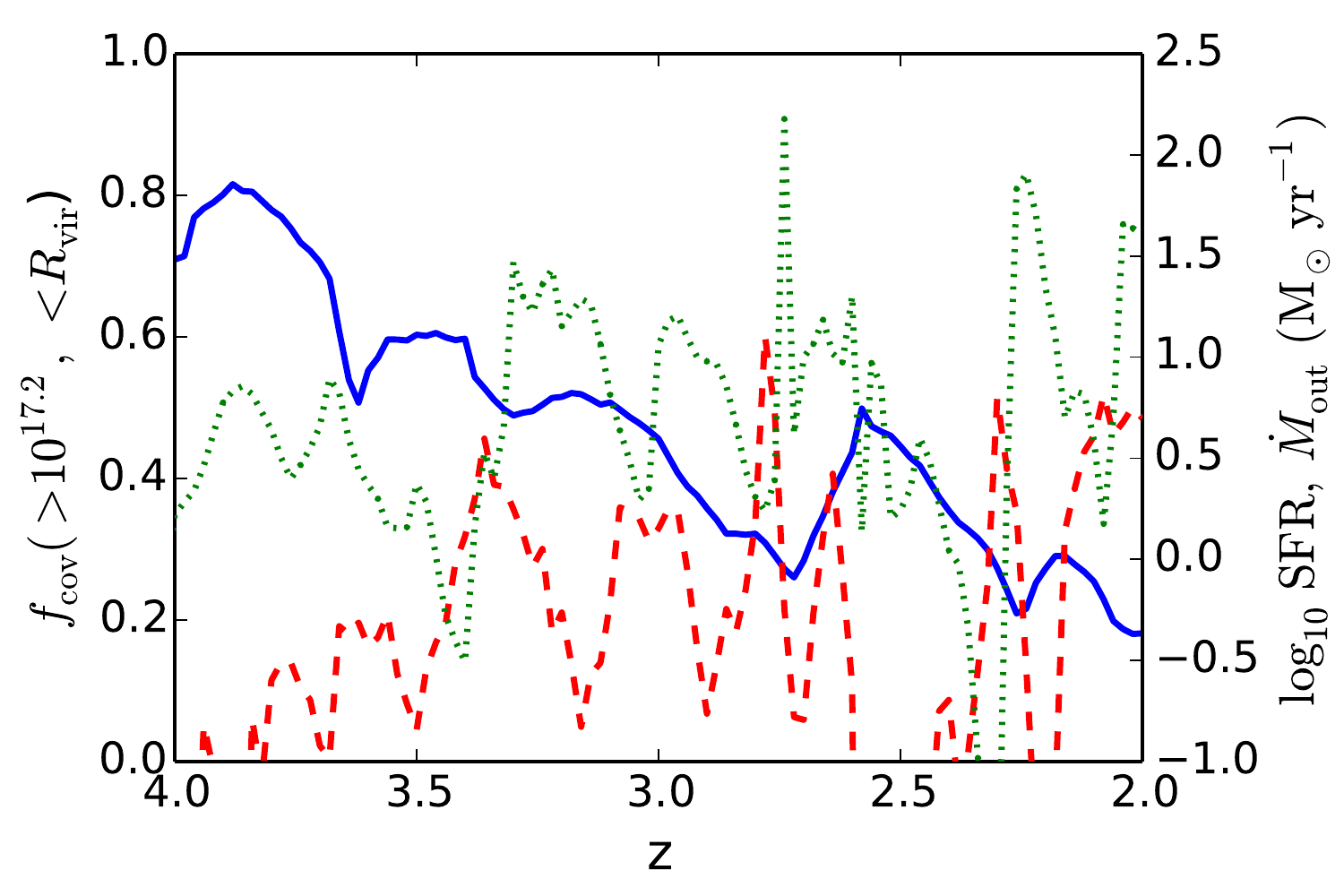}
\includegraphics[width=0.5\textwidth]{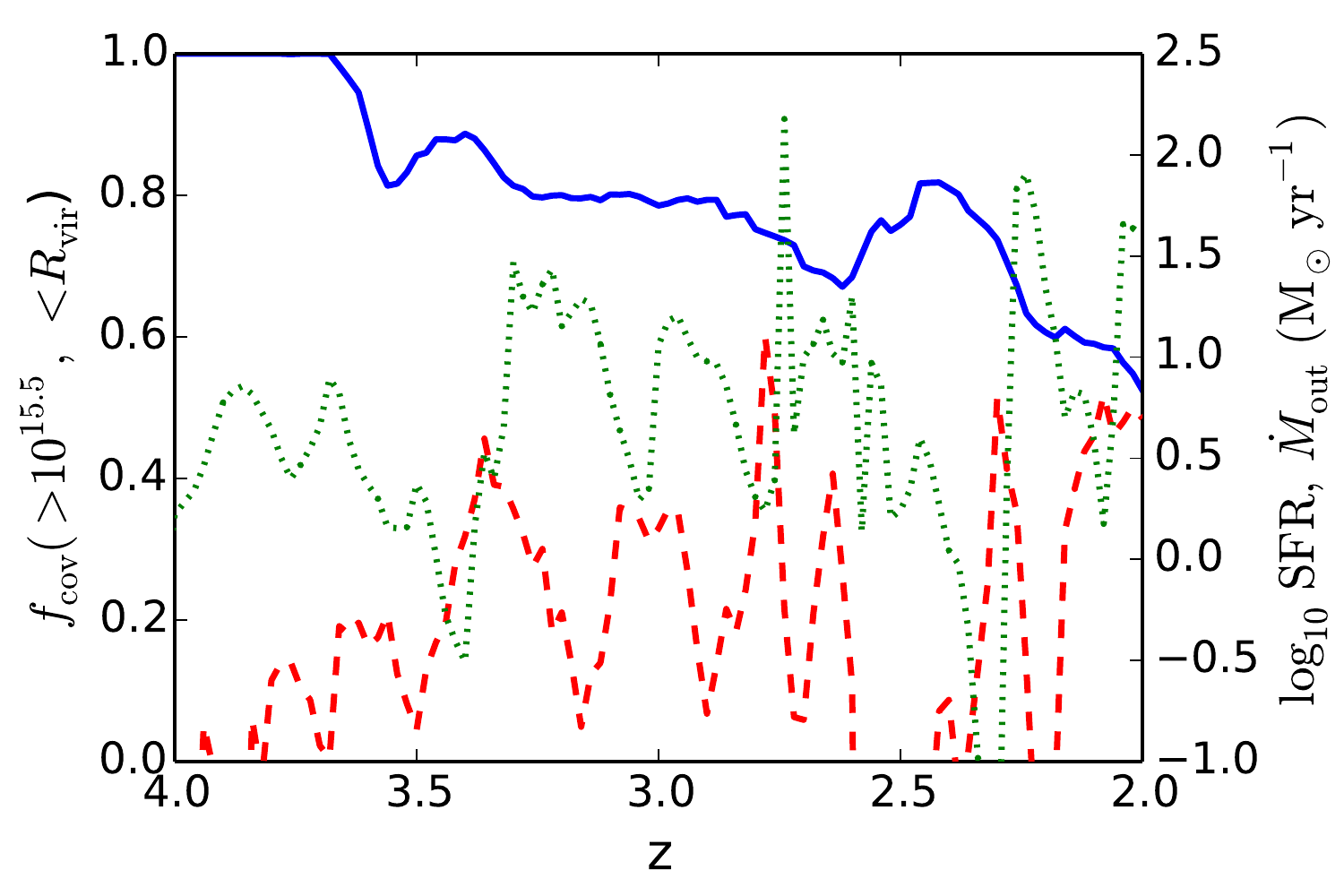}
}
\end{center}
\caption[]{\emph{Solid blue:} HI covering fractions within a projected virial radius as a function of redshift for ${\bf m12i}$ and the same  column density intervals as in Figures \ref{fig:covering_fractions_summary}-\ref{fig:fcov_vs_Mh_Rmax}.  
\emph{Dashed red:} Star formation rate within $0.2 R_{\rm vir}$. 
\emph{Dotted green:} Gas mass outflow rate measured at 0.25$R_{\rm vir}$. 
The covering fractions and star formation rate are both time dependent but do not correlate significantly in spite of the (time delayed) correlation between star formation and outflow events \citep[][]{2015arXiv150103155M}. 
For clarity, we show only quantities evaluated for one sky projection. 
The covering fractions are shown on a linear scale, while the star formation and mass outflow rates are shown on a logarithmic scale.}
\label{fig:SFR_correlation} 
\end{figure*}

\subsection{Inflows vs. outflows}
Figure \ref{fig:inflow_fraction} quantifies the fraction of absorbers in each column density interval that is inflowing vs. outflowing for three representative LBG halos ({\bf m12i}, {\bf z2h506}, and {\bf z2h350}). 
For each pixel in our radiative transfer grids corresponding to a certain minimum HI column density, we weighted the radial velocity (relative to the center of the halo) of each grid cell along the line of sight by gas mass. 
Pixels for which the mass-weighted radial velocity is positive are identified as outflowing (regardless of whether they are part of a galactic wind or, e.g., of tidally-stripped material); the rest as inflowing. 
Overall, inflowing gas contributes $\sim50\%$ of all covering fractions within $R_{\rm vir}$, but the inflowing fraction varies strongly with time, stochastically ranging from $\sim10\%$ to $\sim90\%$. 
Thus, the large covering fractions that we find relative to previous work including cosmological inflows only (i.e., neglecting galactic winds though not tidally-stripped material; Faucher-Gigu\`ere \& Kere\v{s} 2011\nocite{2011MNRAS.412L.118F}, Fumagalli et al. 2011\nocite{2011MNRAS.418.1796F}) arise in significant part from cool outflowing gas \citep[see also][]{2012MNRAS.421.2809V}.  

The $\sim50\%$ fraction of outflowing LLSs is consistent with the fact that simulations without strong outflows previously underestimated the covering fraction of LLSs observed within a projected viral radius of $z\sim2$ LBGs by a factor $\sim2$ \citep[][]{2012ApJ...750...67R}. 
At $z\sim0.5$, \cite{2013ApJ...770..138L} measured a bi-modal metallicity distribution for cosmological selected LLSs. 
They interpret the low-metallicity branch as arising from cosmological inflows, and the high-metallicity branch as arising from a combination of galactic outflows and gas stripped from galaxies by dynamical processes. 
The metallicity distribution they measure has approximately the same number of LLSs in the low-metallicity and high-metallicity branches, and is thus also consistent with our finding that approximately half of LLSs are inflowing and the other half outflowing (albeit at a different redshift).

\subsection{Time variability vs. star formation rate}
\label{sec:time_variability}
Figure \ref{fig:covering_fractions_summary} highlighted the time variability of covering fractions. 
Figure \ref{fig:SFR_correlation} directly plots the covering fractions within a virial radius and the star formation rate versus redshift for ${\bf m12i}$. 
We also show the gas mass outflow rate measured though 0.25$R_{\rm vir}$ as in \citet{2015arXiv150103155M}.  
The star formation rate is in fact much more time variable than the covering fractions, so that overall the star formation rate does not correlate significantly with the covering fractions. 
As illustrated by the $z=2$ row of Figure \ref{fig:halo350_summary}, outflowing clumps of cool gas contribute significantly to the LLS covering fractions. 
Time sequences of HI maps produced from our simulations also clearly show that the distribution of HI within halos vary in concert with galactic outflow events. 
Furthermore, \citet{2015arXiv150103155M} show that outflows correlate well with star formation bursts (albeit with a time delay $\sim30-100$ Myr). 
The lack of a significant correlation that we find here between HI covering fractions and the instantaneous star formation rate is because while outflows move the cool dense gas around in galactic halos, the covering fractions average over large areas. 
We have also verified that the covering fractions evaluated using outflowing gas only do not correlate well with the instantaneous star formation or mass outflow rates, and that these findings apply to other halos as well.

The good news for comparisons with observations is that it is not essential to compare predicted covering fractions with galaxies in the same stage of star formation activity (i.e., whether the galaxies are momentarily near a peak or trough in star formation rate). 
While it is essential to use a statistical sample of simulated halos to compare, for example, with measurements around LBGs owing to halo-to-halo variance, comparisons of covering fractions at fixed halo mass should be reliable even if LBGs are selected preferentially near peaks in the instantaneous star formation rate.

\begin{figure*}
\begin{center}
\includegraphics[width=1.0\textwidth]{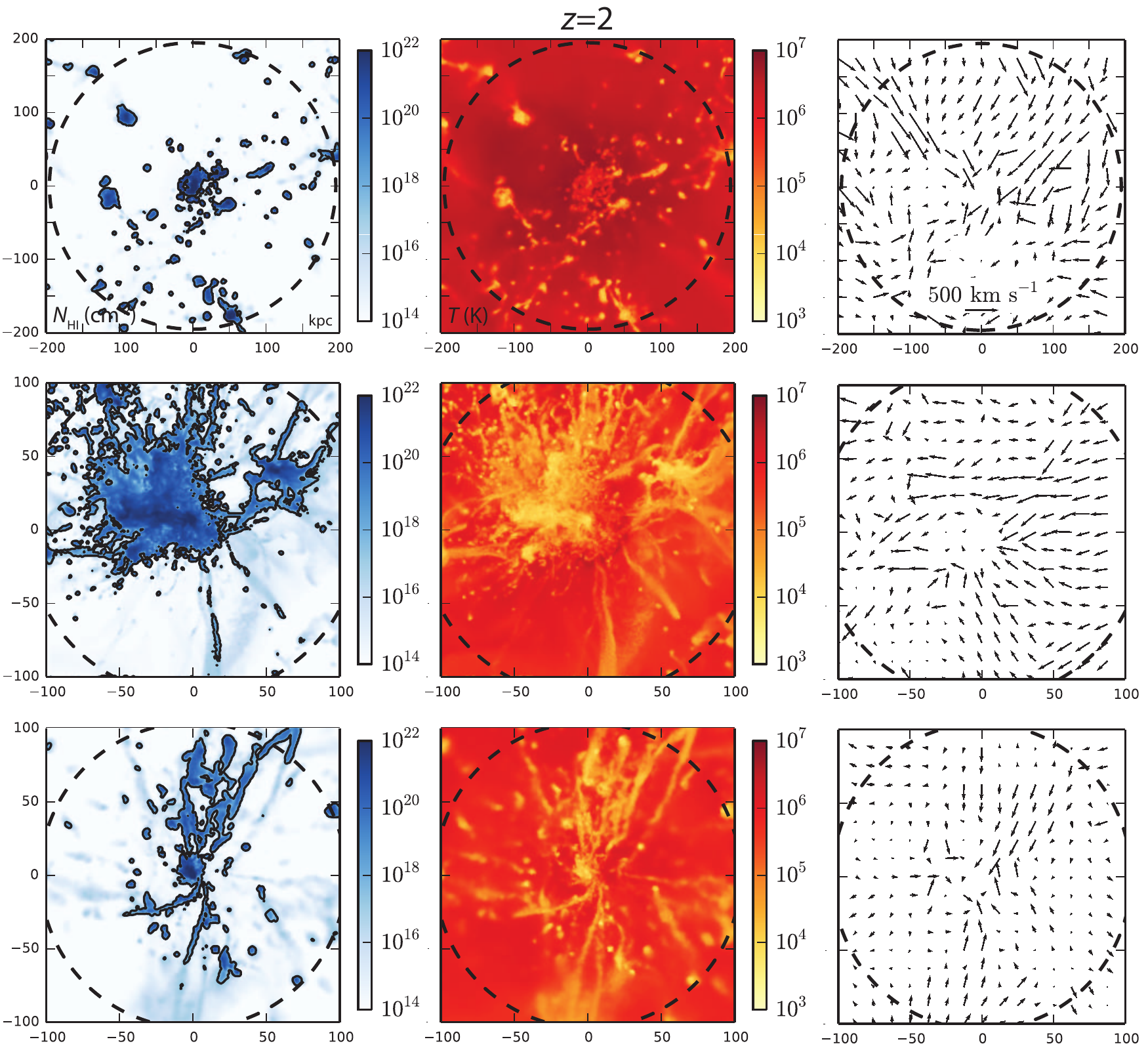}
\end{center}
\caption[]{Our three most massive halos at $z=2$ ({\bf m14}, {\bf z2h506}, {\bf m13} from top to bottom; $M_{\rm h}=5.9\times10^{12},~1.2\times10^{12},~8.7\times10^{11}$ M$_{\odot}$, respectively),
 bracketing the characteristic mass $M_{\rm QSO}\approx3\times10^{12}$  M$_{\odot}$ of quasar-hosting halos at $z\approx2$. 
\emph{Left:} Neutral hydrogen column density map. \emph{Center}: Temperature map (weighted by gas density squared to emphasize dense gas), \emph{Right:} Gas kinematics. 
The virial radius of the halo is indicated in each panel by the dashed circles and Lyman limit systems ($N_{\rm HI}>10^{17.2}$ cm$^{-2}$) are indicated by solid contours. 
Measurements around luminous quasars at $z\sim2$ indicate a LLS covering fraction within a projected virial radius $f_{\rm cov}(>10^{17.2}~{\rm cm^{-2}};~<R_{\rm vir})=64^{+6}_{-7} \%$, much larger than is found in any of our simulated halos with stellar feedback only, suggesting that the presence of the quasar affects halo gas on $\gtrsim100$ kpc scales (\S \ref{sec:quasars}).
}
\label{fig:m14} 
\end{figure*}

\begin{figure}
\begin{center}
\includegraphics[width=0.49\textwidth]{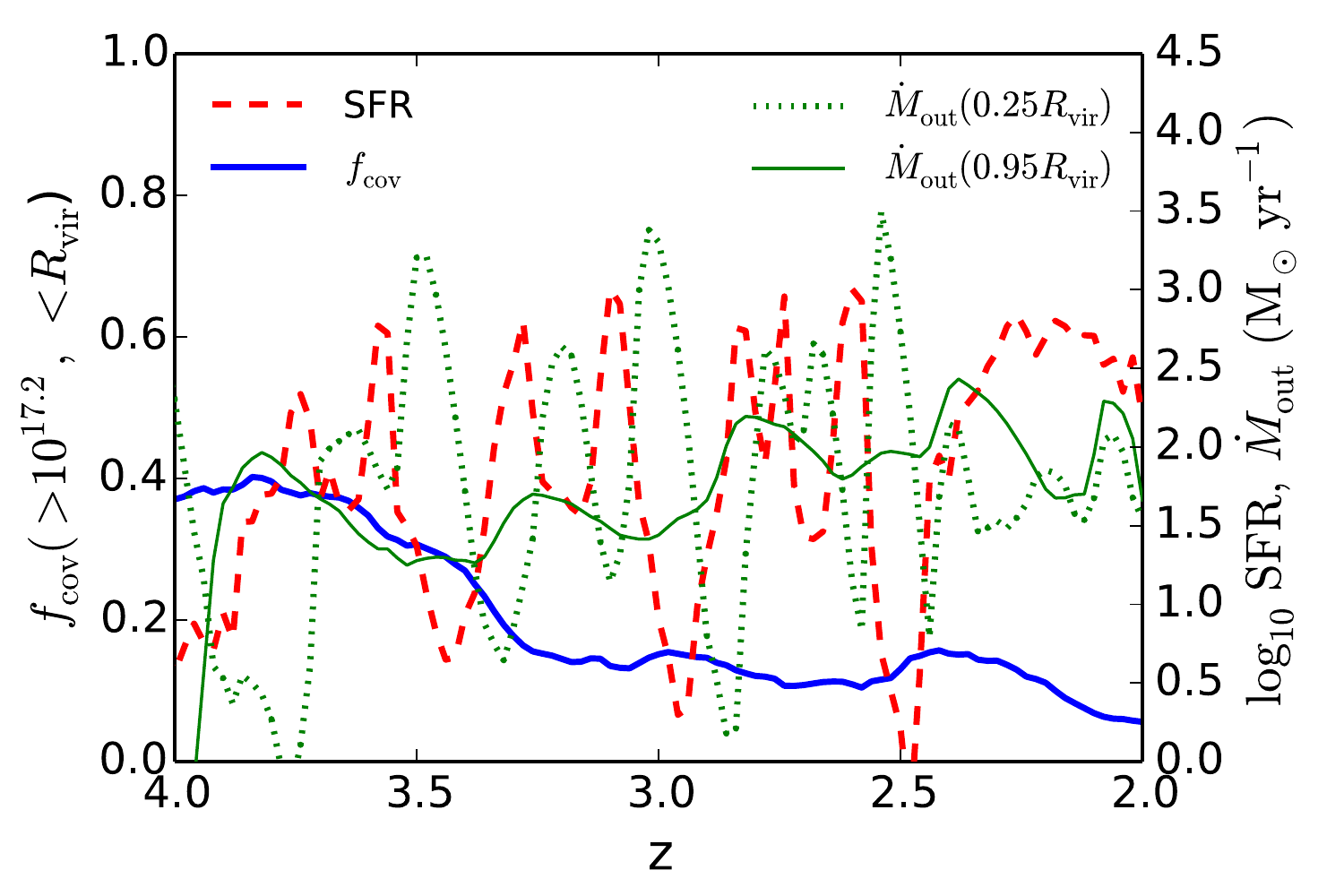}
\end{center}
\caption[]{Similar to Figure \ref{fig:SFR_correlation} but for {\bf m14} (our most massive halo, $M_{\rm h}=5.9\times10^{12}$ M$_{\odot}$ at $z=2$, shown in the top panel of Figure \ref{fig:m14}) and LLSs only. We also show the mass outflow rate through $0.95 R_{\rm vir}$. 
Despite the star formation rate (within $0.2 R_{\rm vir}$) reaching peaks $\sim1,000$ M$_{\odot}$ yr$^{-1}$, the LLS covering fraction decreases steadily with decreasing redshift. 
At $z=2$, the LLS covering fraction within $R_{\rm vir}$ is $12\%$ for this projection, well below the value $64^{+6}_{-7} \%$ measured by \cite{2013ApJ...762L..19P} for quasar-hosting halos at $z\sim2$. 
Comparing mass outflow rates at $0.25 R_{\rm vir}$ and at $0.95 R_{\rm vir}$ shows that galactic winds typically stall at a small fraction of the virial radius and that large mass outflow rates never reach $\sim R_{\rm vir}$.
}
\label{fig:SFR_correlation_m14} 
\end{figure}

\section{Quasar-hosting halos}
\label{sec:quasars}
Figure \ref{fig:m14} shows maps of neutral hydrogen, gas temperature, and kinematics for our three most massive halos at $z=2$ (${\bf m14}$, ${\bf z2h506}$, ${\bf m13}$; $M_{\bf h}=5.9\times10^{12},~1.2\times10^{12},~8.7\times10^{11}$ M$_{\odot}$, respectively). 
These halos bracket the characteristic mass $M_{\rm QSO}=3\times10^{12}$ M$_{\odot}$ of quasar-hosting halos at $z\sim2$ \citep[e.g.,][]{2005MNRAS.356..415C, 2012MNRAS.424..933W}. 
Measurements around luminous quasars at $z\sim2$ indicate a LLS covering fraction within a projected virial radius $f_{\rm cov}(>10^{17.2}~{\rm cm}^{-2};~<R_{\rm vir})=64^{+6}_{-7} \%$ \citep[][]{2013ApJ...762L..19P}, much larger than is found in any of our simulated halos with stellar feedback only, for which $f_{\rm cov}(>10^{17.2}~{\rm cm^{-2}};~<R_{\rm vir})<30\%$ for $M_{\rm h}>10^{12}$ M$_{\odot}$ at $z\approx2$ (Fig. \ref{fig:covering_fractions_summary} and \ref{fig:fcov_vs_Mh_1Rvir}). 
As Figure \ref{fig:m14} shows, our three most massive halos are dominated by hot gas with temperature $T\sim10^{6}-10^{7}$ K. 
For the most massive halo ({\bf m14}), the filaments of cool gas infalling from the IGM are absent within the virial radius and the cool gas morphology is more clumpy (owing in part to the presence of satellite galaxies). 
Nevertheless, coherent large-scale inflows from $\sim R_{\rm vir}$ to the central galaxy are seen in the kinematics map. 

The fact that our simulations with only stellar feedback dramatically under-predict the covering fraction of dense neutral gas observed around quasars in halos of similar mass points to the cool gas being connected to the presence of the central luminous AGN. 
In Figure \ref{fig:SFR_correlation_m14}, we plot the star formation rate (within $0.2R_{\rm vir}$) and the mass outflow rates at $0.25 R_{\rm vir}$ and at $0.95R_{\rm vir}$ as a function of redshift for {\bf m14}. 
Even though the star formation rate reaches peaks $\sim 1,000$ M$_{\odot}$ yr$^{-1}$, the LLS covering fraction within a virial radius decreases steadily with decreasing redshift and is only $12\%$ at $z=2$. 
Thus, neither the large gas mass accreting onto the central galaxy nor the stellar feedback provided by the intense star formation in our simulations is sufficient to explain the large LLS covering fraction observed around quasars.  
Galactic winds are driven by star formation from the central galaxy of ${\bf m14}$ but stall at a small fraction of the virial radius. 
As Figure \ref{fig:SFR_correlation_m14} indicates, the mass outflow rate at $0.95R_{\rm vir}$ following a burst of star formation is typically lower than the mass outflow rate at $0.25R_{\rm vir}$ by more than one order of magnitude.

Recently, direct observational evidence for powerful, wide-angle, galaxy-scale, quasar-driven outflows has emerged \citep[e.g.,][]{2009ApJ...706..525M, 2010A&A...518L.155F, 2011ApJ...733L..16S, 2011ApJ...729L..27R, 2012ApJ...746...86G, 2012MNRAS.425L..66M, 2014MNRAS.441.3306H, 2014A&A...562A..21C}. 
These outflows are observed on scales of several kpc from the nucleus and have been detected in both atomic and molecular gas. 
In luminous quasars, the mass outflow rates are $\dot{M}_{\rm out}\sim1,000$ M$_{\odot}$ yr$^{-1}$, the kinematic luminosities $L_{\rm kin} \equiv 0.5 \dot{M}_{\rm out} v_{\rm out}^{2} \sim0.01-0.05$ $L_{\rm AGN}$ (where $L_{\rm AGN}$ is the AGN bolometic luminosity), and the outflows reach velocities $v_{\rm out}>1,000$ km s$^{-1}$. 
In many cases, the energetics of these winds exceed what star formation can provide and the winds have a higher characteristic velocity than typical star formation-driven outflows. 
Thus, AGN-powered galactic winds may eject more cool gas farther out into quasar-hosting halos than star formation can.

A simple order-of-magnitude estimate suggests that quasar-driven outflows could inject sufficient cool gas in their halos to explain the order unity covering fraction measured for LLSs. 
Consider a time-steady, spherical outflow and suppose that the outflow is primarily mass loaded through its interaction with the ISM of the host galaxy. Then, after escaping the galaxy, the outflow can be approximated as having a constant velocity and a constant mass outflow rate. 
The number density of the outflowing gas as a function of radius $R$ is
\begin{equation}
n_{\rm H,out}(R) = \frac{X \dot{M}_{\rm out}}{4 \pi R^{2} m_{\rm p} v_{\rm out}},
\end{equation}
where $X=0.75$ is the hydrogen mass fraction. 
Integrating along the line of sight, we can convert this to a column density as a function of impact parameter $b$,
\begin{align}
N_{\rm H}(b) & = \frac{X \dot{M}_{\rm out}}{4 m_{\rm p} b v_{\rm out}}.
\end{align}
In photoionization equilibrium, the neutral fraction $x_{\rm HI} \approx \alpha(T) n_{\rm e} / \Gamma$, where $n_{\rm e}$ is the free electron density, $\alpha(T)$ is the hydrogen recombination coefficient, and $\Gamma$ is the photoionization rate. 

A combination of absorption and emission constraints \citep[][]{2007ApJ...655..735H, 2013ApJ...766...58H} indicate that gas transverse to quasars is typically shielded from the intense radiation along the line-of-sight to the observer, presumably because of beaming and/or obscuration effects often invoked in AGN unification models \citep[e.g.,][]{1995PASP..107..803U}. 
We therefore assume that the density and temperature of wind clumps are representative of typical self-shielded IGM absorbers, $n_{\rm H}\approx0.01$ cm$^{-3}$ and $T\approx 10^{4}$ K \citep[e.g.,][]{2009ApJ...703.1416F, 2013MNRAS.430.2427R}, and that the photoionization rate is dominated by the cosmic background, $\Gamma \approx 10^{-12}$ s$^{-1}$ \citep[e.g.,][]{2009ApJ...703.1416F}. 
Since LLSs are nearly fully ionized, $n_{\rm e} \approx 1.2 n_{\rm H}$, and the neutral fraction is therefore $x_{\rm HI}\approx 0.005$. 
For this neutral fraction,
\begin{align}
N_{\rm HI}(b) \approx 10^{18}~{\rm cm^{-2}} & \left( \frac{\dot{M}_{\rm out}}{\rm 1,000~M_{\odot}~yr^{-1}} \right) \left( \frac{b}{\rm 100~kpc} \right)^{-1} \\ 
&\times \left( \frac{v_{\rm out}}{\rm 1,000~km~s^{-1}} \right)^{-1} \left( \frac{x_{\rm HI}}{0.005} \right),
\end{align}
i.e., consistent with a LLS column. This consistency check assumes that the gas in multiphase outflows clumps in such a way as to achieve sufficient density to produce LLSs while remaining relatively area-filling, an assumption that self-consistent simulations of AGN-driven galactic winds could test. 
For the fiducial parameters above, $n_{\rm H,out}(R=100~{\rm kpc})=2\times10^{-4}$ cm$^{-3}$, so reaching a characteristic LLS density of $n_{\rm H} \approx 0.01$ cm$^{-3}$ requires clumping by a factor $\sim50$. 
The time for gas ejected from the galaxy to travel a distance $R$ is
\begin{equation}
t_{\rm flow} = R/v_{\rm out} \approx 10^{8}~{\rm yr} \left( \frac{R}{\rm 100~kpc} \right) \left( \frac{v}{\rm 1,000~km~s^{-1}} \right)^{-1},
\end{equation}
which for the fiducial parameters is comparable to the estimated lifetime of quasars \citep[e.g.,][]{2004cbhg.symp..169M, 2005ApJ...625L..71H, 2011ApJ...735..117F}. 
Thus, quasars can in principle eject in their halos enough cool gas to explain the LLSs observed around them.

Several caveats are warranted. 
First, it is unclear how cool gas ejected in a quasar outflow can survive out to $R\gtrsim100$ kpc without being destroyed by shocks. 
Cool gas however appears quite ubiquitous in galactic outflows \citep[e.g.,][]{2010ApJ...717..289S} and molecular gas is observed to be accelerated to $>1,000$ km s$^{-1}$ by quasars \citep[e.g.,][]{2010A&A...518L.155F, 2012MNRAS.425L..66M}, so we regard it as plausible that a quasar outflow could deliver cool gas to the required radii. 
On smaller scales, \cite{2012MNRAS.420.1347F} showed that AGN-driven outflows can produce dense cool gas \emph{in situ} as they encounter inhomogeneities in the ISM; an analogous process could occur on larger scales in quasar-hosting halos.  
Recent simulations of cool gas clouds permeated by a tangled magnetic field moving through a hot medium also indicate that such clouds survive much longer than purely hydrodynamical calculations suggest \citep{2014arXiv1409.6719M}. 
Second, \cite{2013ApJ...762L..19P} do not find evidence of extreme gas kinematics that might be associated with quasar outflows in their data set. 
The large equivalent widths of low-ionization absorbers indicate motions on the order of a few hundred km s$^{-1}$, but systems with width $\approx1,000$ km s$^{-1}$ are relatively rare. 
It may be that gravity and halo gas pressure, effects that we neglected in the simple model above, decelerate quasar-driven outflows sufficiently by the time they reach $\sim 100$ kpc. 
In the future, we plan to include black holes and AGN-driven outflows in our cosmological simulations, which will allow us to test this self-consistently. 
Another possibility is that the cool halo gas observed $\sim 100$ kpc from quasars is not part of an outflow but rather a signature of how AGN feedback affects the thermal stability of gas in massive halos \citep[e.g.,][]{2012MNRAS.420.3174S, 2012MNRAS.419.3319M}. 
It is also conceivable that quasars select special halos that are not represented in our sample of zoom-in simulations, or that our simulations underpredict the effects of stellar feedback in massive halos. 
It will therefore be important to extend the present study to better understand the effects of quasars on halo gas.

It is noteworthy that radio-quiet quasars are more common than radio-loud quasars by a factor $\sim 10$ \citep[e.g.,][]{2007ApJ...658..815S}. 
Wide-angle outflows from the accretion disks of radio-quiet quasars \citep[e.g.,][]{1995ApJ...451..498M, 2012MNRAS.425..605F} may thus play an important feedback role in a larger range of halos ($M_{\rm h} \gtrsim 10^{12}$ M$_{\odot}$ at $z=2$) than jets from radio-loud AGN, which are observationally inferred to prevent the development of strong cooling flows in galaxy clusters \citep[e.g.,][]{2007ARA&A..45..117M}. 

\section{Conclusions}
\label{sec:discussion}
We have used cosmological zoom-in simulations with resolved ISM and stellar feedback physics to make predictions for the covering fractions of neutral hydrogen in $z=2-4$ galaxy halos. 
Most of our simulations correspond to halos hosting Lyman break galaxies at $z=2$ ($M_{\rm h}\sim10^{11}-10^{12}$ M$_{\odot}$), but our simulation sample covers the halo mass range $M_{\rm h}=10^{9}-6\times10^{12}$ M$_{\odot}$ at $z=2$. 
These simulations have been shown to be in excellent agreement with existing constraints on the star formation histories of galaxies \citep[][]{2014MNRAS.445..581H}, as well as with observed mass-metallicity relations (Ma et al., in prep.).  
In total, we analyzed 16 main halos which we post-processed with ionizing radiative transfer. 
For each simulation, we analyzed 100 time slices from $z=4$ to $z=2$ and for each time slice, we computed statistics for three orthogonal sky projections. 
For our quantitative predictions, we considered four ranges of HI column density, $N_{\rm HI} >10^{15.5}$ cm$^{-2}$, $N_{\rm HI} >10^{17.2}$ cm$^{-2}$ 
(LLSs), $N_{\rm HI} >10^{19}$ cm$^{-2}$ (SLLSs), and $N_{\rm HI} >10^{20.3}$ cm$^{-2}$ (DLAs). 

Our main conclusions are as follows:
\begin{enumerate}
\item For DLAs, SLLSs, and LLSs, our predicted covering fractions within a projected virial radius around LBGs at $z=2-2.5$ are consistent with measurements around actual galaxies occupying halos of the same mass at this redshift \citep[][]{2012ApJ...750...67R}. Our predictions for LLS covering fractions exceed our previous calculations without galactic winds by a factor $\sim2$ (Faucher-Gigu\`ere \& Kere\v{s} 2011\nocite{2011MNRAS.412L.118F}; see also Fumagalli et al. 2011\nocite{2011MNRAS.418.1796F}). 
This enhancement arises both because of cool gas ejected into halos by star formation-driven galactic winds and through interactions of outflowing gas with cosmological inflows. 
The fractions of HI absorbers of all columns arising in inflows and in outflows exhibit significant time dependence, varying from $\sim10\%$ to $\sim90\%$, with an average value of $\sim50\%$ each. 
For $M_{\rm h}\sim10^{11}-10^{12}$ M$_{\odot}$ halos at $z=2-3$, our predicted LLS covering fractions within a projected virial radius are higher than those of \cite{2014ApJ...780...74F} by $\sim30$\%, most likely because the strong stellar feedback in the FIRE simulations is more effective at expelling gas from galaxies. 
\item The covering fractions generally decrease with time. At a given redshift, the covering fractions of the denser systems (LLSs, SLLSs, DLAs) do not vary strongly with halo mass for $M_{\rm h}\sim10^{11}-10^{12}$ M$_{\odot}$, either when evaluated within $R_{\rm vir}$ or within a fixed impact parameter of $100$ proper kpc, but the covering fractions are distinctly lower for lower-mass halos ($M_{\rm h}\lesssim10^{10.5}$ M$_{\odot}$) at all redshifts considered. 
This is consistent with observations that the covering fraction of LLSs around DLAs (spanning a range of halo mass) is similar to that around LBGs within an impact parameter $\sim100$ proper kpc at $z\sim2$ \citep[][]{2014arXiv1411.6016R}. The much smaller covering fractions that we find for $M_{\rm h}\lesssim10^{10.5}$ M$_{\odot}$ indicate that the overall DLA population cannot be dominated by such low-mass halos. 
Interestingly, the covering fractions of cool gas within a virial radius in our simulations peak for halos with masses $M_{\rm h}\sim10^{11}-10^{12}$ M$_{\odot}$ comparable to those of maximum star formation efficiency in dark matter halos \citep[e.g.,][]{2010ApJ...717..379B, 2010ApJ...710..903M}. 
\item For the lowest HI columns that we consider, $N_{\rm HI}>10^{15.5}$ cm$^{-2}$, measurements around $z=2-2.5$ LBGs indicate covering fractions $<R_{\rm vir}$ consistent with unity \citep[][]{2012ApJ...750...67R}; our calculations underestimate this covering fraction by a factor $\approx2$. 
This suggests that our simulations underpredict the amount of low-density neutral gas in LBG halos and that they may be missing at least one physical process capable of accelerating low-density neutral gas in galactic winds without destroying it. 
Magneto-centrifugal forces and cosmic ray pressure may help in this regard and are worthwhile exploring further. 
Tangled magnetic fields may also preserve the integrity of cool gas clouds in galactic winds out to larger radii than purely hydrodynamical simulations suggest \citep{2014arXiv1409.6719M}. 
It is also possible that the resolution of our simulations is simply not sufficient to accurately capture the ejection of low-density gas from galaxies.
\item Covering fractions vary significantly as halos experience accretion-star formation-outflow cycles. However, the fluctuations in covering fractions are much weaker than those in star formation rate owing to areal averaging. 
As a result, the instantaneous star formation rate (for a given halo) does not correlate significantly with the predicted HI covering fractions. 
The covering fractions as viewed from different angles can vary by as much as a factor of $\approx2$, but the scatter is typically $\lesssim20\%$. 
\item Our most massive simulated halos, $M_{\rm h}=8.7\times10^{11}-5.9\times10^{12}$ M$_{\odot}$ at $z=2$, bracket the characteristic halo mass $M_{\rm QSO}\approx3\times10^{12}$ M$_{\odot}$ of luminous quasars but fall short of reproducing the LLS covering fraction $\sim 65\%$ within $R_{\rm vir}$ measured around $z\sim2$ quasars by \cite{2013ApJ...762L..19P} by a factor $\gtrsim3$. 
Our simulations do not include black holes but our most massive halo hosts a galaxy with star formation rate reaching $\sim1,000$ M$_{\odot}$ yr$^{-1}$, suggesting that massive cosmological inflows and strong stellar feedback alone are insufficient to explain the large measured covering fraction of dense neutral gas in quasar-hosting halos. 
This further suggests that the presence of a luminous AGN can significantly alter the state of halo gas on $\sim100$ kpc scales in such halos. 
If so, wide-angle quasar-driven outflows -- which observations have recently revealed are common in luminous AGN -- may play a significant feedback role akin to radio jets in galaxy clusters but over a much wider range of halo mass, $M_{\rm h} \gtrsim10^{12}$ M$_{\odot}$. 
However, further studies are needed to better establish the role of black holes in explaining the cool gas observed in massive halos (\S \ref{sec:quasars}).
\end{enumerate}

There are several opportunities to extend the present study. 
The most straightforward include predictions for metal absorption lines and gas kinematics, and at lower redshifts probed by the Cosmic Origins Spectrograph on the Hubble Space Telescope \citep[e.g.,][]{2011Sci...334..948T, 2013ApJ...770..138L}. 
It will also be important to extend our predictions to larger physical scales. 
Indeed, \cite{2012ApJ...750...67R} find a significant enhancement of IGM absorption as far as $\sim2$ proper Mpc from foreground galaxies. 
\cite{2013ApJ...776..136P} also detect excess HI absorption out to 1 proper Mpc from $z\sim2$ quasars. 
Such impact parameters are well outside the halos of galaxies with which the absorption is measured to be correlated, so this must reflect large-scale correlations in the cosmic matter distribution (the ``2-halo term''). 
Due to their small high-resolution volumes, zoom-in simulations are not well suited to study such correlations. 
In the future, we plan to run simulations of resolution comparable to the zoom-ins presented herein in full cosmological volumes 
which will allow us to address this question more directly. 
We will also extend our simulations to include black holes and AGN feedback, which will allow us to explicitly model the effects of luminous AGN on halo gas. 

\section*{Acknowledgments}
We acknowledge useful discussions with Gwen Rudie, Xavier Prochaska, Joe Hennawi, Kate Rubin, Joop Schaye, Michele Fumagalli, and Freeke van de Voort. 
We also thank the anonymous referee for a constructive review. 
CAFG was supported by a fellowship from the Miller Institute for Basic Research in Science, by NASA through Einstein Postdoctoral Fellowship Award PF3-140106 and grant 10-ATP10-0187, by NSF through grant AST-1412836, and by Northwestern University funds. 
Support for PFH was provided by the Gordon and Betty Moore Foundation through Grant 776 to the Caltech Moore Center for Theoretical Cosmology and Physics, by the Alfred P. Sloan Foundation through Sloan Research Fellowship BR2014-022, and by NSF through grant AST-1411920. 
DK was supported by an Hellman Fellowship and NSF grant AST-1412153. 
EQ was supported by NASA ATP grant 12-APT12-0183, a Simons Investigator award from the Simons Foundation, the David and Lucile Packard Foundation, and the Thomas Alison Schneider Chair in Physics at UC Berkeley. 
The simulations analyzed in this paper were run on XSEDE computational resources (allocations TG-AST120025, TG-AST130039, and TG-AST140023).

\bibliography{references} 

\begin{thebibliography}{152}
\expandafter\ifx\csname natexlab\endcsname\relax\def\natexlab#1{#1}\fi

\bibitem[{{Adelberger} {et~al.}(2005){Adelberger}, {Steidel}, {Pettini},
  {Shapley}, {Reddy}, \& {Erb}}]{2005ApJ...619..697A}
{Adelberger}, K.~L., {Steidel}, C.~C., {Pettini}, M., {Shapley}, A.~E.,
  {Reddy}, N.~A., \& {Erb}, D.~K. 2005, \apj, 619, 697

\bibitem[{{Adelberger et al.}(2003)}]{2003ApJ...584...45A}
{Adelberger et al.} 2003, \apj, 584, 45

\bibitem[{{Agertz} \& {Kravtsov}(2014)}]{2014arXiv1404.2613A}
{Agertz}, O., \& {Kravtsov}, A.~V. 2014, arXiv:1404.2613

\bibitem[{{Agertz} {et~al.}(2013){Agertz}, {Kravtsov}, {Leitner}, \&
  {Gnedin}}]{2013ApJ...770...25A}
{Agertz}, O., {Kravtsov}, A.~V., {Leitner}, S.~N., \& {Gnedin}, N.~Y. 2013,
  \apj, 770, 25

\bibitem[{{Agertz} {et~al.}(2007){Agertz}, {Moore}, {Stadel}, {Potter},
  {Miniati}, {Read}, {Mayer}, {Gawryszczak}, {Kravtsov}, {Nordlund}, {Pearce},
  {Quilis}, {Rudd}, {Springel}, {Stone}, {Tasker}, {Teyssier}, {Wadsley}, \&
  {Walder}}]{2007MNRAS.380..963A}
{Agertz}, O., {Moore}, B., {Stadel}, J., {Potter}, D., {Miniati}, F., {Read},
  J., {Mayer}, L., {Gawryszczak}, A., {Kravtsov}, A., {Nordlund}, {\AA}.,
  {Pearce}, F., {Quilis}, V., {Rudd}, D., {Springel}, V., {Stone}, J.,
  {Tasker}, E., {Teyssier}, R., {Wadsley}, J., \& {Walder}, R. 2007, \mnras,
  380, 963

\bibitem[{{Agertz} {et~al.}(2009){Agertz}, {Teyssier}, \&
  {Moore}}]{2009MNRAS.397L..64A}
{Agertz}, O., {Teyssier}, R., \& {Moore}, B. 2009, \mnras, 397, L64

\bibitem[{{Barnes}(2012)}]{2012MNRAS.425.1104B}
{Barnes}, J.~E. 2012, \mnras, 425, 1104

\bibitem[{{Barnes} \& {Haehnelt}(2014)}]{2014MNRAS.440.2313B}
{Barnes}, L.~A., \& {Haehnelt}, M.~G. 2014, \mnras, 440, 2313

\bibitem[{{Bauermeister} {et~al.}(2010){Bauermeister}, {Blitz}, \&
  {Ma}}]{2010ApJ...717..323B}
{Bauermeister}, A., {Blitz}, L., \& {Ma}, C. 2010, \apj, 717, 323

\bibitem[{{Behroozi} {et~al.}(2010){Behroozi}, {Conroy}, \&
  {Wechsler}}]{2010ApJ...717..379B}
{Behroozi}, P.~S., {Conroy}, C., \& {Wechsler}, R.~H. 2010, \apj, 717, 379

\bibitem[{{Birnboim} \& {Dekel}(2003)}]{2003MNRAS.345..349B}
{Birnboim}, Y., \& {Dekel}, A. 2003, \mnras, 345, 349

\bibitem[{{Booth} {et~al.}(2013){Booth}, {Agertz}, {Kravtsov}, \&
  {Gnedin}}]{2013ApJ...777L..16B}
{Booth}, C.~M., {Agertz}, O., {Kravtsov}, A.~V., \& {Gnedin}, N.~Y. 2013,
  \apjl, 777, L16

\bibitem[{{Bordoloi et al.}(2014)}]{2014ApJ...794..130B}
{Bordoloi et al.} 2014, \apj, 794, 130

\bibitem[{{Bouwens} {et~al.}(2007){Bouwens}, {Illingworth}, {Franx}, \&
  {Ford}}]{2007ApJ...670..928B}
{Bouwens}, R.~J., {Illingworth}, G.~D., {Franx}, M., \& {Ford}, H. 2007, \apj,
  670, 928

\bibitem[{{Brooks} {et~al.}(2011){Brooks}, {Solomon}, {Governato}, {McCleary},
  {MacArthur}, {Brook}, {Jonsson}, {Quinn}, \& {Wadsley}}]{2011ApJ...728...51B}
{Brooks}, A.~M., {Solomon}, A.~R., {Governato}, F., {McCleary}, J.,
  {MacArthur}, L.~A., {Brook}, C.~B.~A., {Jonsson}, P., {Quinn}, T.~R., \&
  {Wadsley}, J. 2011, \apj, 728, 51

\bibitem[{{Bryan} \& {Norman}(1998)}]{1998ApJ...495...80B}
{Bryan}, G.~L., \& {Norman}, M.~L. 1998, \apj, 495, 80

\bibitem[{{Ceverino} {et~al.}(2010){Ceverino}, {Dekel}, \&
  {Bournaud}}]{2010MNRAS.404.2151C}
{Ceverino}, D., {Dekel}, A., \& {Bournaud}, F. 2010, \mnras, 404, 2151

\bibitem[{{Ceverino} {et~al.}(2012){Ceverino}, {Dekel}, {Mandelker},
  {Bournaud}, {Burkert}, {Genzel}, \& {Primack}}]{2012MNRAS.420.3490C}
{Ceverino}, D., {Dekel}, A., {Mandelker}, N., {Bournaud}, F., {Burkert}, A.,
  {Genzel}, R., \& {Primack}, J. 2012, \mnras, 420, 3490

\bibitem[{{Ceverino} {et~al.}(2014){Ceverino}, {Klypin}, {Klimek},
  {Trujillo-Gomez}, {Churchill}, {Primack}, \& {Dekel}}]{2014MNRAS.442.1545C}
{Ceverino}, D., {Klypin}, A., {Klimek}, E.~S., {Trujillo-Gomez}, S.,
  {Churchill}, C.~W., {Primack}, J., \& {Dekel}, A. 2014, \mnras, 442, 1545

\bibitem[{{Cicone et al.}(2014)}]{2014A&A...562A..21C}
{Cicone et al.} 2014, \aap, 562, A21

\bibitem[{{Crighton} {et~al.}(2013){Crighton}, {Hennawi}, \&
  {Prochaska}}]{2013ApJ...776L..18C}
{Crighton}, N.~H.~M., {Hennawi}, J.~F., \& {Prochaska}, J.~X. 2013, \apjl, 776,
  L18

\bibitem[{{Croom} {et~al.}(2005){Croom}, {Boyle}, {Shanks}, {Smith}, {Miller},
  {Outram}, {Loaring}, {Hoyle}, \& {da {\^A}ngela}}]{2005MNRAS.356..415C}
{Croom}, S.~M., {Boyle}, B.~J., {Shanks}, T., {Smith}, R.~J., {Miller}, L.,
  {Outram}, P.~J., {Loaring}, N.~S., {Hoyle}, F., \& {da {\^A}ngela}, J. 2005,
  \mnras, 356, 415

\bibitem[{{Cullen} \& {Dehnen}(2010)}]{2010MNRAS.408..669C}
{Cullen}, L., \& {Dehnen}, W. 2010, \mnras, 408, 669

\bibitem[{{Dalla Vecchia} \& {Schaye}(2008)}]{2008MNRAS.387.1431D}
{Dalla Vecchia}, C., \& {Schaye}, J. 2008, \mnras, 387, 1431

\bibitem[{{Dav{\'e}} {et~al.}(2011){Dav{\'e}}, {Oppenheimer}, \&
  {Finlator}}]{2011MNRAS.415...11D}
{Dav{\'e}}, R., {Oppenheimer}, B.~D., \& {Finlator}, K. 2011, \mnras, 415, 11

\bibitem[{{Dehnen} \& {Aly}(2012)}]{2012MNRAS.425.1068D}
{Dehnen}, W., \& {Aly}, H. 2012, \mnras, 425, 1068

\bibitem[{{Dekel} \& {Silk}(1986)}]{1986ApJ...303...39D}
{Dekel}, A., \& {Silk}, J. 1986, \apj, 303, 39

\bibitem[{{Dekel} {et~al.}(2013){Dekel}, {Zolotov}, {Tweed}, {Cacciato},
  {Ceverino}, \& {Primack}}]{2013MNRAS.435..999D}
{Dekel}, A., {Zolotov}, A., {Tweed}, D., {Cacciato}, M., {Ceverino}, D., \&
  {Primack}, J.~R. 2013, \mnras, 435, 999

\bibitem[{{Durier} \& {Dalla Vecchia}(2012)}]{2012MNRAS.419..465D}
{Durier}, F., \& {Dalla Vecchia}, C. 2012, \mnras, 419, 465

\bibitem[{{Efstathiou}(1992)}]{1992MNRAS.256P..43E}
{Efstathiou}, G. 1992, \mnras, 256, 43P

\bibitem[{{Faucher-Gigu{\`e}re} {et~al.}(2010){Faucher-Gigu{\`e}re}, {Kere{\v
  s}}, {Dijkstra}, {Hernquist}, \& {Zaldarriaga}}]{2010ApJ...725..633F}
{Faucher-Gigu{\`e}re}, C., {Kere{\v s}}, D., {Dijkstra}, M., {Hernquist}, L.,
  \& {Zaldarriaga}, M. 2010, \apj, 725, 633

\bibitem[{{Faucher-Gigu{\`e}re} \& {Kere{\v s}}(2011)}]{2011MNRAS.412L.118F}
{Faucher-Gigu{\`e}re}, C.-A., \& {Kere{\v s}}, D. 2011, \mnras, 412, L118

\bibitem[{{Faucher-Gigu{\`e}re} {et~al.}(2011){Faucher-Gigu{\`e}re}, {Kere{\v
  s}}, \& {Ma}}]{2011MNRAS.417.2982F}
{Faucher-Gigu{\`e}re}, C.-A., {Kere{\v s}}, D., \& {Ma}, C.-P. 2011, \mnras,
  417, 2982

\bibitem[{{Faucher-Gigu{\`e}re} {et~al.}(2009){Faucher-Gigu{\`e}re}, {Lidz},
  {Zaldarriaga}, \& {Hernquist}}]{2009ApJ...703.1416F}
{Faucher-Gigu{\`e}re}, C.-A., {Lidz}, A., {Zaldarriaga}, M., \& {Hernquist}, L.
  2009, \apj, 703, 1416

\bibitem[{{Faucher-Gigu{\`e}re} \& {Quataert}(2012)}]{2012MNRAS.425..605F}
{Faucher-Gigu{\`e}re}, C.-A., \& {Quataert}, E. 2012, \mnras, 425, 605

\bibitem[{{Faucher-Gigu{\`e}re} {et~al.}(2013){Faucher-Gigu{\`e}re},
  {Quataert}, \& {Hopkins}}]{2013MNRAS.433.1970F}
{Faucher-Gigu{\`e}re}, C.-A., {Quataert}, E., \& {Hopkins}, P.~F. 2013, \mnras,
  433, 1970

\bibitem[{{Faucher-Gigu{\`e}re} {et~al.}(2012){Faucher-Gigu{\`e}re},
  {Quataert}, \& {Murray}}]{2012MNRAS.420.1347F}
{Faucher-Gigu{\`e}re}, C.-A., {Quataert}, E., \& {Murray}, N. 2012, \mnras,
  420, 1347

\bibitem[{{Feruglio} {et~al.}(2010){Feruglio}, {Maiolino}, {Piconcelli},
  {Menci}, {Aussel}, {Lamastra}, \& {Fiore}}]{2010A&A...518L.155F}
{Feruglio}, C., {Maiolino}, R., {Piconcelli}, E., {Menci}, N., {Aussel}, H.,
  {Lamastra}, A., \& {Fiore}, F. 2010, \aap, 518, L155+

\bibitem[{{Font-Ribera} {et~al.}(2012){Font-Ribera}, {Miralda-Escud{\'e}},
  {Arnau}, {Carithers}, {Lee}, {Noterdaeme}, {P{\^a}ris}, {Petitjean}, {Rich},
  {Rollinde}, {Ross}, {Schneider}, {White}, \& {York}}]{2012JCAP...11..059F}
{Font-Ribera}, A., {Miralda-Escud{\'e}}, J., {Arnau}, E., {Carithers}, B.,
  {Lee}, K.-G., {Noterdaeme}, P., {P{\^a}ris}, I., {Petitjean}, P., {Rich}, J.,
  {Rollinde}, E., {Ross}, N.~P., {Schneider}, D.~P., {White}, M., \& {York},
  D.~G. 2012, \jcap, 11, 59

\bibitem[{{Ford} {et~al.}(2013){Ford}, {Oppenheimer}, {Dav{\'e}}, {Katz},
  {Kollmeier}, \& {Weinberg}}]{2013MNRAS.432...89F}
{Ford}, A.~B., {Oppenheimer}, B.~D., {Dav{\'e}}, R., {Katz}, N., {Kollmeier},
  J.~A., \& {Weinberg}, D.~H. 2013, \mnras, 432, 89

\bibitem[{{Frenk et al.}(1999)}]{1999ApJ...525..554F}
{Frenk et al.} 1999, \apj, 525, 554

\bibitem[{{Fumagalli} {et~al.}(2014){Fumagalli}, {Hennawi}, {Prochaska},
  {Kasen}, {Dekel}, {Ceverino}, \& {Primack}}]{2014ApJ...780...74F}
{Fumagalli}, M., {Hennawi}, J.~F., {Prochaska}, J.~X., {Kasen}, D., {Dekel},
  A., {Ceverino}, D., \& {Primack}, J. 2014, \apj, 780, 74

\bibitem[{{Fumagalli} {et~al.}(2011){Fumagalli}, {Prochaska}, {Kasen}, {Dekel},
  {Ceverino}, \& {Primack}}]{2011MNRAS.418.1796F}
{Fumagalli}, M., {Prochaska}, J.~X., {Kasen}, D., {Dekel}, A., {Ceverino}, D.,
  \& {Primack}, J.~R. 2011, \mnras, 418, 1796

\bibitem[{{Furlanetto} \& {Lidz}(2011)}]{2011ApJ...735..117F}
{Furlanetto}, S.~R., \& {Lidz}, A. 2011, \apj, 735, 117

\bibitem[{{Gauthier} {et~al.}(2014){Gauthier}, {Chen}, {Cooksey}, {Simcoe},
  {Seyffert}, \& {O'Meara}}]{2014MNRAS.439..342G}
{Gauthier}, J.-R., {Chen}, H.-W., {Cooksey}, K.~L., {Simcoe}, R.~A.,
  {Seyffert}, E.~N., \& {O'Meara}, J.~M. 2014, \mnras, 439, 342

\bibitem[{{Genel} {et~al.}(2014){Genel}, {Vogelsberger}, {Springel}, {Sijacki},
  {Nelson}, {Snyder}, {Rodriguez-Gomez}, {Torrey}, \&
  {Hernquist}}]{2014MNRAS.445..175G}
{Genel}, S., {Vogelsberger}, M., {Springel}, V., {Sijacki}, D., {Nelson}, D.,
  {Snyder}, G., {Rodriguez-Gomez}, V., {Torrey}, P., \& {Hernquist}, L. 2014,
  \mnras, 445, 175

\bibitem[{{Genzel et al.}(2010)}]{2010MNRAS.407.2091G}
{Genzel et al.} 2010, \mnras, 407, 2091

\bibitem[{{Genzel et al.}(2011)}]{2011ApJ...733..101G}
---. 2011, \apj, 733, 101

\bibitem[{{Goerdt} {et~al.}(2012){Goerdt}, {Dekel}, {Sternberg}, {Gnat}, \&
  {Ceverino}}]{2012MNRAS.424.2292G}
{Goerdt}, T., {Dekel}, A., {Sternberg}, A., {Gnat}, O., \& {Ceverino}, D. 2012,
  \mnras, 424, 2292

\bibitem[{{Governato} {et~al.}(2010){Governato}, {Brook}, {Mayer}, {Brooks},
  {Rhee}, {Wadsley}, {Jonsson}, {Willman}, {Stinson}, {Quinn}, \&
  {Madau}}]{2010Natur.463..203G}
{Governato}, F., {Brook}, C., {Mayer}, L., {Brooks}, A., {Rhee}, G., {Wadsley},
  J., {Jonsson}, P., {Willman}, B., {Stinson}, G., {Quinn}, T., \& {Madau}, P.
  2010, \nat, 463, 203

\bibitem[{{Governato} {et~al.}(2007){Governato}, {Willman}, {Mayer}, {Brooks},
  {Stinson}, {Valenzuela}, {Wadsley}, \& {Quinn}}]{2007MNRAS.374.1479G}
{Governato}, F., {Willman}, B., {Mayer}, L., {Brooks}, A., {Stinson}, G.,
  {Valenzuela}, O., {Wadsley}, J., \& {Quinn}, T. 2007, \mnras, 374, 1479

\bibitem[{{Greene} {et~al.}(2012){Greene}, {Zakamska}, \&
  {Smith}}]{2012ApJ...746...86G}
{Greene}, J.~E., {Zakamska}, N.~L., \& {Smith}, P.~S. 2012, \apj, 746, 86

\bibitem[{{Guedes} {et~al.}(2011){Guedes}, {Callegari}, {Madau}, \&
  {Mayer}}]{2011ApJ...742...76G}
{Guedes}, J., {Callegari}, S., {Madau}, P., \& {Mayer}, L. 2011, \apj, 742, 76

\bibitem[{{Hahn} \& {Abel}(2011)}]{2011MNRAS.415.2101H}
{Hahn}, O., \& {Abel}, T. 2011, \mnras, 415, 2101

\bibitem[{{Harrison} {et~al.}(2014){Harrison}, {Alexander}, {Mullaney}, \&
  {Swinbank}}]{2014MNRAS.441.3306H}
{Harrison}, C.~M., {Alexander}, D.~M., {Mullaney}, J.~R., \& {Swinbank}, A.~M.
  2014, \mnras, 441, 3306

\bibitem[{{Hennawi} \& {Prochaska}(2007)}]{2007ApJ...655..735H}
{Hennawi}, J.~F., \& {Prochaska}, J.~X. 2007, \apj, 655, 735

\bibitem[{{Hennawi} \& {Prochaska}(2013)}]{2013ApJ...766...58H}
---. 2013, \apj, 766, 58

\bibitem[{{Hennawi} {et~al.}(2006){Hennawi}, {Prochaska}, {Burles}, {Strauss},
  {Richards}, {Schlegel}, {Fan}, {Schneider}, {Zakamska}, {Oguri}, {Gunn},
  {Lupton}, \& {Brinkmann}}]{2006ApJ...651...61H}
{Hennawi}, J.~F., {Prochaska}, J.~X., {Burles}, S., {Strauss}, M.~A.,
  {Richards}, G.~T., {Schlegel}, D.~J., {Fan}, X., {Schneider}, D.~P.,
  {Zakamska}, N.~L., {Oguri}, M., {Gunn}, J.~E., {Lupton}, R.~H., \&
  {Brinkmann}, J. 2006, \apj, 651, 61

\bibitem[{{Henriques} {et~al.}(2013){Henriques}, {White}, {Thomas}, {Angulo},
  {Guo}, {Lemson}, \& {Springel}}]{2013MNRAS.431.3373H}
{Henriques}, B.~M.~B., {White}, S.~D.~M., {Thomas}, P.~A., {Angulo}, R.~E.,
  {Guo}, Q., {Lemson}, G., \& {Springel}, V. 2013, \mnras, 431, 3373

\bibitem[{{Hernquist} \& {Springel}(2003)}]{2003MNRAS.341.1253H}
{Hernquist}, L., \& {Springel}, V. 2003, \mnras, 341, 1253

\bibitem[{{Hopkins}(2013)}]{2013MNRAS.428.2840H}
{Hopkins}, P.~F. 2013, \mnras, 428, 2840

\bibitem[{{Hopkins}(2014)}]{2014arXiv1409.7395H}
---. 2014, arXiv:1409.7395

\bibitem[{{Hopkins} {et~al.}(2013{\natexlab{a}}){Hopkins}, {Cox}, {Hernquist},
  {Narayanan}, {Hayward}, \& {Murray}}]{2013MNRAS.430.1901H}
{Hopkins}, P.~F., {Cox}, T.~J., {Hernquist}, L., {Narayanan}, D., {Hayward},
  C.~C., \& {Murray}, N. 2013{\natexlab{a}}, \mnras, 430, 1901

\bibitem[{{Hopkins} {et~al.}(2005){Hopkins}, {Hernquist}, {Martini}, {Cox},
  {Robertson}, {Di Matteo}, \& {Springel}}]{2005ApJ...625L..71H}
{Hopkins}, P.~F., {Hernquist}, L., {Martini}, P., {Cox}, T.~J., {Robertson},
  B., {Di Matteo}, T., \& {Springel}, V. 2005, \apjl, 625, L71

\bibitem[{{Hopkins} {et~al.}(2012{\natexlab{a}}){Hopkins}, {Kere{\v s}},
  {Murray}, {Quataert}, \& {Hernquist}}]{2012MNRAS.427..968H}
{Hopkins}, P.~F., {Kere{\v s}}, D., {Murray}, N., {Quataert}, E., \&
  {Hernquist}, L. 2012{\natexlab{a}}, \mnras, 427, 968

\bibitem[{{Hopkins} {et~al.}(2014){Hopkins}, {Kere{\v s}}, {O{\~n}orbe},
  {Faucher-Gigu{\`e}re}, {Quataert}, {Murray}, \&
  {Bullock}}]{2014MNRAS.445..581H}
{Hopkins}, P.~F., {Kere{\v s}}, D., {O{\~n}orbe}, J., {Faucher-Gigu{\`e}re},
  C.-A., {Quataert}, E., {Murray}, N., \& {Bullock}, J.~S. 2014, \mnras, 445,
  581

\bibitem[{{Hopkins} {et~al.}(2013{\natexlab{b}}){Hopkins}, {Narayanan}, \&
  {Murray}}]{2013MNRAS.432.2647H}
{Hopkins}, P.~F., {Narayanan}, D., \& {Murray}, N. 2013{\natexlab{b}}, \mnras,
  432, 2647

\bibitem[{{Hopkins} {et~al.}(2011){Hopkins}, {Quataert}, \&
  {Murray}}]{2011MNRAS.417..950H}
{Hopkins}, P.~F., {Quataert}, E., \& {Murray}, N. 2011, \mnras, 417, 950

\bibitem[{{Hopkins} {et~al.}(2012{\natexlab{b}}){Hopkins}, {Quataert}, \&
  {Murray}}]{2012MNRAS.421.3522H}
---. 2012{\natexlab{b}}, \mnras, 421, 3522

\bibitem[{{Hopkins} {et~al.}(2012{\natexlab{c}}){Hopkins}, {Quataert}, \&
  {Murray}}]{2012MNRAS.421.3488H}
---. 2012{\natexlab{c}}, \mnras, 421, 3488

\bibitem[{{Jones} {et~al.}(2012){Jones}, {Stark}, \&
  {Ellis}}]{2012ApJ...751...51J}
{Jones}, T., {Stark}, D.~P., \& {Ellis}, R.~S. 2012, \apj, 751, 51

\bibitem[{{Kacprzak} {et~al.}(2012){Kacprzak}, {Churchill}, \&
  {Nielsen}}]{2012ApJ...760L...7K}
{Kacprzak}, G.~G., {Churchill}, C.~W., \& {Nielsen}, N.~M. 2012, \apjl, 760, L7

\bibitem[{{Kannan} {et~al.}(2014){Kannan}, {Stinson}, {Macci{\`o}}, {Brook},
  {Weinmann}, {Wadsley}, \& {Couchman}}]{2014MNRAS.437.3529K}
{Kannan}, R., {Stinson}, G.~S., {Macci{\`o}}, A.~V., {Brook}, C., {Weinmann},
  S.~M., {Wadsley}, J., \& {Couchman}, H.~M.~P. 2014, \mnras, 437, 3529

\bibitem[{{Katz} {et~al.}(1996){Katz}, {Weinberg}, \&
  {Hernquist}}]{1996ApJS..105...19K}
{Katz}, N., {Weinberg}, D.~H., \& {Hernquist}, L. 1996, \apjs, 105, 19

\bibitem[{{Katz} \& {White}(1993)}]{1993ApJ...412..455K}
{Katz}, N., \& {White}, S.~D.~M. 1993, \apj, 412, 455

\bibitem[{{Kaufmann} {et~al.}(2006){Kaufmann}, {Mayer}, {Wadsley}, {Stadel}, \&
  {Moore}}]{2006MNRAS.370.1612K}
{Kaufmann}, T., {Mayer}, L., {Wadsley}, J., {Stadel}, J., \& {Moore}, B. 2006,
  \mnras, 370, 1612

\bibitem[{{Kennicutt}(1998)}]{1998ApJ...498..541K}
{Kennicutt}, Jr., R.~C. 1998, \apj, 498, 541

\bibitem[{{Kere{\v s}} \& {Hernquist}(2009)}]{2009ApJ...700L...1K}
{Kere{\v s}}, D., \& {Hernquist}, L. 2009, \apjl, 700, L1

\bibitem[{{Kere{\v s}} {et~al.}(2009){Kere{\v s}}, {Katz}, {Fardal},
  {Dav{\'e}}, \& {Weinberg}}]{2009MNRAS.395..160K}
{Kere{\v s}}, D., {Katz}, N., {Fardal}, M., {Dav{\'e}}, R., \& {Weinberg},
  D.~H. 2009, \mnras, 395, 160

\bibitem[{{Kere{\v s}} {et~al.}(2005){Kere{\v s}}, {Katz}, {Weinberg}, \&
  {Dav{\'e}}}]{2005MNRAS.363....2K}
{Kere{\v s}}, D., {Katz}, N., {Weinberg}, D.~H., \& {Dav{\'e}}, R. 2005,
  \mnras, 363, 2

\bibitem[{{Kere{\v s}} {et~al.}(2012){Kere{\v s}}, {Vogelsberger}, {Sijacki},
  {Springel}, \& {Hernquist}}]{2012MNRAS.425.2027K}
{Kere{\v s}}, D., {Vogelsberger}, M., {Sijacki}, D., {Springel}, V., \&
  {Hernquist}, L. 2012, \mnras, 425, 2027

\bibitem[{{Kim et al.}(2014)}]{2014ApJS..210...14K}
{Kim et al.} 2014, \apjs, 210, 14

\bibitem[{{Kimm} {et~al.}(2011){Kimm}, {Slyz}, {Devriendt}, \&
  {Pichon}}]{2011MNRAS.413L..51K}
{Kimm}, T., {Slyz}, A., {Devriendt}, J., \& {Pichon}, C. 2011, \mnras, 413, L51

\bibitem[{{Knollmann} \& {Knebe}(2009)}]{2009ApJS..182..608K}
{Knollmann}, S.~R., \& {Knebe}, A. 2009, \apjs, 182, 608

\bibitem[{{Lehner} {et~al.}(2014){Lehner}, {Howk}, \&
  {Wakker}}]{2014arXiv1404.6540L}
{Lehner}, N., {Howk}, C., \& {Wakker}, B. 2014, arXiv:1404.6540

\bibitem[{{Lehner} {et~al.}(2013){Lehner}, {Howk}, {Tripp}, {Tumlinson},
  {Prochaska}, {O'Meara}, {Thom}, {Werk}, {Fox}, \&
  {Ribaudo}}]{2013ApJ...770..138L}
{Lehner}, N., {Howk}, J.~C., {Tripp}, T.~M., {Tumlinson}, J., {Prochaska},
  J.~X., {O'Meara}, J.~M., {Thom}, C., {Werk}, J.~K., {Fox}, A.~J., \&
  {Ribaudo}, J. 2013, \apj, 770, 138

\bibitem[{{Leitherer} {et~al.}(1999){Leitherer}, {Schaerer}, {Goldader},
  {Delgado}, {Robert}, {Kune}, {de Mello}, {Devost}, \&
  {Heckman}}]{1999ApJS..123....3L}
{Leitherer}, C., {Schaerer}, D., {Goldader}, J.~D., {Delgado}, R.~M.~G.,
  {Robert}, C., {Kune}, D.~F., {de Mello}, D.~F., {Devost}, D., \& {Heckman},
  T.~M. 1999, \apjs, 123, 3

\bibitem[{{Lu} {et~al.}(2012){Lu}, {Mo}, {Katz}, \&
  {Weinberg}}]{2012MNRAS.421.1779L}
{Lu}, Y., {Mo}, H.~J., {Katz}, N., \& {Weinberg}, M.~D. 2012, \mnras, 421, 1779

\bibitem[{{Lu} {et~al.}(2014){Lu}, {Mo}, {Lu}, {Katz}, \&
  {Weinberg}}]{2014MNRAS.443.1252L}
{Lu}, Y., {Mo}, H.~J., {Lu}, Z., {Katz}, N., \& {Weinberg}, M.~D. 2014, \mnras,
  443, 1252

\bibitem[{{Lundgren} {et~al.}(2011){Lundgren}, {Wake}, {Padmanabhan}, {Coil},
  \& {York}}]{2011MNRAS.417..304L}
{Lundgren}, B.~F., {Wake}, D.~A., {Padmanabhan}, N., {Coil}, A., \& {York},
  D.~G. 2011, \mnras, 417, 304

\bibitem[{{Maiolino} {et~al.}(2012){Maiolino}, {Gallerani}, {Neri}, {Cicone},
  {Ferrara}, {Genzel}, {Lutz}, {Sturm}, {Tacconi}, {Walter}, {Feruglio},
  {Fiore}, \& {Piconcelli}}]{2012MNRAS.425L..66M}
{Maiolino}, R., {Gallerani}, S., {Neri}, R., {Cicone}, C., {Ferrara}, A.,
  {Genzel}, R., {Lutz}, D., {Sturm}, E., {Tacconi}, L.~J., {Walter}, F.,
  {Feruglio}, C., {Fiore}, F., \& {Piconcelli}, E. 2012, \mnras, 425, L66

\bibitem[{{Marinacci} {et~al.}(2014){Marinacci}, {Pakmor}, {Springel}, \&
  {Simpson}}]{2014MNRAS.442.3745M}
{Marinacci}, F., {Pakmor}, R., {Springel}, V., \& {Simpson}, C.~M. 2014,
  \mnras, 442, 3745

\bibitem[{{Martin}(2005)}]{2005ApJ...621..227M}
{Martin}, C.~L. 2005, \apj, 621, 227

\bibitem[{{Martin} {et~al.}(2012){Martin}, {Shapley}, {Coil}, {Kornei},
  {Bundy}, {Weiner}, {Noeske}, \& {Schiminovich}}]{2012ApJ...760..127M}
{Martin}, C.~L., {Shapley}, A.~E., {Coil}, A.~L., {Kornei}, K.~A., {Bundy}, K.,
  {Weiner}, B.~J., {Noeske}, K.~G., \& {Schiminovich}, D. 2012, \apj, 760, 127

\bibitem[{{Martini}(2004)}]{2004cbhg.symp..169M}
{Martini}, P. 2004, Coevolution of Black Holes and Galaxies, 169

\bibitem[{{McCourt} {et~al.}(2014){McCourt}, {O'Leary}, {Madigan}, \&
  {Quataert}}]{2014arXiv1409.6719M}
{McCourt}, M., {O'Leary}, R.~M., {Madigan}, A.-M., \& {Quataert}, E. 2014,
  arXiv:1409.6719

\bibitem[{{McCourt} {et~al.}(2012){McCourt}, {Sharma}, {Quataert}, \&
  {Parrish}}]{2012MNRAS.419.3319M}
{McCourt}, M., {Sharma}, P., {Quataert}, E., \& {Parrish}, I.~J. 2012, \mnras,
  419, 3319

\bibitem[{{McNamara} \& {Nulsen}(2007)}]{2007ARA&A..45..117M}
{McNamara}, B.~R., \& {Nulsen}, P.~E.~J. 2007, \araa, 45, 117

\bibitem[{{Moe} {et~al.}(2009){Moe}, {Arav}, {Bautista}, \&
  {Korista}}]{2009ApJ...706..525M}
{Moe}, M., {Arav}, N., {Bautista}, M.~A., \& {Korista}, K.~T. 2009, \apj, 706,
  525

\bibitem[{{Moster} {et~al.}(2010){Moster}, {Somerville}, {Maulbetsch}, {van den
  Bosch}, {Macci{\`o}}, {Naab}, \& {Oser}}]{2010ApJ...710..903M}
{Moster}, B.~P., {Somerville}, R.~S., {Maulbetsch}, C., {van den Bosch}, F.~C.,
  {Macci{\`o}}, A.~V., {Naab}, T., \& {Oser}, L. 2010, \apj, 710, 903

\bibitem[{{Muratov} {et~al.}(2015){Muratov}, {Keres}, {Faucher-Gigu{\`e}re},
  {Hopkins}, {Quataert}, \& {Murray}}]{2015arXiv150103155M}
{Muratov}, A.~L., {Keres}, D., {Faucher-Gigu{\`e}re}, C.-A., {Hopkins}, P.~F.,
  {Quataert}, E., \& {Murray}, N. 2015, arXiv:1501.03155

\bibitem[{{Murray} {et~al.}(1995){Murray}, {Chiang}, {Grossman}, \&
  {Voit}}]{1995ApJ...451..498M}
{Murray}, N., {Chiang}, J., {Grossman}, S.~A., \& {Voit}, G.~M. 1995, \apj,
  451, 498

\bibitem[{{Murray} {et~al.}(2005){Murray}, {Quataert}, \&
  {Thompson}}]{2005ApJ...618..569M}
{Murray}, N., {Quataert}, E., \& {Thompson}, T.~A. 2005, \apj, 618, 569

\bibitem[{{Nelson} {et~al.}(2013){Nelson}, {Vogelsberger}, {Genel}, {Sijacki},
  {Kere{\v s}}, {Springel}, \& {Hernquist}}]{2013MNRAS.429.3353N}
{Nelson}, D., {Vogelsberger}, M., {Genel}, S., {Sijacki}, D., {Kere{\v s}}, D.,
  {Springel}, V., \& {Hernquist}, L. 2013, \mnras, 429, 3353

\bibitem[{{Newman et al.}(2012)}]{2012ApJ...761...43N}
{Newman et al.} 2012, \apj, 761, 43

\bibitem[{{Oppenheimer} \& {Dav{\'e}}(2006)}]{2006MNRAS.373.1265O}
{Oppenheimer}, B.~D., \& {Dav{\'e}}, R. 2006, \mnras, 373, 1265

\bibitem[{{Porter}(1985)}]{1985PhDT.........7P}
{Porter}, D.~H. 1985, PhD thesis, California Univ., Berkeley.

\bibitem[{{Power} {et~al.}(2014){Power}, {Read}, \&
  {Hobbs}}]{2014MNRAS.440.3243P}
{Power}, C., {Read}, J.~I., \& {Hobbs}, A. 2014, \mnras, 440, 3243

\bibitem[{{Price}(2008)}]{2008JCoPh.22710040P}
{Price}, D.~J. 2008, Journal of Computational Physics, 227, 10040

\bibitem[{{Price} \& {Monaghan}(2007)}]{2007MNRAS.374.1347P}
{Price}, D.~J., \& {Monaghan}, J.~J. 2007, \mnras, 374, 1347

\bibitem[{{Prochaska} {et~al.}(2013{\natexlab{a}}){Prochaska}, {Hennawi},
  {Lee}, {Cantalupo}, {Bovy}, {Djorgovski}, {Ellison}, {Lau}, {Martin},
  {Myers}, {Rubin}, \& {Simcoe}}]{2013ApJ...776..136P}
{Prochaska}, J.~X., {Hennawi}, J.~F., {Lee}, K.-G., {Cantalupo}, S., {Bovy},
  J., {Djorgovski}, S.~G., {Ellison}, S.~L., {Lau}, M.~W., {Martin}, C.~L.,
  {Myers}, A., {Rubin}, K.~H.~R., \& {Simcoe}, R.~A. 2013{\natexlab{a}}, \apj,
  776, 136

\bibitem[{{Prochaska} {et~al.}(2013{\natexlab{b}}){Prochaska}, {Hennawi}, \&
  {Simcoe}}]{2013ApJ...762L..19P}
{Prochaska}, J.~X., {Hennawi}, J.~F., \& {Simcoe}, R.~A. 2013{\natexlab{b}},
  \apjl, 762, L19

\bibitem[{{Prochaska} \& {Wolfe}(2009)}]{2009ApJ...696.1543P}
{Prochaska}, J.~X., \& {Wolfe}, A.~M. 2009, \apj, 696, 1543

\bibitem[{{Puchwein} \& {Springel}(2013)}]{2013MNRAS.428.2966P}
{Puchwein}, E., \& {Springel}, V. 2013, \mnras, 428, 2966

\bibitem[{{Rahmati} {et~al.}(2013){Rahmati}, {Pawlik}, {Raicevic}, \&
  {Schaye}}]{2013MNRAS.430.2427R}
{Rahmati}, A., {Pawlik}, A.~H., {Raicevic}, M., \& {Schaye}, J. 2013, \mnras,
  430, 2427

\bibitem[{{Rakic} {et~al.}(2012){Rakic}, {Schaye}, {Steidel}, \&
  {Rudie}}]{2012ApJ...751...94R}
{Rakic}, O., {Schaye}, J., {Steidel}, C.~C., \& {Rudie}, G.~C. 2012, \apj, 751,
  94

\bibitem[{{Rubin} {et~al.}(2014{\natexlab{a}}){Rubin}, {Hennawi}, {Prochaska},
  {Simcoe}, {Myers}, \& {Wingyee Lau}}]{2014arXiv1411.6016R}
{Rubin}, K.~H.~R., {Hennawi}, J.~F., {Prochaska}, J.~X., {Simcoe}, R.~A.,
  {Myers}, A., \& {Wingyee Lau}, M. 2014{\natexlab{a}}, arXiv:1411.6016

\bibitem[{{Rubin} {et~al.}(2012){Rubin}, {Prochaska}, {Koo}, \&
  {Phillips}}]{2012ApJ...747L..26R}
{Rubin}, K.~H.~R., {Prochaska}, J.~X., {Koo}, D.~C., \& {Phillips}, A.~C. 2012,
  \apjl, 747, L26

\bibitem[{{Rubin} {et~al.}(2014{\natexlab{b}}){Rubin}, {Prochaska}, {Koo},
  {Phillips}, {Martin}, \& {Winstrom}}]{2014ApJ...794..156R}
{Rubin}, K.~H.~R., {Prochaska}, J.~X., {Koo}, D.~C., {Phillips}, A.~C.,
  {Martin}, C.~L., \& {Winstrom}, L.~O. 2014{\natexlab{b}}, \apj, 794, 156

\bibitem[{{Rudie} {et~al.}(2012){Rudie}, {Steidel}, {Trainor}, {Rakic},
  {Bogosavljevi{\'c}}, {Pettini}, {Reddy}, {Shapley}, {Erb}, \&
  {Law}}]{2012ApJ...750...67R}
{Rudie}, G.~C., {Steidel}, C.~C., {Trainor}, R.~F., {Rakic}, O.,
  {Bogosavljevi{\'c}}, M., {Pettini}, M., {Reddy}, N., {Shapley}, A.~E., {Erb},
  D.~K., \& {Law}, D.~R. 2012, \apj, 750, 67

\bibitem[{{Rupke} \& {Veilleux}(2011)}]{2011ApJ...729L..27R}
{Rupke}, D.~S.~N., \& {Veilleux}, S. 2011, \apjl, 729, L27

\bibitem[{{Saitoh} \& {Makino}(2013)}]{2013ApJ...768...44S}
{Saitoh}, T.~R., \& {Makino}, J. 2013, \apj, 768, 44

\bibitem[{{Salem} \& {Bryan}(2014)}]{2014MNRAS.437.3312S}
{Salem}, M., \& {Bryan}, G.~L. 2014, \mnras, 437, 3312

\bibitem[{{Schaye}(2001)}]{2001ApJ...562L..95S}
{Schaye}, J. 2001, \apjl, 562, L95

\bibitem[{{Schaye} {et~al.}(2015){Schaye}, {Crain}, {Bower}, {Furlong},
  {Schaller}, {Theuns}, {Dalla Vecchia}, {Frenk}, {McCarthy}, {Helly},
  {Jenkins}, {Rosas-Guevara}, {White}, {Baes}, {Booth}, {Camps}, {Navarro},
  {Qu}, {Rahmati}, {Sawala}, {Thomas}, \& {Trayford}}]{2015MNRAS.446..521S}
{Schaye}, J., {Crain}, R.~A., {Bower}, R.~G., {Furlong}, M., {Schaller}, M.,
  {Theuns}, T., {Dalla Vecchia}, C., {Frenk}, C.~S., {McCarthy}, I.~G.,
  {Helly}, J.~C., {Jenkins}, A., {Rosas-Guevara}, Y.~M., {White}, S.~D.~M.,
  {Baes}, M., {Booth}, C.~M., {Camps}, P., {Navarro}, J.~F., {Qu}, Y.,
  {Rahmati}, A., {Sawala}, T., {Thomas}, P.~A., \& {Trayford}, J. 2015, \mnras,
  446, 521

\bibitem[{{Shapley} {et~al.}(2003){Shapley}, {Steidel}, {Pettini}, \&
  {Adelberger}}]{2003ApJ...588...65S}
{Shapley}, A.~E., {Steidel}, C.~C., {Pettini}, M., \& {Adelberger}, K.~L. 2003,
  \apj, 588, 65

\bibitem[{{Sharma} {et~al.}(2012){Sharma}, {McCourt}, {Quataert}, \&
  {Parrish}}]{2012MNRAS.420.3174S}
{Sharma}, P., {McCourt}, M., {Quataert}, E., \& {Parrish}, I.~J. 2012, \mnras,
  420, 3174

\bibitem[{{Shen} {et~al.}(2013){Shen}, {Madau}, {Guedes}, {Mayer}, {Prochaska},
  \& {Wadsley}}]{2013ApJ...765...89S}
{Shen}, S., {Madau}, P., {Guedes}, J., {Mayer}, L., {Prochaska}, J.~X., \&
  {Wadsley}, J. 2013, \apj, 765, 89

\bibitem[{{Shen} {et~al.}(2010){Shen}, {Wadsley}, \&
  {Stinson}}]{2010MNRAS.407.1581S}
{Shen}, S., {Wadsley}, J., \& {Stinson}, G. 2010, \mnras, 407, 1581

\bibitem[{{Sikora} {et~al.}(2007){Sikora}, {Stawarz}, \&
  {Lasota}}]{2007ApJ...658..815S}
{Sikora}, M., {Stawarz}, {\L}., \& {Lasota}, J.-P. 2007, \apj, 658, 815

\bibitem[{{Sommer-Larsen}(2006)}]{2006ApJ...644L...1S}
{Sommer-Larsen}, J. 2006, \apjl, 644, L1

\bibitem[{{Springel}(2005)}]{2005MNRAS.364.1105S}
{Springel}, V. 2005, \mnras, 364, 1105

\bibitem[{{Springel} \& {Hernquist}(2003{\natexlab{a}})}]{2003MNRAS.339..289S}
{Springel}, V., \& {Hernquist}, L. 2003{\natexlab{a}}, \mnras, 339, 289

\bibitem[{{Springel} \& {Hernquist}(2003{\natexlab{b}})}]{2003MNRAS.339..312S}
---. 2003{\natexlab{b}}, \mnras, 339, 312

\bibitem[{{Steidel et al.}(2010)}]{2010ApJ...717..289S}
{Steidel et al.} 2010, \apj, 717, 289

\bibitem[{{Stinson} {et~al.}(2006){Stinson}, {Seth}, {Katz}, {Wadsley},
  {Governato}, \& {Quinn}}]{2006MNRAS.373.1074S}
{Stinson}, G., {Seth}, A., {Katz}, N., {Wadsley}, J., {Governato}, F., \&
  {Quinn}, T. 2006, \mnras, 373, 1074

\bibitem[{{Stinson} {et~al.}(2013){Stinson}, {Brook}, {Macci{\`o}}, {Wadsley},
  {Quinn}, \& {Couchman}}]{2013MNRAS.428..129S}
{Stinson}, G.~S., {Brook}, C., {Macci{\`o}}, A.~V., {Wadsley}, J., {Quinn},
  T.~R., \& {Couchman}, H.~M.~P. 2013, \mnras, 428, 129

\bibitem[{{Stinson} {et~al.}(2007){Stinson}, {Dalcanton}, {Quinn}, {Kaufmann},
  \& {Wadsley}}]{2007ApJ...667..170S}
{Stinson}, G.~S., {Dalcanton}, J.~J., {Quinn}, T., {Kaufmann}, T., \&
  {Wadsley}, J. 2007, \apj, 667, 170

\bibitem[{{Sturm et al.}(2011)}]{2011ApJ...733L..16S}
{Sturm et al.} 2011, \apjl, 733, L16+

\bibitem[{{Thoul} \& {Weinberg}(1996)}]{1996ApJ...465..608T}
{Thoul}, A.~A., \& {Weinberg}, D.~H. 1996, \apj, 465, 608

\bibitem[{{Tinker} \& {Chen}(2008)}]{2008ApJ...679.1218T}
{Tinker}, J.~L., \& {Chen}, H.-W. 2008, \apj, 679, 1218

\bibitem[{{Torrey} {et~al.}(2014){Torrey}, {Vogelsberger}, {Genel}, {Sijacki},
  {Springel}, \& {Hernquist}}]{2014MNRAS.tmp...38T}
{Torrey}, P., {Vogelsberger}, M., {Genel}, S., {Sijacki}, D., {Springel}, V.,
  \& {Hernquist}, L. 2014, \mnras

\bibitem[{{Trainor} \& {Steidel}(2012)}]{2012ApJ...752...39T}
{Trainor}, R.~F., \& {Steidel}, C.~C. 2012, \apj, 752, 39

\bibitem[{{Tumlinson} {et~al.}(2011){Tumlinson}, {Thom}, {Werk}, {Prochaska},
  {Tripp}, {Weinberg}, {Peeples}, {O'Meara}, {Oppenheimer}, {Meiring}, {Katz},
  {Dav{\'e}}, {Ford}, \& {Sembach}}]{2011Sci...334..948T}
{Tumlinson}, J., {Thom}, C., {Werk}, J.~K., {Prochaska}, J.~X., {Tripp}, T.~M.,
  {Weinberg}, D.~H., {Peeples}, M.~S., {O'Meara}, J.~M., {Oppenheimer}, B.~D.,
  {Meiring}, J.~D., {Katz}, N.~S., {Dav{\'e}}, R., {Ford}, A.~B., \& {Sembach},
  K.~R. 2011, Science, 334, 948

\bibitem[{{Turner} {et~al.}(2014){Turner}, {Schaye}, {Steidel}, {Rudie}, \&
  {Strom}}]{2014arXiv1403.0942T}
{Turner}, M.~L., {Schaye}, J., {Steidel}, C.~C., {Rudie}, G.~C., \& {Strom},
  A.~L. 2014, arXiv:1410.8214

\bibitem[{{Urry} \& {Padovani}(1995)}]{1995PASP..107..803U}
{Urry}, C.~M., \& {Padovani}, P. 1995, \pasp, 107, 803

\bibitem[{{van Daalen} {et~al.}(2011){van Daalen}, {Schaye}, {Booth}, \& {Dalla
  Vecchia}}]{2011MNRAS.415.3649V}
{van Daalen}, M.~P., {Schaye}, J., {Booth}, C.~M., \& {Dalla Vecchia}, C. 2011,
  \mnras, 415, 3649

\bibitem[{{van de Voort} {et~al.}(2012){van de Voort}, {Schaye}, {Altay}, \&
  {Theuns}}]{2012MNRAS.421.2809V}
{van de Voort}, F., {Schaye}, J., {Altay}, G., \& {Theuns}, T. 2012, \mnras,
  421, 2809

\bibitem[{{Vogelsberger} {et~al.}(2013){Vogelsberger}, {Genel}, {Sijacki},
  {Torrey}, {Springel}, \& {Hernquist}}]{2013MNRAS.436.3031V}
{Vogelsberger}, M., {Genel}, S., {Sijacki}, D., {Torrey}, P., {Springel}, V.,
  \& {Hernquist}, L. 2013, \mnras, 436, 3031

\bibitem[{{Vogelsberger} {et~al.}(2014){Vogelsberger}, {Genel}, {Springel},
  {Torrey}, {Sijacki}, {Xu}, {Snyder}, {Nelson}, \&
  {Hernquist}}]{2014MNRAS.444.1518V}
{Vogelsberger}, M., {Genel}, S., {Springel}, V., {Torrey}, P., {Sijacki}, D.,
  {Xu}, D., {Snyder}, G., {Nelson}, D., \& {Hernquist}, L. 2014, \mnras, 444,
  1518

\bibitem[{{Weiner} {et~al.}(2009){Weiner}, {Coil}, {Prochaska}, {Newman},
  {Cooper}, {Bundy}, {Conselice}, {Dutton}, {Faber}, {Koo}, {Lotz}, {Rieke}, \&
  {Rubin}}]{2009ApJ...692..187W}
{Weiner}, B.~J., {Coil}, A.~L., {Prochaska}, J.~X., {Newman}, J.~A., {Cooper},
  M.~C., {Bundy}, K., {Conselice}, C.~J., {Dutton}, A.~A., {Faber}, S.~M.,
  {Koo}, D.~C., {Lotz}, J.~M., {Rieke}, G.~H., \& {Rubin}, K.~H.~R. 2009, \apj,
  692, 187

\bibitem[{{White et al.}(2012)}]{2012MNRAS.424..933W}
{White et al.} 2012, \mnras, 424, 933

\end{thebibliography}

\appendix

\section{Convergence with resolution and artificial conductivity}
\label{sec:convergence}
\begin{figure*}
\begin{center}
\mbox{
\includegraphics[width=0.5\textwidth]{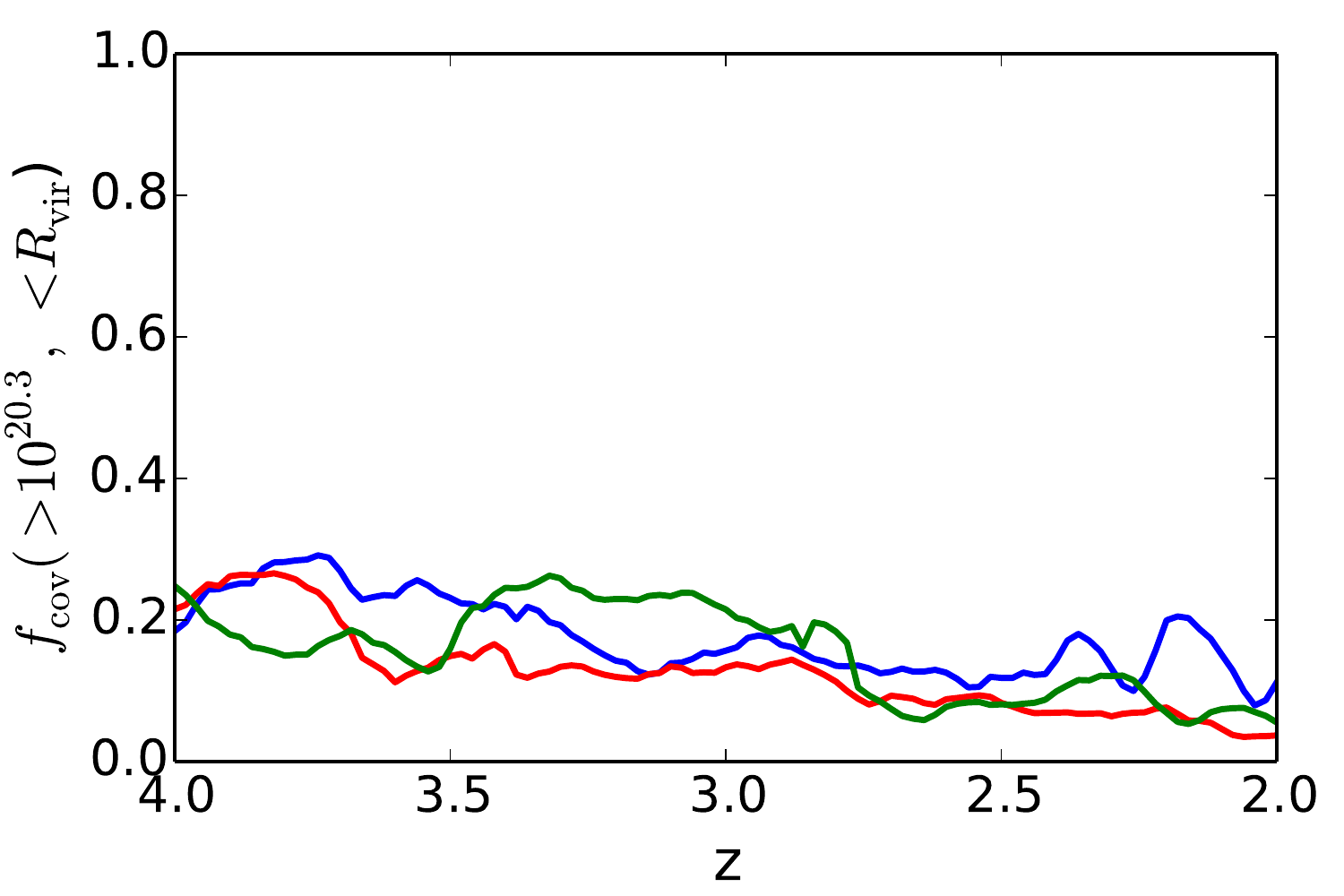}
\includegraphics[width=0.5\textwidth]{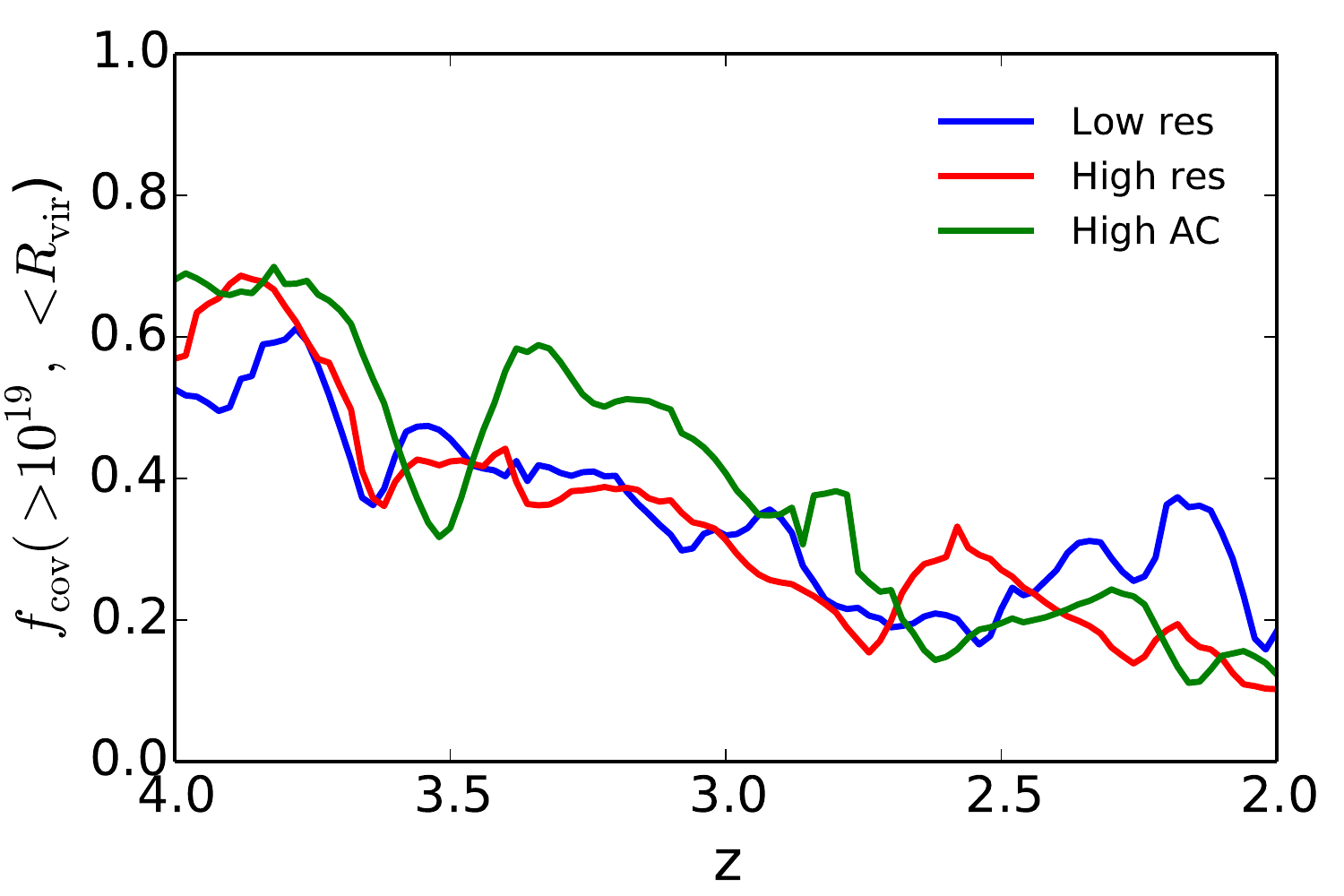}
}
\mbox{
\includegraphics[width=0.5\textwidth]{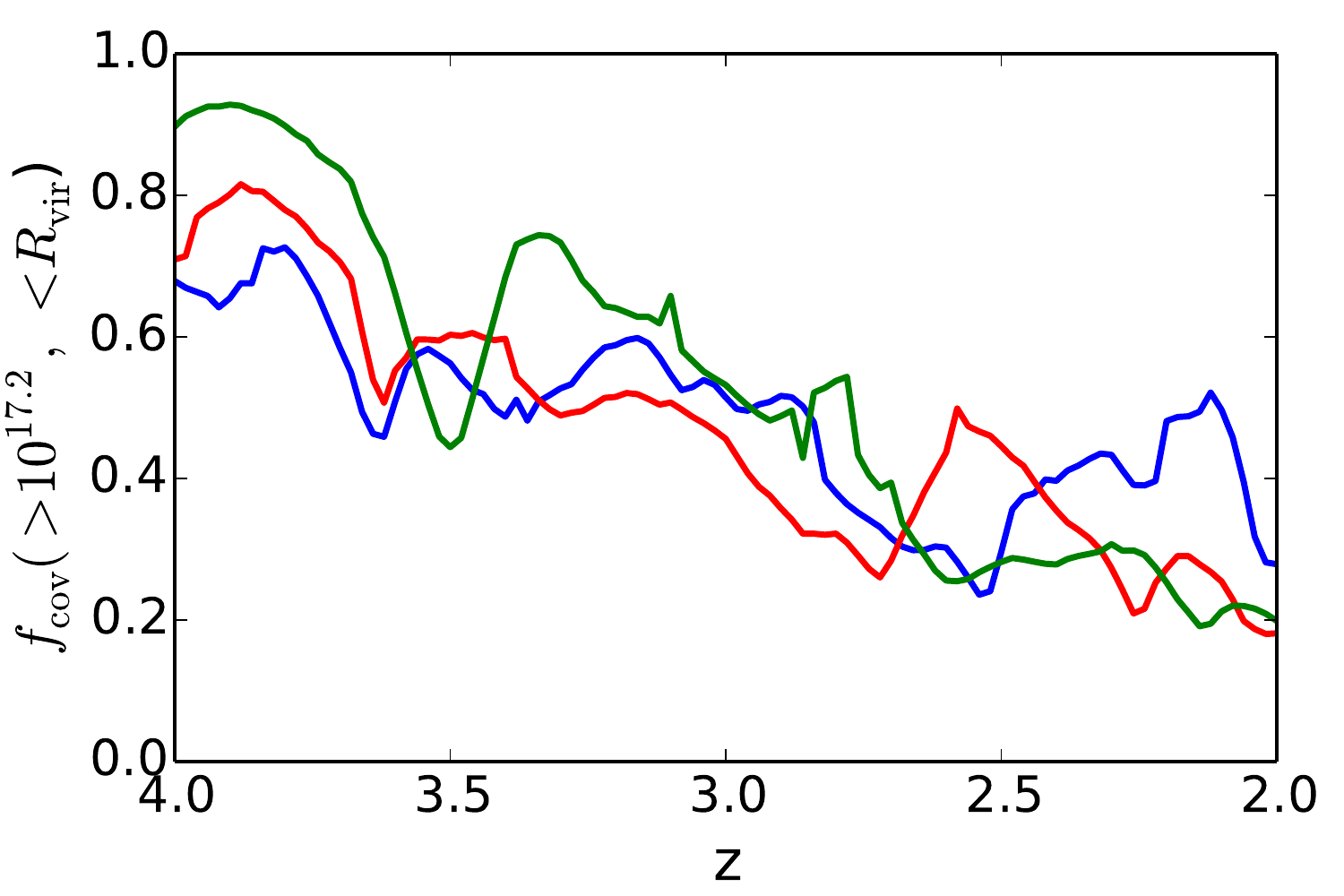}
\includegraphics[width=0.5\textwidth]{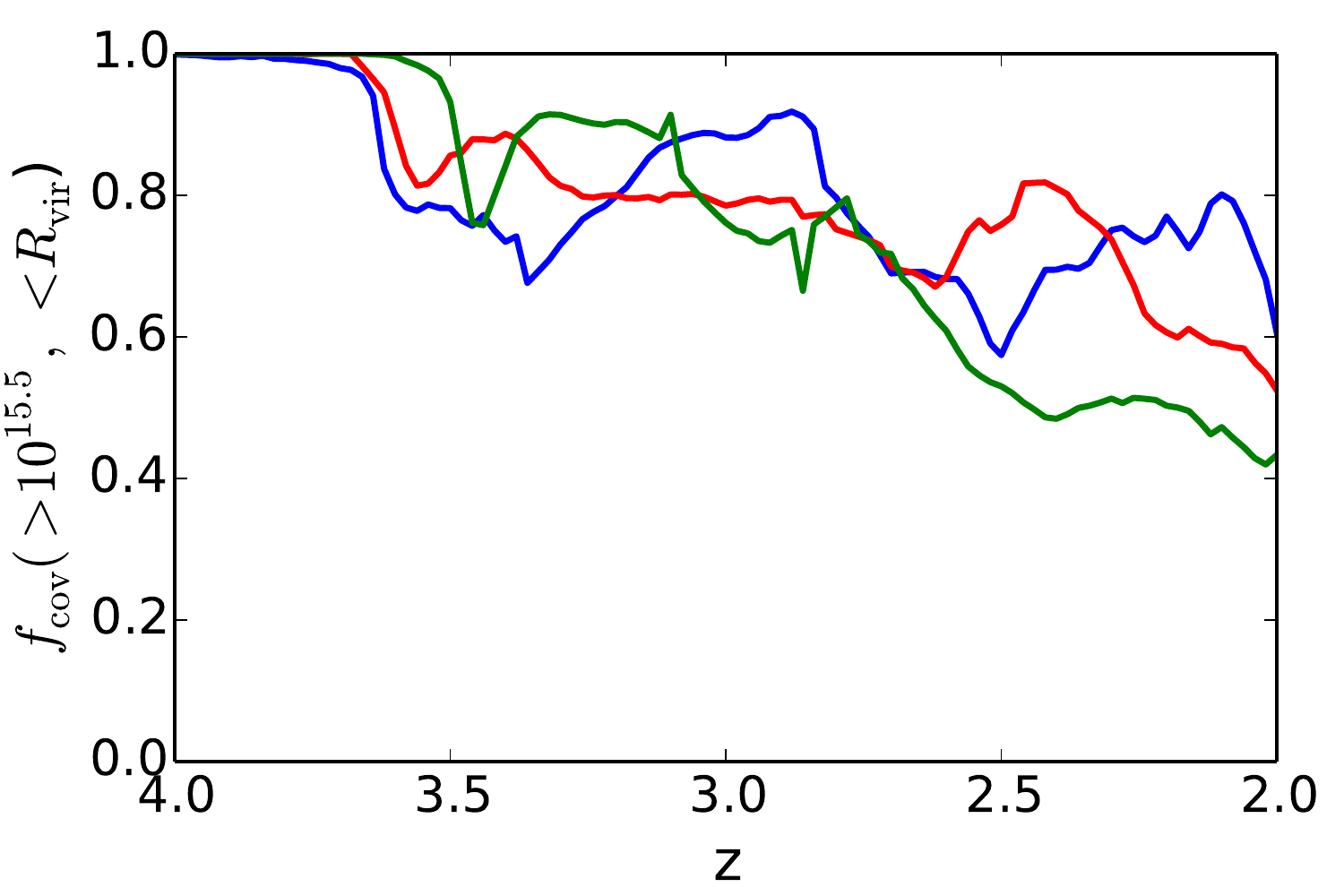}
}
\end{center}
\caption[]{Convergence tests for covering fractions with respect to simulation resolution and artificial conductivity normalization. \emph{Red:} Predicted covering fractions from $z=4$ to $z=2$ for our standard (`high res') {\bf m12i} simulation (Table \ref{tbl:sims_Hopkins}). 
\emph{Blue:} Same but for a realization of {\bf m12i} with 8 times coarser gas mass resolution and 5 times larger minimum gas softening length (`low res'). 
\emph{Green:} Same but for a standard resolution {\bf m12i} realization with normalization of the artificial conductivity increased by a factor of four (`high AC'). 
Averaging over stochastic time fluctuations, the agreement between the three runs is very good. 
For clarity, we show covering fraction for only one sky projection for each simulation. 
}
\label{fig:m12i_convergence} 
\end{figure*}
We test the convergence of our HI covering fractions with respect to resolution of the SPH calculation using two realizations of the {\bf m12i} halo. 
We label the standard {\bf m12i} simulation in Table \ref{tbl:sims_Hopkins} `high res' and consider a lower resolution `low res' version. The `low res' version has a gas particle mass 8 times larger and a minimum gas softening length 5 times larger than the `high res' version. 
We also ran a `high res' simulation with normalization of the artificial conductivity (entropy mixing parameter) increased by a factor of four, but otherwise identical parameters (`high AC'). 
As the maps show, the dense HI gas in source simulations is often clumpy so it is important to test that the clumps are not evaporated when the artificial conductivity is increased. 
Figure \ref{fig:m12i_convergence} shows the HI covering fractions within $R_{\rm vir}$ as a function of redshift for the four different ranges of $N_{\rm HI}$ studied in this paper. 
In a statistical sense (i.e., after averaging over stochastic time fluctuations), the agreement between the `high res', `low res', and `high AC' runs is very good. 

\begin{figure*}
\begin{center}
\includegraphics[width=0.99\textwidth]{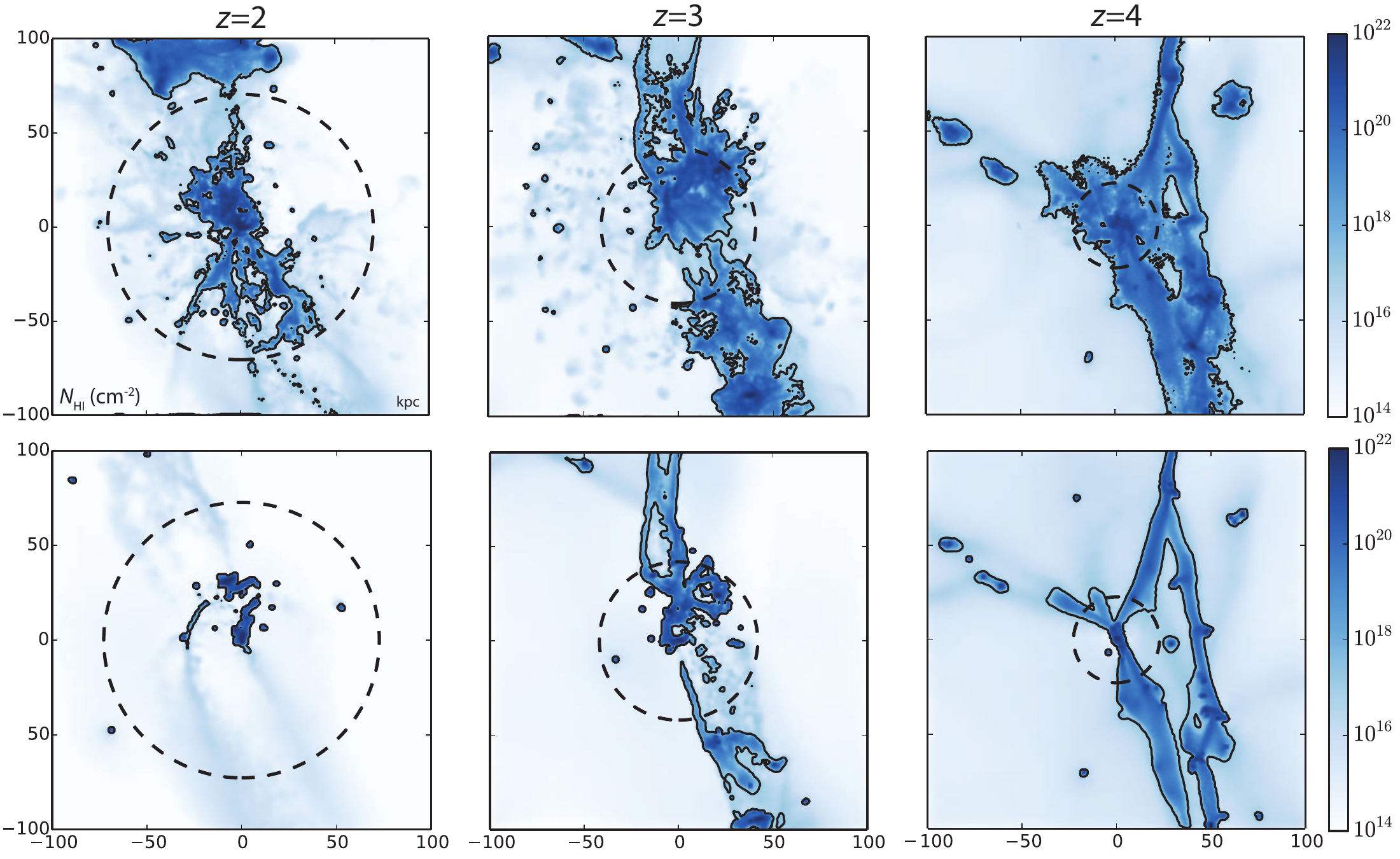}
\end{center}
\caption[]{\emph{Top:} HI maps for our {\bf m12i} simulation with full stellar feedback at $z=2,~3,~{\rm and}~4$. 
\emph{Bottom:} Simulation from the same initial conditions but with a sub-resolution ISM model and no galactic winds. 
The virial radius of the halo is indicated in each panel by the dashed circles and 
Lyman limit systems ($N_{\rm HI}>10^{17.2}$ cm$^{-2}$) are indicated by solid contours. 
Stellar feedback increases the HI covering fractions in galaxy halos both by directly ejecting cool gas from galaxies and through the interaction of galactic winds with cosmological inflows. 
At $z=2$, LLSs and DLAs ($N_{\rm HI}>2\times10^{20}$ cm$^{-2}$) in this example are almost exclusively restricted to galaxies and their immediate vicinity but with stellar feedback areas covered by LLSs and DLAs are enhanced owing to galactic winds. Length scales are consistent across rows and columns.
}
\label{fig:m12i_nofbk_fbk} 
\end{figure*}

\section{Effects of feedback}
\label{sec:feedback_effects} 
Figure \ref{fig:m12i_nofbk_fbk} compares our ${\bf m12i}$ simulation with full stellar feedback with a simulation from the same initial conditions and SPH implementation but with the \cite{2003MNRAS.339..289S} sub-resolution ISM model and no galactic winds. 
Runs without feedback become very computationally expensive as a large stellar mass accumulates at high density, so our no-feedback run has a gas mass resolution lower by a factor of 8 and a gas softening length of 570 proper pc. 
The comparison is nevertheless meaningful because the ISM physics converges at much lower resolution with the \cite{2003MNRAS.339..289S} ISM model. 
Stellar feedback increases the HI covering fractions in galaxy halos both by directly ejecting cool gas from galaxies and through the interaction of galactic winds with cosmological inflows. 
At $z=2$, LLSs and DLAs in this example of a no feedback simulation are almost exclusively restricted to galaxies because the dense cold filaments infalling from the IGM at $z=4$ and $z=3$ have disappeared as virial shocks are no longer cooling rapidly relative to gas infall \citep[e.g.,][]{2003MNRAS.345..349B, 2005MNRAS.363....2K, 2009MNRAS.395..160K, 2011MNRAS.417.2982F, 2013MNRAS.429.3353N}. 
With stellar feedback, areas covered by LLSs and DLAs are increased owing to galactic winds. 
 
\end{document}